\documentclass[prd,twocolumn,showpacs,superscriptaddress,nofootinbib,floatfix,showkeys,10pt]{revtex4-2}

\usepackage{graphicx}
\usepackage{amsmath}
\usepackage{bm}
\usepackage{yhmath}
\usepackage{mathtools}
\usepackage{wasysym}
\usepackage{color,xcolor}
\usepackage{subfigure}
\usepackage{cases}
\usepackage{subfigure}
\usepackage{times}
\usepackage{dcolumn,booktabs,bm}
\usepackage{slashed}
\usepackage{amsfonts,amssymb,stmaryrd,latexsym}
\usepackage{mathrsfs}
\usepackage{textcomp}
\usepackage{multirow}
\usepackage{cancel}
\usepackage{array}
\usepackage{enumerate,enumitem}
\usepackage{orcidlink}
\usepackage{dashrule}
\usepackage{multirow}
\usepackage{times}
\usepackage{mdframed}
\usepackage{hyperref}

\definecolor{mycolor}{RGB}{45,48,146}
\hypersetup{colorlinks=true,citecolor=mycolor,urlcolor=mycolor,linkcolor=mycolor}

\allowdisplaybreaks[4]
\newcounter{cnt}
\makeatletter
\let\oldhypertarget\hypertarget
\renewcommand{\hypertarget}[2]{%
  \oldhypertarget{#1}{#2}%
    \protected@write\@mainaux{}{%
        \string\expandafter\string\gdef
          \string\csname\string\detokenize{#1}\string\endcsname{#2}%
    }%
  }
\newcommand{\myhyperlink}[1]{%
  \hyperlink{#1}{\csname #1\endcsname}%
  }
\makeatother

\newcommand{\etal}{\textit{et al}.}
\newcommand{\heavy}{\textit{heavy}~}

\begin{document}

\title{Spectrum of the molecular hexaquarks}

\author{Bo Wang\,\orcidlink{0000-0003-0985-2958}}
\affiliation{College of Physics Science and Technology, Hebei University, Baoding, 071002, China}
\affiliation{Hebei Key Laboratory of High-precision Computation and Application of Quantum Field Theory, Baoding, 071002, China}
\affiliation{Hebei Research Center of the Basic Discipline for Computational Physics, Baoding, 071002, China}

\author{Kan Chen\,\orcidlink{0000-0002-1435-6564}}\email{chenk10@nwu.edu.cn}
\affiliation{School of Physics, Northwest University, Xi’an, 710127, China}

\author{Lu Meng\,\orcidlink{0000-0001-9791-7138}}\email{lu.meng@rub.de}
\affiliation{ Institut f\"ur Theoretische Physik II, Ruhr-Universit\"at Bochum,  D-44780 Bochum, Germany}

\author{Shi-Lin Zhu\,\orcidlink{0000-0002-4055-6906}}\email{zhusl@pku.edu.cn}
\affiliation{School of Physics and Center of High Energy Physics, Peking University, Beijing, 100871, China}

\begin{abstract}
We investigate the mass spectra of molecular-type hexaquark states in the dibaryon systems. These systems are composed of the charmed baryons $[\Sigma_c^{(\ast)}$, $\Xi_c^{(\prime,\ast)}]$, doubly charmed baryons $[\Xi_{cc}^{(\ast)}]$, and hyperons $[\Sigma^{(\ast)}$, $\Xi^{(\ast)}]$. We consider all possible combinations of particle-particle and particle-antiparticle pairs, including the S-wave spin multiplets in each combination. We establish the underlying connections among the molecular tetraquarks, pentaquarks, and hexaquarks with the effective quark-level interactions. We find that the existence of molecular states in $DD^\ast$, $D\bar{D}^\ast$, and $\Sigma_c\bar{D}^{(\ast)}$ systems leads to the emergence of a large number of deuteron-like hexaquarks in the heavy flavor sectors. 
Currently, there have been several experimental candidates for molecular tetraquarks and pentaquarks. 
The experimental search for near-threshold hexaquarks will further advance the establishment of the underlying dynamical picture of hadronic molecules and deepen our understanding of the role of spin-flavor symmetry in near-threshold residual strong interactions.
\end{abstract}

\maketitle

\section{Introduction}\label{sec:intro}

The nature manifests an elaborately organized hierarchical system. The matter world is not simply formed by the mere agglomeration of elementary particles, but rather comprises distinct structural units that are effective at each level, which is exemplified by the progression from quarks to hadrons, from nucleons to atomic nuclei, and from atoms to molecules. Analogous to the formation of molecules via residual electromagnetic interactions between electrically neutral atoms, which constitute our diverse and vibrant daily matter world, the color-neutral hadrons can also form hadronic molecules through residual strong interactions, establishing novel matter structures in the quantum chromodynamics (QCD) world. The most archetypal example is the deuteron, a molecular configuration formed by the proton and neutron via the nuclear force (residual strong interaction). Over the past two decades, a series of ground-breaking discoveries in experiments~\cite{Workman:2022ynf} have significantly facilitated the cross-fertilization between nuclear physics and hadron physics, leading to the growing recognition that deuteron-like hadronic molecules should be widespread~\cite{Chen:2016qju,Guo:2017jvc,Liu:2019zoy,Lebed:2016hpi,Esposito:2016noz,Brambilla:2019esw,Yang:2020atz,Chen:2021ftn,Chen:2022asf,Meng:2022ozq,Liu:2024uxn}.

In the past two decades, the field of hadron physics has witnessed a flourishing growth in both experimental investigations and theoretical studies. One significant factor contributing to this progress is the abundant discovery of novel (exotic) hadronic states near threshold energies. The peculiar hadrons such as the XYZ states~\cite{Belle:2003nnu,BaBar:2005hhc,BESIII:2013ris,BESIII:2013ouc,Belle:2011aa}, hidden-charm pentaquarks~\cite{LHCb:2019kea,LHCb:2020jpq,LHCb:2022ogu}, doubly charmed tetraquarks~\cite{LHCb:2021vvq,LHCb:2021auc}, and charmed strange tetraquarks~\cite{LHCb:2020bls,LHCb:2020pxc,LHCb:2022sfr,LHCb:2022lzp} have emerged as prominent examples. Theorists have extensively studied the mass spectra, decays, production mechanisms, and electromagnetic properties of these states~\cite{Chen:2016qju,Guo:2017jvc,Liu:2019zoy,Lebed:2016hpi,Esposito:2016noz,Brambilla:2019esw,Yang:2020atz,Chen:2021ftn,Chen:2022asf,Meng:2022ozq}. However, due to limited experimental data and the reliance on specific models, the nature of these states still lacks a definitive consensus. Current theoretical interpretations include hadronic molecular states, kinematic effects, and compact multi-quark states. The molecular picture is gaining increasing recognition of the  communities, as the aforementioned exotic states are very close to the threshold of a pair of conventional hadrons. The weaker residual strong interaction ensures that the corresponding hadronic molecules naturally reside near the threshold, similar to the deuteron with its mass just about $2.2$ MeV below the two-nucleon threshold. 

These experimental discoveries indicate that the hadronic molecules are not exclusive to the nucleon-nucleon systems alone, while there are intricate processes and interactions even in the heavy flavor region that give rise to the emergence of higher-level structures and entities---the heavy flavor hadronic molecules.~The experimental evidences suggest that these hadronic molecule candidates should not be regarded as individual entities but rather tightly interconnected through underlying interactions, such as the tetraquark state $T_{cc}$~\cite{LHCb:2021vvq,LHCb:2021auc}, the pentaquark states $P_\psi^N$~\cite{LHCb:2019kea} and $P_{\psi s}^\Lambda$~\cite{LHCb:2020jpq,LHCb:2022ogu} were observed in the lowest isospin channels.

In Ref.~\cite{Wang:2023hpp}, our quantitative analysis shows that for the residual strong interaction near the threshold energy, its strength is not sufficient to excite the strange quarks inside the hadrons, so the strange quarks can also be approximately treated as heavy quarks. In this paper, we will use \heavy to collectively refer to the systems containing $c$ and/or $s$ quarks. We proposed a concise model based on quark-level interactions, which enables us to effectively interpret the recently observed $T_{cs0}(2900)$ and $T_{c\bar{s}0}^a(2900)$ as the molecular states of $\bar{D}^\ast K^\ast$ and $D^\ast K^\ast$, respectively~\cite{Wang:2023hpp}. Furthermore, we find that they can be respectively associated with the $T_{cc}$ and $Z_c$ states under the \heavy quark symmetry. This provides insights into the underlying symmetries and relationships between different types of exotic hadrons. Subsequently, we calculated the mass spectra of various possible \heavy pentaquarks and naturally explained why the $P_{\psi s}^\Lambda(4338)$ is located very close to the $\Xi_c\bar{D}$ threshold~\cite{Wang:2023eng}. In this work, we will further investigate the mass spectra of possible \heavy molecular hexaquark states. Below, we will briefly review the research status of hexaquarks.

At the beginning of the quark model, Dyson \etal~predicted several dibaryon states based on the SU(6) symmetry, including a $\Delta\Delta$ resonance with isospin-$0$ and spin-$3$, with a mass of approximately $2376$ MeV~\cite{Dyson:1964xwa}. This state was considered to possibly correspond to the $d^\ast(2380)$~\cite{Gal:2013dca} observed experimentally around $2010$~\cite{Bashkanov:2008ih,WASA-at-COSY:2011bjg}. Jaffe predicted the existence of a dihyperon composed of $\Lambda\Lambda$, as well as possible bound states of $\Sigma\Sigma$~\cite{Jaffe:1976yi}. Numerous studies have been conducted to calculate the properties of these configurations~\cite{Balachandran:1983dj,Mackenzie:1985vv,Rosner:1985yh,Polinder:2007mp,NPLQCD:2010ocs,Inoue:2010es,Li:2018tbt,Chen:2019vdh,Clement:2020mab,ALICE:2018ysd,ALICE:2019eol}, as well as the nucleon-hyperon systems~\cite{Oka:1986fr,Haidenbauer:2013oca,Haidenbauer:2019boi,Gal:2016boi,Ren:2019qow}, nucleon-charmed baryon systems~\cite{Meng:2019nzy,Liu:2011xc,Huang:2013zva,Miyamoto:2017tjs,Haidenbauer:2017dua,Maeda:2018xcl,Garcilazo:2019ryw,Song:2020isu}, and nucleon-doubly charmed baryon systems~\cite{Meng:2017udf}.

Inspired by the recent experimental observation of a large number of hadronic molecule candidates, a significant amount of research has been devoted to investigating possible molecular-type hexaquark states in heavy-heavy systems. For example, Lee \etal~investigated potential molecular states in systems composed of two charmed baryons using the one-boson exchange (OBE) model~\cite{Lee:2011rka}. Their results indicated the absence of bound states in the $\Lambda_c\Lambda_c$ system, while molecular states were found to exist in the $\Xi_c\Xi_c$ and $\Xi_c^\prime\Xi_c^\prime$ systems. This line of research has been further extended to the systems containing $b$ quarks~\cite{Li:2012bt}. Other research works, such as those based on the OBE model~\cite{Meguro:2011nr,Chen:2017vai,Yang:2018amd,Pan:2020xek,Ling:2021asz,Dong:2021bvy,Cheng:2022vgy,Kong:2022rvd,Kong:2022rvd,Song:2022svi,Kong:2023dwz,Wu:2024trh}, quark potential models~\cite{Huang:2013rla,Carames:2015sya,Vijande:2016nzk}, effective field theories~\cite{Chen:2013sba,Lu:2017dvm,Chen:2021cfl,Chen:2021spf,Chen:2022iil,Chen:2022wkh,Chen:2024tuu}, QCD sum rules~\cite{Chen:2016ymy,Wan:2019ake,Wang:2019gal,Wang:2021qmn}, chromomagnetic interaction model~\cite{Liu:2021gva,Liu:2022rzu}, and lattice QCD calculations~\cite{Liu:2022gxf}, have also extensively studied similar systems. In addition, Meng and Yang \etal~have also studied the possible molecular states composed of two doubly charmed baryons~\cite{Meng:2017fwb,Yang:2019rgw}.

As a natural extension of our previous two research works~\cite{Wang:2023hpp,Wang:2023eng}, in this paper, we will further systematically investigate possible molecular states in all systems formed by the combinations of hyperons, charmed baryons, and doubly charmed baryons (including both the particle-particle and particle-antiparticle combinations). See Table~\ref{tab:systems} for a list of the considered systems.

The structure of this article is arranged as follows: In Sec.~\ref{sec:effp}, we will present the effective potentials at the quark level, the definition of parity for the neutral systems, and the systems considered in this paper. In Sec.~\ref{sec:spec}, we will provide the mass spectra of various possible molecular hexaquarks, and conduct a detailed comparison and discussion with existing results in the aforementioned works. In Sec.~\ref{sec:sum}, we will summarize and provide an outlook for this work.

\section{Framework}\label{sec:effp}

\subsection{Effective potentials for the S-wave dihadron systems}

In Ref.~\cite{Wang:2023hpp}, we utilized the basic ideas of the one-boson exchange model and effective field theory to construct the near-threshold interacting potentials for the S-wave dihadron systems.~Compared to conventional approaches based on the hadronic degrees of freedom, starting from the quark level enables us to establish connections between different {\it heavy}-flavor dihadron systems.~We also noticed that for the near-threshold two-body interactions, the involvement of strange quarks inside hadrons is actually very limited. Therefore, we can further extend the range of heavy quarks to include strange quarks.~This allows us to clearly clarify the role of spin-flavor symmetry in hadronic molecular states, and describe the mass spectra of different molecular states using a few parameters. In our framework, the near-threshold interactions in {\it heavy}-flavor dihadron systems is primarily dominated by the correlations of $q_1q_2$ and $\bar{q}_1q_2$, where $q = u, d$, and $q_1$ and $q_2$ belong to different {\it heavy}-flavor hadrons. The nonrelativistic effective potentials describing the $q_1q_2$ and $\bar{q}_1q_2$ correlations are respectively given as follows~\cite{Wang:2023hpp}:
\begin{eqnarray}
V_{qq}&=&\left(\bm{\tau}_{1}\cdot\bm{\tau}_{2}+\frac{1}{2}\tau_{0,1}\cdot\tau_{0,2}\right)(c_{s}+c_{a}\bm{\sigma}_{1}\cdot\bm{\sigma}_{2}),\label{eq:Vqq}\\
V_{\bar{q}q}&=&\left(-\bm{\tau}_{1}^\ast\cdot\bm{\tau}_{2}+\frac{1}{2}\tau_{0,1}\cdot\tau_{0,2}\right)(\tilde{c}_{s}+\tilde{c}_{a}\bm{\sigma}_{1}\cdot\bm{\sigma}_{2}),\label{eq:Vqqbar}
\end{eqnarray}
where $\bm{\tau} (\bm{\sigma})$ are the Pauli matrices denoting the isospin (spin) operators of light quarks, while $\tau_{0}$ is the $2\times2$ identity matrix. The four terms in Eqs.~\eqref{eq:Vqq} and~\eqref{eq:Vqqbar} incorporate the isospin unrelated and related central potentials and spin-spin couplings. The low-energy constants $c_s$ ($\tilde{c}_{s}$) and $c_a$ ($\tilde{c}_{a}$) are determined with the experimentally well-established states in the molecular scenario. For example, using the masses of $P_\psi^N$ and $T_{cc}$ states as the input, we obtained the values of $c_s$ and $c_a$ as
\begin{eqnarray}
c_{s}&=&146.4\pm10.8\textrm{ GeV}^{-2},\nonumber\\
c_{a}&=&-7.3\pm10.5\textrm{ GeV}^{-2},
\end{eqnarray}
while utilizing the bound state and virtual state pictures respectively for $X(3872)$ and $Z_c(3900)$ gives the ranges of $\tilde{c}_{s}$ and $\tilde{c}_{a}$ as
\begin{eqnarray}
184.3\textrm{ GeV}^{-2}<&\tilde{c}_{a}+\tilde{c}_{s}&<187.5\textrm{ GeV}^{-2},\nonumber\\
78.1\textrm{ GeV}^{-2}<&\tilde{c}_{a}-\tilde{c}_{s}&<180.3\textrm{ GeV}^{-2}.
\end{eqnarray}

We ultimately need to translate the quark-level effective potentials to the hadron level. Here, we take $V_{qq}$ as an example, and the translation of $V_{\bar{q}q}$ can be done using the same form.
\begin{eqnarray}\label{eq:Vsys}
V_{\mathcal{H}_1\mathcal{H}_2}^{I,J}&=&\left\langle [\mathcal{H}_1\mathcal{H}_2]_J^I\left|V_{qq}\right|[\mathcal{H}_1\mathcal{H}_2]_J^I \right\rangle,
\end{eqnarray}
where $[\mathcal{H}_1\mathcal{H}_2]_J^I$ represents the spin-flavor wave function of the dihadron system $\mathcal{H}_1\mathcal{H}_2$ with isospin $I$ and spin $J$. These wave functions are expanded in terms of the spin-flavor wave functions of individual hadrons $\mathcal{H}_1$ and $\mathcal{H}_2$. For more specific details, we refer to Refs.~\cite{Chen:2022wkh,Chen:2024tuu}.

With the hadron-level effective potentials in hand, we can insert them into the following Lippmann-Shiwinger (LS) equation to solve for the existence of bound or virtual states in dihadron systems with specific quantum numbers,
\begin{eqnarray}\label{eq:intLSE}
t=v+vGt,
\end{eqnarray}
where $v$ is the $V_{\mathcal{H}_1\mathcal{H}_2}^{I,J}$ obtained in Eq.~\eqref{eq:Vsys}. The scattering information is encoded in the T-matrix $t$. The nonrelativistic form of the two-body propagator $G$ is adopted,
\begin{eqnarray}\label{eq:Ge}
  G(E+i\epsilon) &=& \int_{0}^{\Lambda}\frac{k^{2}dk}{(2\pi)^{3}}\frac{2\mu}{p^{2}-k^{2}+i\epsilon}\nonumber\\
    &=& \frac{2\mu}{(2\pi)^{3}}\left[p\tanh^{-1}\left(\frac{p}{\Lambda}\right)-\Lambda-\frac{i\pi}{2}p\right],
\end{eqnarray}
where we use a sharp cutoff $\Lambda$ to regularize the divergent integral. The momentum $p$ is related to the center of mass energy $E$ via $p=[2\mu(E-m_{\text{th}})]^{1/2}$, in which $\mu$ and $m_{\text{th}}$ denote the reduced mass and threshold of the dihadron $\mathcal{H}_1\mathcal{H}_2$. To maintain consistency with Refs.~\cite{Wang:2023hpp,Wang:2023eng}, in this calculation, the cutoff value for $\Lambda$ will also be taken as $0.4$ GeV.

For the constant potential $v$, the integral equation~\eqref{eq:intLSE} can be further simplified into the following algebraic equation,
\begin{eqnarray}\label{eq:invt}
  t^{-1} &=& v^{-1}-G.
\end{eqnarray}
For single-channel scattering, the branch cut in Eq.~\eqref{eq:Ge} gives rise to two Riemann sheets, known as the physical sheet and the unphysical sheet. Bound and virtual states correspond to the below-threshold poles of the T-matrix locating at the real axis of the physical and unphysical Riemann sheets, respectively. These poles in $t$ correspond to the zeros of $t^{-1}$. The physical and unphysical Riemann sheets can be reached by the following substitution of the $G$ function,
\begin{eqnarray}
\text{Physical}&:& G(E+i\epsilon),\nonumber\\
\text{Unphysical}&:& G(E+i\epsilon)+i\frac{\mu}{4\pi^2} p,
\end{eqnarray}
where the $G(E+i\epsilon)$ is given by Eq.~\eqref{eq:Ge}.

\subsection{Parities of the neutral systems}

We will encounter three types of neutral systems when studying the particle-antiparticle combinations, for which we can define their $C$ parity.
\begin{itemize}
  \item [(i)] The neutral systems composed of a particle and its own antiparticle, such as the $\Sigma_c\bar{\Sigma}_c$ and $\Sigma_c^\ast\bar{\Sigma}_c^\ast$, etc. Their $P$, $C$ and $G$ parities are conventionally defined as 
  \begin{eqnarray}
    P&=&(-1)^{L+1},\label{eq:Pparity}\\
    C&=&(-1)^{L+S},\\
    G&=&(-1)^{L+S+I},
  \end{eqnarray}
  where $L$, $S$ and $I$ represent the orbital angular momentum, spin and isospin of the dihadron systems $\mathcal{H}_1\mathcal{H}_2$. For the S-wave case ($L=0$), the total angular momentum $J$ equals to $S$.
  \item [(ii)] The neutral systems composed of two particles with different spins, such as the $\Sigma_c\bar{\Sigma}_c^\ast$, $\Xi_c^{(\prime)}\bar{\Xi}_c^\ast$, etc. Their $P$ parity is still given by Eq.~\eqref{eq:Pparity}, while the $C$ parity depends on the relative phase factor $\eta$ in the wave function, as well as the convention for the transformation of the spin-$\frac{3}{2}$ field under charge conjugation. We take the $\Sigma_c\bar{\Sigma}_c^\ast$ as an example, and its wave function is given by
  \begin{eqnarray}
    |\Sigma_c\bar{\Sigma}_c^\ast\rangle&=&\frac{1}{\sqrt{2}}\left[\Sigma_{c}\bar{\Sigma}_{c}^{\ast}+\eta \bar{\Sigma}_{c}\Sigma_{c}^{\ast}\right].
  \end{eqnarray}
  We adopt the following convention
  \begin{eqnarray}
  \hat{C}\Sigma_{c}^{\ast}&=&\bar{\Sigma}_{c}^{\ast},
  \end{eqnarray}
  which is similar to the one used in Ref.~\cite{Thomas:2008ja}, but differs from the convention in Ref.~\cite{Lu:2017dvm} by a negative sign. Then the $C$ and $G$ parities for such systems will be defined as
   \begin{eqnarray}
   C&=&\eta(-1)^{S}(-1)(-1)^{L}=\eta(-1)^{L+S+1},\\
   G&=&\eta(-1)^{L+S+I+1}.
   \end{eqnarray}
  \item [(iii)] The neutral system composed of $\Xi_c$ and $\bar{\Xi}_c^\prime$. Its $P$ parity is also given by Eq.~\eqref{eq:Pparity}, while the $C$ parity will depend on the relative phase factor $\eta$ in its wave function,
  \begin{eqnarray}
  |\Xi_{c}\bar{\Xi}_{c}^{\prime}\rangle&=&\frac{1}{\sqrt{2}}\left[\Xi_{c}\bar{\Xi}_{c}^{\prime}+\eta \bar{\Xi}_{c} \Xi_{c}^{\prime}\right],
  \end{eqnarray}
  and its $C$ and $G$ parities are defined as
  \begin{eqnarray}
  C&=&\eta(-1)^{S+1}(-1)(-1)^{L}=\eta(-1)^{L+S},\\
  G&=&\eta(-1)^{L+S+I}.
  \end{eqnarray}
\end{itemize}

For the dihadron systems composed of generalized identical particles, e.g., the $\Sigma_c\Sigma_c$ and $\Sigma_c^\ast\Sigma_c^\ast$, where $\Sigma_c^{(\ast)}=(\Sigma_c^{(\ast)++},\Sigma_c^{(\ast)+},\Sigma_c^{(\ast)0})$, their quantum numbers must satisfy the following selection rule:
\begin{eqnarray}
L+S+I+2i&=&\mathrm{Even~number},
\end{eqnarray}
where $i$ represents the isospin of the component particle, such as $i=1$ for the $\Sigma_c\Sigma_c$ and $\Sigma_c^\ast\Sigma_c^\ast$ systems.

The various systems considered in this study are shown in Table~\ref{tab:systems}. We do not consider the systems that include the isospin singlet $\Lambda_c$ and $\Lambda$ due to the absence of isospin related interactions.~As a result, the interactions in $\Lambda_{(c)}$-containing systems are often very weak, making it difficult to form molecular states in a single-channel scenario, such as the $\Lambda_c\bar{D}^{(\ast)}$~\cite{Wang:2011rga,Yang:2011wz,Wang:2019nvm} and $\Lambda_c\Lambda_c$ systems~\cite{Lee:2011rka}.

\begin{table}[htbp]
\centering
\caption{The molecular hexaquark systems under consideration and their quark compositions. The Type-I and Type-II systems' interactions are governed by the $V_{qq}$ and $V_{\bar{q}q}$, respectively.\label{tab:systems}}
\setlength{\tabcolsep}{3.7mm}
{
\begin{tabular}{cccc}
\hline
\hline
\multicolumn{2}{c}{Type-I: $V_{qq}$} & \multicolumn{2}{c}{Type-II: $V_{\bar{q}q}$}\tabularnewline
\hline 
\multicolumn{4}{c}{Charmed baryon-(anti)charmed baryon~(\ref{sec:BQBQ})}\tabularnewline
$\Sigma_{c}^{(\ast)}\Sigma_{c}^{(\ast)}$ & $\left[cqq\right]\left[cqq\right]$ & $\Sigma_{c}^{(\ast)}\bar{\Sigma}_{c}^{(\ast)}$ & $\left[cqq\right]\left[\bar{c}\bar{q}\bar{q}\right]$\tabularnewline
$\Sigma_{c}^{(\ast)}\Xi_{c}^{(\prime,\ast)}$ & $\left[cqq\right]\left[csq\right]$ & $\Sigma_{c}^{(\ast)}\bar{\Xi}_{c}^{(\prime,\ast)}$ & $\left[cqq\right]\left[\bar{c}\bar{s}\bar{q}\right]$\tabularnewline
$\Xi_{c}^{(\prime,\ast)}\Xi_{c}^{(\prime,\ast)}$ & $\left[csq\right]\left[csq\right]$ & $\Xi_{c}^{(\prime,\ast)}\bar{\Xi}_{c}^{(\prime,\ast)}$ & $\left[csq\right]\left[\bar{c}\bar{s}\bar{q}\right]$\tabularnewline
\multicolumn{4}{c}{Doubly charmed baryon-(anti)doubly charmed baryon~(\ref{sec:BQQBQQ})}\tabularnewline
$\Xi_{cc}^{(\ast)}\Xi_{cc}^{(\ast)}$ & $\left[ccq\right]\left[ccq\right]$ & $\Xi_{cc}^{(\ast)}\bar{\Xi}_{cc}^{(\ast)}$ & $\left[ccq\right]\left[\bar{c}\bar{c}\bar{q}\right]$\tabularnewline
\multicolumn{4}{c}{Hyperon-(anti)hyperon~(\ref{sec:HH})}\tabularnewline
$\Sigma^{(\ast)}\Sigma^{(\ast)}$ & $\left[sqq\right]\left[sqq\right]$ & $\Sigma^{(\ast)}\bar{\Sigma}^{(\ast)}$ & $\left[sqq\right]\left[\bar{s}\bar{q}\bar{q}\right]$\tabularnewline
$\Sigma^{(\ast)}\Xi^{(\ast)}$ & $\left[sqq\right]\left[ssq\right]$ & $\Sigma^{(\ast)}\bar{\Xi}^{(\ast)}$ & $\left[sqq\right]\left[\bar{s}\bar{s}\bar{q}\right]$\tabularnewline
$\Xi^{(\ast)}\Xi^{(\ast)}$ & $\left[ssq\right]\left[ssq\right]$ & $\Xi^{(\ast)}\bar{\Xi}^{(\ast)}$ & $\left[ssq\right]\left[\bar{s}\bar{s}\bar{q}\right]$\tabularnewline
\multicolumn{4}{c}{Charmed baryon-(anti)doubly charmed baryon~(\ref{sec:BQBQQ})}\tabularnewline
$\Sigma_{c}^{(\ast)}\Xi_{cc}^{(\ast)}$ & $\left[cqq\right]\left[ccq\right]$ & $\Sigma_{c}^{(\ast)}\bar{\Xi}_{cc}^{(\ast)}$ & $\left[cqq\right]\left[\bar{c}\bar{c}\bar{q}\right]$\tabularnewline
$\Xi_{c}^{(\prime,\ast)}\Xi_{cc}^{(\ast)}$ & $\left[csq\right]\left[ccq\right]$ & $\Xi_{c}^{(\prime,\ast)}\bar{\Xi}_{cc}^{(\ast)}$ & $\left[csq\right]\left[\bar{c}\bar{c}\bar{q}\right]$\tabularnewline
\multicolumn{4}{c}{Charmed baryon-(anti)hyperon~(\ref{sec:BQH})}\tabularnewline
$\Sigma_{c}^{(\ast)}\Sigma^{(\ast)}$ & $\left[cqq\right]\left[sqq\right]$ & $\Sigma_{c}^{(\ast)}\bar{\Sigma}^{(\ast)}$ & $\left[cqq\right]\left[\bar{s}\bar{q}\bar{q}\right]$\tabularnewline
$\Sigma_{c}^{(\ast)}\Xi^{(\ast)}$ & $\left[cqq\right]\left[ssq\right]$ & $\Sigma_{c}^{(\ast)}\bar{\Xi}^{(\ast)}$ & $\left[cqq\right]\left[\bar{s}\bar{s}\bar{q}\right]$\tabularnewline
$\Xi_{c}^{(\prime,\ast)}\Sigma^{(\ast)}$ & $\left[csq\right]\left[sqq\right]$ & $\Xi_{c}^{(\prime,\ast)}\bar{\Sigma}^{(\ast)}$ & $\left[csq\right]\left[\bar{s}\bar{q}\bar{q}\right]$\tabularnewline
$\Xi_{c}^{(\prime,\ast)}\Xi^{(\ast)}$ & $\left[csq\right]\left[ssq\right]$ & $\Xi_{c}^{(\prime,\ast)}\bar{\Xi}^{(\ast)}$ & $\left[csq\right]\left[\bar{s}\bar{s}\bar{q}\right]$\tabularnewline
\multicolumn{4}{c}{Doubly charmed baryon-(anti)hyperon~(\ref{sec:BQQH})}\tabularnewline
$\Xi_{cc}^{(\ast)}\Sigma^{(\ast)}$ & $\left[ccq\right]\left[sqq\right]$ & $\Xi_{cc}^{(\ast)}\bar{\Sigma}^{(\ast)}$ & $\left[ccq\right]\left[\bar{s}\bar{q}\bar{q}\right]$\tabularnewline
$\Xi_{cc}^{(\ast)}\Xi^{(\ast)}$ & $\left[ccq\right]\left[ssq\right]$ & $\Xi_{cc}^{(\ast)}\bar{\Xi}^{(\ast)}$ & $\left[ccq\right]\left[\bar{s}\bar{s}\bar{q}\right]$\tabularnewline
\hline 
\hline
\end{tabular}
}
\end{table}

\section{Spectrum of the molecular hexaquarks}\label{sec:spec}

\subsection{Charmed baryon-(anti)charmed baryon systems}\label{sec:BQBQ}

The results of the charmed baryon-(anti)charmed baryon systems are presented in Tables~\ref{tab:SigmacSigmac},~\ref{tab:SigmacXic} and~\ref{tab:XicXic}. Before delving into a formal discussion of our results, we provide an overview of the research status on these systems. 

In Ref.~\cite{Froemel:2004ea}, Fr\"o{}emel \etal~investigated the bound states of the $\Sigma_c\Sigma_c$ system using various phenomenological nucleon-nucleon effective potentials. The binding energies range from a few MeV to several hundred MeV, and some potentials even yield unrealistically deep binding solutions. In Ref.~\cite{Lee:2011rka}, Lee \etal~studied the $\Sigma_c\Sigma_c$, $\Xi_c\Xi_c$ and $\Xi_c^\prime\Xi_c^\prime$ systems and their corresponding particle-antiparticle combinations using the OBE model. They also calculated possible molecular states that might exist in different channels. In Ref.~\cite{Yang:2018amd}, Yang \etal~performed systematic calculations in the systems composed of the spin-$\frac{3}{2}$ charmed baryons using the OBE model. They explored the possible molecular states and predicted bound states in several channels. The OBE model and OBE-based approaches have also been employed by the authors in Refs.~\cite{Ling:2021asz,Dong:2021bvy,Cheng:2022vgy,Kong:2022rvd,Kong:2022rvd,Song:2022svi} to investigate such systems. For instance, Ling \etal~simultaneously calculated the mass spectra and strong decays of the molecules in $\Sigma_c^{(\ast)}\Sigma_c^{(\ast)}$ systems~\cite{Ling:2021asz}. Cheng \etal~on the other hand, employed the complex scaling method to calculate the binding energies and widths of these systems~\cite{Cheng:2022vgy}. In Ref.~\cite{Huang:2013rla}, Huang \etal~utilized calculations based on quark potential model to demonstrate that the interaction in the $\Lambda_c\Lambda_c$ system is repulsive. Without considering the coupled-channels, it is not possible to form bound states in this system. In contrast, the $\Sigma_c^{(\ast)}\Sigma_c^{(\ast)}$ systems exhibit strong attractive interactions~\cite{Huang:2013rla}, which is conducive to the formation of molecular states. Other calculations, such as those from QCD sum rules and chromomagnetic interaction models can be found in Refs.~\cite{Chen:2016ymy,Wan:2019ake,Liu:2021gva,Liu:2022rzu}.

Our results for the $\Sigma_c^{(\ast)}\Sigma_c^{(\ast)}$ and $\Sigma_c^{(\ast)}\bar{\Sigma}_c^{(\ast)}$ systems are shown in Table~\ref{tab:SigmacSigmac}. We obtain three (two) molecular states in the $\Sigma_c\bar{\Sigma}_c$ ($\Sigma_c\Sigma_c$) system, six (four) molecular states in the $\Sigma_c\bar{\Sigma}_c^\ast$ ($\Sigma_c\Sigma_c^\ast$) system, and five (four) molecular states in the $\Sigma_c^\ast\bar{\Sigma}_c^\ast$ ($\Sigma_c^\ast\Sigma_c^\ast$) system. It can be seen that the molecular states of the $\Sigma_c^{(\ast)}\bar{\Sigma}_c^{(\ast)}$ systems are often found in the highest isospin or/and highest spin channels, while the molecular states in the $\Sigma_c^{(\ast)}\Sigma_c^{(\ast)}$ systems are found in the low isospin channels, i.e., the channels with $I=0,1$, and there are no molecular states in the channel with $I=2$. This is a typical feature of systems where molecular states are generated by the interaction between light quarks (via the $V_{qq}$), such as the tetraquark state $T_{cc}$, the pentaquark states $P_\psi^N$ and $P_{\psi s}^\Lambda$, etc., which all follow this pattern. The systems we considered in the following sections will mostly follow this pattern. Meanwhile, we notice that the attractive interaction of most isotensor channels in $\Sigma_c^{(\ast)}\bar{\Sigma}_c^{(\ast)}$ systems is stronger than that of other channels, such as the $2^+(0^{-+})$ $\Sigma_c\bar{\Sigma}_c$, $2^{\mp}(1^{-\mp})$ $\Sigma_c\bar{\Sigma}_c^\ast$, and $2^+(0^{-+})$, $2^-(1^{--})$ $\Sigma_c^\ast\bar{\Sigma}_c^\ast$. The results from the OBE model in Ref.~\cite{Lee:2011rka} are consistent with our results, for example, the attractive potential of the isotensor $\Sigma_c\bar{\Sigma}_c$ is significantly stronger than that of other channels. The calculations of the molecular states of the $\Sigma_c^{(\ast)}\Sigma_c^{(\ast)}$ systems in Ref.~\cite{Ling:2021asz} are also qualitatively consistent with ours.

The results for $\Sigma_c^{(\ast)}\Xi_c^{(\prime,\ast)}$, $\Sigma_c^{(\ast)}\bar{\Xi}_c^{(\prime,\ast)}$ and $\Xi_c^{(\prime,\ast)}\Xi_c^{(\prime,\ast)}$, $\Xi_c^{(\prime,\ast)}\bar{\Xi}_c^{(\prime,\ast)}$ systems are presented in Tables~\ref{tab:SigmacXic} and~\ref{tab:XicXic}, respectively. The systems such as $\Xi_c\Xi_c$, $\Xi_c^\prime\Xi_c^\prime$, $\Xi_c^\ast\Xi_c^\ast$ as well as the corresponding particle-antiparticle pairs were also considered in Refs.~\cite{Lee:2011rka,Yang:2018amd}. Their results in most channels are in line with ours. It is worth noting that in our framework, the effective potentials of channels $0^-(1^{--})$ $\Xi_c\bar{\Xi}_c$, $0^{\pm}(0^{-\pm})$ $\Xi_c\bar{\Xi}_c^\prime$, $0^{\mp}(2^{-\mp})$ $\Xi_c\bar{\Xi}_c^\ast$, $0^{+}(1^{-+})$ $\Xi_c^\prime\bar{\Xi}_c^\ast$ and , $0^{-}(3^{--})$ $\Xi_c^\ast\bar{\Xi}_c^\ast$ are exactly the same as that of $0^+(1^{++})$ $D\bar{D}^\ast$ (see Table IV of Ref.~\cite{Wang:2023hpp}), which means that if $X(3872)$ is a weakly bound state of $D\bar{D}^\ast$, there must be corresponding molecular states in these systems as well.

\begin{table*}[htbp]
\centering
\caption{The $I^{(G)}(J^{P(C)})$ quantum numbers, effective potentials, and bound/virtual state solutions of the $\Sigma_c^{(\ast)}\Sigma_c^{(\ast)}$ and $\Sigma_c^{(\ast)}\bar{\Sigma}_c^{(\ast)}$ systems. The masses/poles are denoted as $E_B/E_V$ for the bound/virtual states, respectively. The thresholds of corresponding systems are given in the brackets in the first column with the form like ``$\Sigma_{c}\bar{\Sigma}_{c}$ $\left[4906.9\right]$". The superscript $\dagger$ represents that the state can also be a near-threshold virtual/bound state if it is labeled as a bound/virtual state. The masses (poles) are given in units of MeV.\label{tab:SigmacSigmac}}
\setlength{\tabcolsep}{3.0mm}
{
\begin{tabular}{cccccccc}
\hline
\hline  
Systems $[m_{\text{th}}]$ & $I^{G}(J^{PC})$ & $V_{\mathcal{H}_1\mathcal{H}_2}^{I,J}$ & $E_{B}/E_{V}$ & Systems & $I(J^{P})$ & $V_{\mathcal{H}_1\mathcal{H}_2}^{I,J}$ & $E_{B}/E_{V}$\tabularnewline
\hline 
\multirow{6}{*}{$\Sigma_{c}\bar{\Sigma}_{c}$ $\left[4906.9\right]$} & $0^{+}(0^{-+})$ & $8\tilde{c}_{a}-6\tilde{c}_{s}$ & $\cdots$ & \multirow{6}{*}{$\Sigma_{c}\Sigma_{c}$} & \multirow{2}{*}{$0(0^{+})$} & \multirow{2}{*}{$8c_{a}-6c_{s}$} & \multirow{2}{*}{$\left[4861.6_{-12.2}^{+12.0}\right]_{B}$}\tabularnewline
 & $0^{-}(1^{--})$ & $-\frac{8}{3}\tilde{c}_{a}-6\tilde{c}_{s}$ & $\left[4882.0,4895.2\right]_{B}$ &  &  &  & \tabularnewline
 & $1^{-}(0^{-+})$ & $\frac{8}{3}\tilde{c}_{a}-2\tilde{c}_{s}$ & $\cdots$ &  & \multirow{2}{*}{$1(1^{+})$} & \multirow{2}{*}{$-\frac{8}{9}c_{a}-2c_{s}$} & \multirow{2}{*}{$\left[4906.5_{-0.9}^{+0.4}\right]_{B}$}\tabularnewline
 & $1^{+}(1^{--})$ & $-\frac{8}{9}\tilde{c}_{a}-2\tilde{c}_{s}$ & $\left[4901.8,4906.6\right]_{V}$ &  &  &  & \tabularnewline
 & $2^{+}(0^{-+})$ & $-8\tilde{c}_{a}+6\tilde{c}_{s}$ & $\left[4819.0,4878.0\right]_{B}$ &  & \multirow{2}{*}{$2(0^{+})$} & \multirow{2}{*}{$-8c_{a}+6c_{s}$} & \multirow{2}{*}{$\cdots$}\tabularnewline
 & $2^{-}(1^{--})$ & $\frac{8}{3}\tilde{c}_{a}+6\tilde{c}_{s}$ & $\cdots$ &  &  &  & \tabularnewline
\hline 
\multirow{12}{*}{$\Sigma_{c}\bar{\Sigma}_{c}^{\ast}$ $\left[4971.6\right]$} & $0^{-}(1^{--})$ & $\frac{22}{3}\tilde{c}_{a}-6\tilde{c}_{s}$ & $\cdots$ & \multirow{12}{*}{$\Sigma_{c}\Sigma_{c}^{\ast}$} & \multirow{2}{*}{$0(1^{+})$} & \multirow{2}{*}{$6c_{a}-6c_{s}$} & \multirow{2}{*}{$\left[4927.0_{-10.5}^{+10.3}\right]_{B}$}\tabularnewline
 & $0^{+}(1^{-+})$ & $6\tilde{c}_{a}-6\tilde{c}_{s}$ & $\cdots$ &  &  &  & \tabularnewline
 & $0^{+}(2^{-+})$ & $-2\tilde{c}_{a}-6\tilde{c}_{s}$ & $\left[4953.0,4967.2\right]_{B}$ &  & \multirow{2}{*}{$0(2^{+})$} & \multirow{2}{*}{$-2c_{a}-6c_{s}$} & \multirow{2}{*}{$\left[4932.0_{-7.0}^{+7.0}\right]_{B}$}\tabularnewline
 & $0^{-}(2^{--})$ & $-6\tilde{c}_{a}-6\tilde{c}_{s}$ & $\left[4910.5,4911.7\right]_{B}$ &  &  &  & \tabularnewline
 & $1^{+}(1^{--})$ & $\frac{22}{9}\tilde{c}_{a}-2\tilde{c}_{s}$ & $\cdots$ &  & \multirow{2}{*}{$1(1^{+})$} & \multirow{2}{*}{$\frac{22}{9}c_{a}-2c_{s}$} & \multirow{2}{*}{$\left[4970.3_{-2.0}^{+1.1}\right]_{B}$}\tabularnewline
 & $1^{-}(1^{-+})$ & $2\tilde{c}_{a}-2\tilde{c}_{s}$ & $\cdots$ &  &  &  & \tabularnewline
 & $1^{-}(2^{-+})$ & $-\frac{2}{3}\tilde{c}_{a}-2\tilde{c}_{s}$ & $\left[4956.6,4970.1\right]_{V}$ &  & \multirow{2}{*}{$1(2^{+})$} & \multirow{2}{*}{$-2c_{a}-2c_{s}$} & \multirow{2}{*}{$\left[4971.3_{-1.3}^{+0.3}\right]_{B}^{\dagger}$}\tabularnewline
 & $1^{+}(2^{--})$ & $-2\tilde{c}_{a}-2\tilde{c}_{s}$ & $\left[4967.5,4967.7\right]_{B}$ &  &  &  & \tabularnewline
 & $2^{-}(1^{--})$ & $-\frac{22}{3}\tilde{c}_{a}+6\tilde{c}_{s}$ & $\left[4893.5,4949.1\right]_{B}$ &  & \multirow{2}{*}{$2(1^{+})$} & \multirow{2}{*}{$-6c_{a}+6c_{s}$} & \multirow{2}{*}{$\cdots$}\tabularnewline
 & $2^{+}(1^{-+})$ & $-6\tilde{c}_{a}+6\tilde{c}_{s}$ & $\left[4914.0,4961.7\right]_{B}$ &  &  &  & \tabularnewline
 & $2^{+}(2^{-+})$ & $2\tilde{c}_{a}+6\tilde{c}_{s}$ & $\cdots$ &  & \multirow{2}{*}{$2(2^{+})$} & \multirow{2}{*}{$2c_{a}+6c_{s}$} & \multirow{2}{*}{$\cdots$}\tabularnewline
 & $2^{-}(2^{--})$ & $6\tilde{c}_{a}+6\tilde{c}_{s}$ & $\cdots$ &  &  &  & \tabularnewline
\hline 
\multirow{12}{*}{$\Sigma_{c}^{\ast}\bar{\Sigma}_{c}^{\ast}$ $\left[5036.2\right]$} & $0^{+}(0^{-+})$ & $10\tilde{c}_{a}-6\tilde{c}_{s}$ & $\cdots$ & \multirow{12}{*}{$\Sigma_{c}^{\ast}\Sigma_{c}^{\ast}$} & \multirow{2}{*}{$0(0^{+})$} & \multirow{2}{*}{$10c_{a}-6c_{s}$} & \multirow{2}{*}{$\left[4989.0_{-14.0}^{+13.8}\right]_{B}$}\tabularnewline
 & $0^{-}(1^{--})$ & $\frac{22}{3}\tilde{c}_{a}-6\tilde{c}_{s}$ & $\cdots$ &  &  &  & \tabularnewline
 & $0^{+}(2^{-+})$ & $2\tilde{c}_{a}-6\tilde{c}_{s}$ & $\cdots$ &  & \multirow{2}{*}{$0(2^{+})$} & \multirow{2}{*}{$2c_{a}-6c_{s}$} & \multirow{2}{*}{$\left[4994.0_{-7.0}^{+7.0}\right]_{B}$}\tabularnewline
 & $0^{-}(3^{--})$ & $-6\tilde{c}_{a}-6\tilde{c}_{s}$ & $\left[4974.7,4975.9\right]_{B}$ &  &  &  & \tabularnewline
 & $1^{-}(0^{-+})$ & $\frac{10}{3}\tilde{c}_{a}-2\tilde{c}_{s}$ & $\cdots$ &  & \multirow{2}{*}{$1(1^{+})$} & \multirow{2}{*}{$\frac{22}{9}c_{a}-2c_{s}$} & \multirow{2}{*}{$\left[5034.9_{-2.1}^{+1.2}\right]_{B}$}\tabularnewline
 & $1^{+}(1^{--})$ & $\frac{22}{9}\tilde{c}_{a}-2\tilde{c}_{s}$ & $\cdots$ &  &  &  & \tabularnewline
 & $1^{-}(2^{-+})$ & $\frac{2}{3}\tilde{c}_{a}-2\tilde{c}_{s}$ & $\cdots$ &  & \multirow{2}{*}{$1(3^{+})$} & \multirow{2}{*}{$-2c_{a}-2c_{s}$} & \multirow{2}{*}{$\left[5035.8_{-1.4}^{+0.3}\right]_{B}^{\dagger}$}\tabularnewline
 & $1^{+}(3^{--})$ & $-2\tilde{c}_{a}-2\tilde{c}_{s}$ & $\left[5031.9,5032.2\right]_{B}$ &  &  &  & \tabularnewline
 & $2^{+}(0^{-+})$ & $-10\tilde{c}_{a}+6\tilde{c}_{s}$ & $\left[4916.3,4985.2\right]_{B}$ &  & \multirow{2}{*}{$2(0^{+})$} & \multirow{2}{*}{$-10c_{a}+6c_{s}$} & \multirow{2}{*}{$\cdots$}\tabularnewline
 & $2^{-}(1^{--})$ & $-\frac{22}{3}\tilde{c}_{a}+6\tilde{c}_{s}$ & $\left[4957.7,5013.3\right]_{B}$ &  &  &  & \tabularnewline
 & $2^{+}(2^{-+})$ & $-2\tilde{c}_{a}+6\tilde{c}_{s}$ & $\left[5033.3,5036.1\right]_{B}^{\dagger}$ &  & \multirow{2}{*}{$2(2^{+})$} & \multirow{2}{*}{$-2c_{a}+6c_{s}$} & \multirow{2}{*}{$\cdots$}\tabularnewline
 & $2^{-}(3^{--})$ & $6\tilde{c}_{a}+6\tilde{c}_{s}$ & $\cdots$ &  &  &  & \tabularnewline
\hline 
\hline 
\end{tabular}
}
\end{table*}

\begin{table*}[htbp]
\centering
\caption{The $I(J^{P})$ quantum numbers, effective potentials, and bound/virtual state solutions of the $\Sigma_c^{(\ast)}\Xi_c^{(\prime,\ast)}$ and $\Sigma_c^{(\ast)}\bar{\Xi}_c^{(\prime,\ast)}$ systems. The superscript $\sharp$ represents that this state might be nonexistent in the range of parameters. Other notations are the same as those in Table~\ref{tab:SigmacSigmac}.\label{tab:SigmacXic}}
\setlength{\tabcolsep}{3.4mm}
{
\begin{tabular}{cccccccc}
\hline
\hline  
Systems $[m_{\text{th}}]$ & $I(J^{P})$ & $V_{\mathcal{H}_1\mathcal{H}_2}^{I,J}$ & $E_{B}/E_{V}$ & Systems & $I(J^{P})$ & $V_{\mathcal{H}_1\mathcal{H}_2}^{I,J}$ & $E_{B}/E_{V}$\tabularnewline
\hline 
\multirow{4}{*}{$\Sigma_{c}\bar{\Xi}_{c}$ $\left[4922.5\right]$} & $\frac{1}{2}(0^{-})$ & $6\tilde{c}_{a}-3\tilde{c}_{s}$ & $\cdots$ & \multirow{4}{*}{$\Sigma_{c}\Xi_{c}$} & $\frac{1}{2}(0^{+})$ & $6c_{a}-3c_{s}$ & $\left[4912.0_{-6.7}^{+6.0}\right]_{B}$\tabularnewline
 & $\frac{1}{2}(1^{-})$ & $-2\tilde{c}_{a}-3\tilde{c}_{s}$ & $\left[4915.5,4918.7\right]_{B}$ &  & $\frac{1}{2}(1^{+})$ & $-2c_{a}-3c_{s}$ & $\left[4915.7_{-3.4}^{+3.0}\right]_{B}$\tabularnewline
 & $\frac{3}{2}(0^{-})$ & $-6\tilde{c}_{a}+3\tilde{c}_{s}$ & $\left[4864.4,4901.4\right]_{B}$ &  & $\frac{3}{2}(0^{+})$ & $-6c_{a}+3c_{s}$ & $\cdots$\tabularnewline
 & $\frac{3}{2}(1^{-})$ & $2\tilde{c}_{a}+3\tilde{c}_{s}$ & $\cdots$ &  & $\frac{3}{2}(1^{+})$ & $2c_{a}+3c_{s}$ & $\cdots$\tabularnewline
\hline 
\multirow{4}{*}{$\Sigma_{c}\bar{\Xi}_{c}^{\prime}$ $\left[5031.7\right]$} & $\frac{1}{2}(0^{-})$ & $-2\tilde{c}_{a}-3\tilde{c}_{s}$ & $\left[5024.2,5027.5\right]_{B}$ & \multirow{4}{*}{$\Sigma_{c}\Xi_{c}^{\prime}$} & $\frac{1}{2}(0^{+})$ & $-2c_{a}-3c_{s}$ & $\left[5024.5_{-3.5}^{+3.1}\right]_{B}$\tabularnewline
 & $\frac{1}{2}(1^{-})$ & $\frac{2}{3}\tilde{c}_{a}-3\tilde{c}_{s}$ & $\cdots$ &  & $\frac{1}{2}(1^{+})$ & $\frac{2}{3}c_{a}-3c_{s}$ & $\left[5023.2_{-2.6}^{+2.4}\right]_{B}$\tabularnewline
 & $\frac{3}{2}(0^{-})$ & $2\tilde{c}_{a}+3\tilde{c}_{s}$ & $\cdots$ &  & $\frac{3}{2}(0^{+})$ & $2c_{a}+3c_{s}$ & $\cdots$\tabularnewline
 & $\frac{3}{2}(1^{-})$ & $-\frac{2}{3}\tilde{c}_{a}+3\tilde{c}_{s}$ & $\left[\sim5010.6\right]_{V}^{\sharp}$ &  & $\frac{3}{2}(1^{+})$ & $-\frac{2}{3}c_{a}+3c_{s}$ & $\cdots$\tabularnewline
\hline 
\multirow{4}{*}{$\Sigma_{c}\bar{\Xi}_{c}^{\ast}$ $\left[5098.6\right]$} & $\frac{1}{2}(1^{-})$ & $\frac{10}{3}\tilde{c}_{a}-3\tilde{c}_{s}$ & $\cdots$ & \multirow{4}{*}{$\Sigma_{c}\Xi_{c}^{\ast}$} & $\frac{1}{2}(1^{+})$ & $\frac{10}{3}c_{a}-3c_{s}$ & $\left[5088.6_{-4.7}^{+4.3}\right]_{B}$\tabularnewline
 & $\frac{1}{2}(2^{-})$ & $-2\tilde{c}_{a}-3\tilde{c}_{s}$ & $\left[5090.9,5094.2\right]_{B}$ &  & $\frac{1}{2}(2^{+})$ & $-2c_{a}-3c_{s}$ & $\left[5091.1_{-3.5}^{+3.1}\right]_{B}$\tabularnewline
 & $\frac{3}{2}(1^{-})$ & $-\frac{10}{3}\tilde{c}_{a}+3\tilde{c}_{s}$ & $\left[5078.7,5098.1\right]_{B}$ &  & $\frac{3}{2}(1^{+})$ & $-\frac{10}{3}c_{a}+3c_{s}$ & $\cdots$\tabularnewline
 & $\frac{3}{2}(2^{-})$ & $2\tilde{c}_{a}+3\tilde{c}_{s}$ & $\cdots$ &  & $\frac{3}{2}(2^{+})$ & $2c_{a}+3c_{s}$ & $\cdots$\tabularnewline
\hline 
\multirow{4}{*}{$\Sigma_{c}^{\ast}\bar{\Xi}_{c}$ $\left[4987.2\right]$} & $\frac{1}{2}(1^{-})$ & $5\tilde{c}_{a}-3\tilde{c}_{s}$ & $\cdots$ & \multirow{4}{*}{$\Sigma_{c}^{\ast}\Xi_{c}$} & $\frac{1}{2}(1^{+})$ & $5c_{a}-3c_{s}$ & $\left[4976.8_{-6.0}^{+5.3}\right]_{B}$\tabularnewline
 & $\frac{1}{2}(2^{-})$ & $-3\tilde{c}_{a}-3\tilde{c}_{s}$ & $\left[4970.8,4971.4\right]_{B}$ &  & $\frac{1}{2}(2^{+})$ & $-3c_{a}-3c_{s}$ & $\left[4980.6_{-4.0}^{+3.5}\right]_{B}$\tabularnewline
 & $\frac{3}{2}(1^{-})$ & $-5\tilde{c}_{a}+3\tilde{c}_{s}$ & $\left[4943.7,4975.3\right]_{B}$ &  & $\frac{3}{2}(1^{+})$ & $-5c_{a}+3c_{s}$ & $\cdots$\tabularnewline
 & $\frac{3}{2}(2^{-})$ & $3\tilde{c}_{a}+3\tilde{c}_{s}$ & $\cdots$ &  & $\frac{3}{2}(2^{+})$ & $3c_{a}+3c_{s}$ & $\cdots$\tabularnewline
\hline 
\multirow{4}{*}{$\Sigma_{c}^{\ast}\bar{\Xi}_{c}^{\prime}$ $\left[5096.3\right]$} & $\frac{1}{2}(1^{-})$ & $-\frac{5}{3}\tilde{c}_{a}-3\tilde{c}_{s}$ & $\left[5091.2,5094.8\right]_{B}$ & \multirow{4}{*}{$\Sigma_{c}^{\ast}\Xi_{c}^{\prime}$} & $\frac{1}{2}(1^{+})$ & $-\frac{5}{3}c_{a}-3c_{s}$ & $\left[5088.7_{-3.3}^{+3.0}\right]_{B}$\tabularnewline
 & $\frac{1}{2}(2^{-})$ & $\tilde{c}_{a}-3\tilde{c}_{s}$ & $\cdots$ &  & $\frac{1}{2}(2^{+})$ & $c_{a}-3c_{s}$ & $\left[5087.4_{-2.9}^{+2.7}\right]_{B}$\tabularnewline
 & $\frac{3}{2}(1^{-})$ & $\frac{5}{3}\tilde{c}_{a}+3\tilde{c}_{s}$ & $\cdots$ &  & $\frac{3}{2}(1^{+})$ & $\frac{5}{3}c_{a}+3c_{s}$ & $\cdots$\tabularnewline
 & $\frac{3}{2}(2^{-})$ & $-\tilde{c}_{a}+3\tilde{c}_{s}$ & $\left[\sim5093.1\right]_{V}^{\sharp}$ &  & $\frac{3}{2}(2^{+})$ & $-c_{a}+3c_{s}$ & $\cdots$\tabularnewline
\hline 
\multirow{8}{*}{$\Sigma_{c}^{\ast}\bar{\Xi}_{c}^{\ast}$ $\left[5163.2\right]$} & $\frac{1}{2}(0^{-})$ & $5\tilde{c}_{a}-3\tilde{c}_{s}$ & $\cdots$ & \multirow{8}{*}{$\Sigma_{c}^{\ast}\Xi_{c}^{\ast}$} & $\frac{1}{2}(0^{+})$ & $5c_{a}-3c_{s}$ & $\left[5152.1_{-6.1}^{+5.4}\right]_{B}$\tabularnewline
 & $\frac{1}{2}(1^{-})$ & $\frac{11}{3}\tilde{c}_{a}-3\tilde{c}_{s}$ & $\cdots$ &  & $\frac{1}{2}(1^{+})$ & $\frac{11}{3}c_{a}-3c_{s}$ & $\left[5153.0_{-5.0}^{+4.5}\right]_{B}$\tabularnewline
 & $\frac{1}{2}(2^{-})$ & $\tilde{c}_{a}-3\tilde{c}_{s}$ & $\cdots$ &  & $\frac{1}{2}(2^{+})$ & $c_{a}-3c_{s}$ & $\left[5154.1_{-2.9}^{+2.7}\right]_{B}$\tabularnewline
 & $\frac{1}{2}(3^{-})$ & $-3\tilde{c}_{a}-3\tilde{c}_{s}$ & $\left[5146.0,5146.6\right]_{B}$ &  & $\frac{1}{2}(3^{+})$ & $-3c_{a}-3c_{s}$ & $\left[5156.0_{-4.0}^{+3.7}\right]_{B}$\tabularnewline
 & $\frac{3}{2}(0^{-})$ & $-5\tilde{c}_{a}+3\tilde{c}_{s}$ & $\left[5118.7,5150.6\right]_{B}$ &  & $\frac{3}{2}(0^{+})$ & $-5c_{a}+3c_{s}$ & $\cdots$\tabularnewline
 & $\frac{3}{2}(1^{-})$ & $-\frac{11}{3}\tilde{c}_{a}+3\tilde{c}_{s}$ & $\left[5138.3,5161.1\right]_{B}$ &  & $\frac{3}{2}(1^{+})$ & $-\frac{11}{3}c_{a}+3c_{s}$ & $\cdots$\tabularnewline
 & $\frac{3}{2}(2^{-})$ & $-\tilde{c}_{a}+3\tilde{c}_{s}$ & $\left[\sim5160.3\right]_{V}^{\sharp}$ &  & $\frac{3}{2}(2^{+})$ & $-c_{a}+3c_{s}$ & $\cdots$\tabularnewline
 & $\frac{3}{2}(3^{-})$ & $3\tilde{c}_{a}+3\tilde{c}_{s}$ & $\cdots$ &  & $\frac{3}{2}(3^{+})$ & $3c_{a}+3c_{s}$ & $\cdots$\tabularnewline
\hline 
\hline 
\end{tabular}
}
\end{table*}

\begin{table*}[htbp]
\centering
\caption{The $I^{(G)}(J^{P(C)})$ quantum numbers, effective potentials, and bound/virtual state solutions of the $\Xi_c^{(\prime,\ast)}\Xi_c^{(\prime,\ast)}$ and $\Xi_c^{(\prime,\ast)}\bar{\Xi}_c^{(\prime,\ast)}$ systems. The notations are the same as those in Tables~\ref{tab:SigmacSigmac} and~\ref{tab:SigmacXic}.\label{tab:XicXic}}
\setlength{\tabcolsep}{3.1mm}
{
\begin{tabular}{cccccccc}
\hline
\hline  
Systems $[m_{\text{th}}]$ & $I^G(J^{PC})$ & $V_{\mathcal{H}_1\mathcal{H}_2}^{I,J}$ & $E_{B}/E_{V}$ & Systems & $I(J^{P})$ & $V_{\mathcal{H}_1\mathcal{H}_2}^{I,J}$ & $E_{B}/E_{V}$\tabularnewline
\hline 
\multirow{4}{*}{$\Xi_{c}\bar{\Xi}_{c}$ $\left[4938.2\right]$} & $0^{+}(0^{-+})$ & $\frac{15}{2}\tilde{c}_{a}-\frac{5}{2}\tilde{c}_{s}$ & $\cdots$ & \multirow{4}{*}{$\Xi_{c}\Xi_{c}$} & \multirow{2}{*}{$0(1^{+})$} & \multirow{2}{*}{$-\frac{5}{2}c_{a}-\frac{5}{2}c_{s}$} & \multirow{2}{*}{$\left[4935.5_{-2.8}^{+2.0}\right]_{B}$}\tabularnewline
 & $0^{-}(1^{--})$ & $-\frac{5}{2}\tilde{c}_{a}-\frac{5}{2}\tilde{c}_{s}$ & $\left[4928.5,4929.0\right]_{B}$ &  &  &  & \tabularnewline
 & $1^{-}(0^{-+})$ & $-\frac{9}{2}\tilde{c}_{a}+\frac{3}{2}\tilde{c}_{s}$ & $\left[4902.0,4925.5\right]_{B}$ &  & \multirow{2}{*}{$1(0^{+})$} & \multirow{2}{*}{$-\frac{9}{2}c_{a}+\frac{3}{2}c_{s}$} & \multirow{2}{*}{$\cdots$}\tabularnewline
 & $1^{+}(1^{--})$ & $\frac{3}{2}\tilde{c}_{a}+\frac{3}{2}\tilde{c}_{s}$ & $\cdots$ &  &  &  & \tabularnewline
\hline 
\multirow{4}{*}{$\Xi_{c}\bar{\Xi}_{c}^{\prime}$ $\left[5047.3\right]$} & $0^{\pm}(0^{-\pm})$ & $-\frac{5}{2}\tilde{c}_{a}-\frac{5}{2}\tilde{c}_{s}$ & $\left[5037.2,5037.6\right]_{B}$ & \multirow{4}{*}{$\Xi_{c}\Xi_{c}^{\prime}$} & $0(0^{+})$ & $-\frac{5}{2}c_{a}-\frac{5}{2}c_{s}$ & $\left[5044.3_{-2.9}^{+2.1}\right]_{B}$\tabularnewline
 & $0^{\mp}(1^{-\mp})$ & $\frac{5}{6}\tilde{c}_{a}-\frac{5}{2}\tilde{c}_{s}$ & $\cdots$ &  & $0(1^{+})$ & $\frac{5}{6}c_{a}-\frac{5}{2}c_{s}$ & $\left[5043.1_{-2.0}^{+1.8}\right]_{B}$\tabularnewline
 & $1^{\mp}(0^{-\pm})$ & $\frac{3}{2}\tilde{c}_{a}+\frac{3}{2}\tilde{c}_{s}$ & $\cdots$ &  & $1(0^{+})$ & $\frac{3}{2}c_{a}+\frac{3}{2}c_{s}$ & $\cdots$\tabularnewline
 & $1^{\pm}(1^{-\mp})$ & $-\frac{1}{2}\tilde{c}_{a}+\frac{3}{2}\tilde{c}_{s}$ & $\left[\sim5005.1\right]_{V}^{\sharp}$ &  & $1(1^{+})$ & $-\frac{1}{2}c_{a}+\frac{3}{2}c_{s}$ & $\cdots$\tabularnewline
\hline 
\multirow{4}{*}{$\Xi_{c}\bar{\Xi}_{c}^{\ast}$ $\left[5114.2\right]$} & $0^{\pm}(1^{-\pm})$ & $\frac{25}{6}\tilde{c}_{a}-\frac{5}{2}\tilde{c}_{s}$ & $\cdots$ & \multirow{4}{*}{$\Xi_{c}\Xi_{c}^{\ast}$} & $0(1^{+})$ & $\frac{25}{6}c_{a}-\frac{5}{2}c_{s}$ & $\left[5108.4_{-4.5}^{+3.7}\right]_{B}$\tabularnewline
 & $0^{\mp}(2^{-\mp})$ & $-\frac{5}{2}\tilde{c}_{a}-\frac{5}{2}\tilde{c}_{s}$ & $\left[5103.8,5104.3\right]_{B}$ &  & $0(2^{+})$ & $-\frac{5}{2}c_{a}-\frac{5}{2}c_{s}$ & $\left[5111.0_{-2.9}^{+2.2}\right]_{B}$\tabularnewline
 & $1^{\mp}(1^{-\pm})$ & $-\frac{5}{2}\tilde{c}_{a}+\frac{3}{2}\tilde{c}_{s}$ & $\left[5104.8,5114.1\right]_{B}$ &  & $1(1^{+})$ & $-\frac{5}{2}c_{a}+\frac{3}{2}c_{s}$ & $\cdots$\tabularnewline
 & $1^{\pm}(2^{-\mp})$ & $\frac{3}{2}\tilde{c}_{a}+\frac{3}{2}\tilde{c}_{s}$ & $\cdots$ &  & $1(2^{+})$ & $\frac{3}{2}c_{a}+\frac{3}{2}c_{s}$ & $\cdots$\tabularnewline
\hline 
\multirow{4}{*}{$\Xi_{c}^{\prime}\bar{\Xi}_{c}^{\prime}$ $\left[5156.4\right]$} & $0^{+}(0^{-+})$ & $\frac{5}{6}\tilde{c}_{a}-\frac{5}{2}\tilde{c}_{s}$ & $\cdots$ & \multirow{4}{*}{$\Xi_{c}^{\prime}\Xi_{c}^{\prime}$} & \multirow{2}{*}{$0(1^{+})$} & \multirow{2}{*}{$-\frac{5}{18}c_{a}-\frac{5}{2}c_{s}$} & \multirow{2}{*}{$\left[5152.3_{-1.7}^{+1.5}\right]_{V}$}\tabularnewline
 & $0^{-}(1^{--})$ & $-\frac{5}{18}\tilde{c}_{a}-\frac{5}{2}\tilde{c}_{s}$ & $\left[5051.6,5153.5\right]_{V}$ &  &  &  & \tabularnewline
 & $1^{-}(0^{-+})$ & $-\frac{1}{2}\tilde{c}_{a}+\frac{3}{2}\tilde{c}_{s}$ & $\left[\sim5117.4\right]_{V}^{\sharp}$ &  & \multirow{2}{*}{$1(0^{+})$} & \multirow{2}{*}{$-\frac{1}{2}c_{a}+\frac{3}{2}c_{s}$} & \multirow{2}{*}{$\cdots$}\tabularnewline
 & $1^{+}(1^{--})$ & $\frac{1}{6}\tilde{c}_{a}+\frac{3}{2}\tilde{c}_{s}$ & $\cdots$ &  &  &  & \tabularnewline
\hline 
\multirow{8}{*}{$\Xi_{c}^{\prime}\bar{\Xi}_{c}^{\ast}$ $\left[5223.3\right]$} & $0^{-}(1^{--})$ & $-\frac{5}{18}\tilde{c}_{a}-\frac{5}{2}\tilde{c}_{s}$ & $\left[5122.7,5220.6\right]_{V}$ & \multirow{8}{*}{$\Xi_{c}^{\prime}\Xi_{c}^{\ast}$} & \multirow{2}{*}{$0(1^{+})$} & \multirow{2}{*}{$-\frac{5}{18}c_{a}-\frac{5}{2}c_{s}$} & \multirow{2}{*}{$\left[5219.0_{-1.7}^{+1.5}\right]_{B}$}\tabularnewline
 & $0^{+}(1^{-+})$ & $-\frac{5}{2}\tilde{c}_{a}-\frac{5}{2}\tilde{c}_{s}$ & $\left[5212.5,5212.9\right]_{B}$ &  &  &  & \tabularnewline
 & $0^{+}(2^{-+})$ & $\frac{25}{6}\tilde{c}_{a}-\frac{5}{2}\tilde{c}_{s}$ & $\cdots$ &  & \multirow{2}{*}{$0(2^{+})$} & \multirow{2}{*}{$-\frac{5}{2}c_{a}-\frac{5}{2}c_{s}$} & \multirow{2}{*}{$\left[5219.9_{-3.0}^{+2.3}\right]_{B}$}\tabularnewline
 & $0^{-}(2^{--})$ & $-\frac{5}{2}\tilde{c}_{a}-\frac{5}{2}\tilde{c}_{s}$ & $\left[5212.5,5212.9\right]_{B}$ &  &  &  & \tabularnewline
 & $1^{+}(1^{--})$ & $\frac{1}{6}\tilde{c}_{a}+\frac{3}{2}\tilde{c}_{s}$ & $\cdots$ &  & \multirow{2}{*}{$1(1^{+})$} & \multirow{2}{*}{$\frac{3}{2}c_{a}+\frac{3}{2}c_{s}$} & \multirow{2}{*}{$\cdots$}\tabularnewline
 & $1^{-}(1^{-+})$ & $\frac{3}{2}\tilde{c}_{a}+\frac{3}{2}\tilde{c}_{s}$ & $\cdots$ &  &  &  & \tabularnewline
 & $1^{-}(2^{-+})$ & $-\frac{5}{2}\tilde{c}_{a}+\frac{3}{2}\tilde{c}_{s}$ & $\left[5213.5,5223.3\right]_{B}$ &  & \multirow{2}{*}{$1(2^{+})$} & \multirow{2}{*}{$-\frac{5}{2}c_{a}+\frac{3}{2}c_{s}$} & \multirow{2}{*}{$\cdots$}\tabularnewline
 & $1^{+}(2^{--})$ & $\frac{3}{2}\tilde{c}_{a}+\frac{3}{2}\tilde{c}_{s}$ & $\cdots$ &  &  &  & \tabularnewline
\hline 
\multirow{8}{*}{$\Xi_{c}^{\ast}\bar{\Xi}_{c}^{\ast}$ $\left[5290.2\right]$} & $0^{+}(0^{-+})$ & $\frac{25}{6}\tilde{c}_{a}-\frac{5}{2}\tilde{c}_{s}$ & $\cdots$ & \multirow{8}{*}{$\Xi_{c}^{\ast}\Xi_{c}^{\ast}$} & \multirow{2}{*}{$0(1^{+})$} & \multirow{2}{*}{$\frac{55}{18}c_{a}-\frac{5}{2}c_{s}$} & \multirow{2}{*}{$\left[5284.3_{-3.7}^{+3.2}\right]_{B}$}\tabularnewline
 & $0^{-}(1^{--})$ & $\frac{55}{18}\tilde{c}_{a}-\frac{5}{2}\tilde{c}_{s}$ & $\cdots$ &  &  &  & \tabularnewline
 & $0^{+}(2^{-+})$ & $\frac{5}{6}\tilde{c}_{a}-\frac{5}{2}\tilde{c}_{s}$ & $\cdots$ &  & \multirow{2}{*}{$0(3^{+})$} & \multirow{2}{*}{$-\frac{5}{2}c_{a}-\frac{5}{2}c_{s}$} & \multirow{2}{*}{$\left[5286.6_{-3.1}^{+2.4}\right]_{B}$}\tabularnewline
 & $0^{-}(3^{--})$ & $-\frac{5}{2}\tilde{c}_{a}-\frac{5}{2}\tilde{c}_{s}$ & $\left[5279.1,5279.6\right]_{B}$ &  &  &  & \tabularnewline
 & $1^{-}(0^{-+})$ & $-\frac{5}{2}\tilde{c}_{a}+\frac{3}{2}\tilde{c}_{s}$ & $\left[5280.1,5290.1\right]_{B}$ &  & \multirow{2}{*}{$1(0^{+})$} & \multirow{2}{*}{$-\frac{5}{2}c_{a}+\frac{3}{2}c_{s}$} & \multirow{2}{*}{$\cdots$}\tabularnewline
 & $1^{+}(1^{--})$ & $-\frac{11}{6}\tilde{c}_{a}+\frac{3}{2}\tilde{c}_{s}$ & $\left[5286.4,5290.2\right]_{V}$ &  &  &  & \tabularnewline
 & $1^{-}(2^{-+})$ & $-\frac{1}{2}\tilde{c}_{a}+\frac{3}{2}\tilde{c}_{s}$ & $\left[\sim5254.6\right]_{V}^{\sharp}$ &  & \multirow{2}{*}{$1(2^{+})$} & \multirow{2}{*}{$-\frac{1}{2}c_{a}+\frac{3}{2}c_{s}$} & \multirow{2}{*}{$\cdots$}\tabularnewline
 & $1^{+}(3^{--})$ & $\frac{3}{2}\tilde{c}_{a}+\frac{3}{2}\tilde{c}_{s}$ & $\cdots$ &  &  &  & \tabularnewline
\hline 
\hline 
\end{tabular}
}
\end{table*}

\subsection{Doubly charmed baryon-(anti)doubly charmed baryon systems}\label{sec:BQQBQQ}

Our results for the $\Xi_{cc}^{(\ast)}\Xi_{cc}^{(\ast)}$ and $\Xi_{cc}^{(\ast)}\bar{\Xi}_{cc}^{(\ast)}$ systems are shown in Table~\ref{tab:XiccXicc}, where $\Xi_{cc}^\ast$ has not been observed. Therefore, we refer to both the predictions based on heavy diquark-antiquark symmetry and the calculations of the quark model to obtain a mass difference of $m_{\Xi_{cc}^\ast}-m_{\Xi_{cc}}$ of $85\pm15$ MeV~\cite{Wang:2023eng}. As can be seen from Table~\ref{tab:XiccXicc}, we use the notation $\pm15$ MeV to represent the mass uncertainty of $\Xi_{cc}^\ast$.

In Ref.~\cite{Meng:2017fwb}, Meng \etal~investigated the possible molecular states composed of two spin-$\frac{1}{2}$ doubly charmed baryons, and they obtained that there is a molecular state in the $0(1^+)$ $\Xi_{cc}\Xi_{cc}$ channel, with a binding energy of about a few MeV to tens of MeV. Our calculation gives a binding energy of about $10$ MeV, which is consistent with their result. As for the isovector $\Xi_{cc}\bar{\Xi}_{cc}$ system, they also obtained a bound state, while in our calculation, the $1^-(0^{-+})$ $\Xi_{cc}\bar{\Xi}_{cc}$ may be a virtual state or may not exist within the parameter range. 

In Ref.~\cite{Yang:2019rgw}, Yang \etal~studied the molecular states composed of doubly charmed baryons with spin-$\frac{3}{2}$ using the OBE model. They obtained bound states in the isoscalar $\Xi_{cc}^{(\ast)}\Xi_{cc}^\ast$ systems, with  binding energies from a few MeV to tens of MeV, which are consistent with our calculations. Furthermore, most of the results of $\Xi_{cc}^{(\ast)}\bar{\Xi}_{cc}^\ast$ systems in Ref.~\cite{Yang:2019rgw} agree with ours.

The $0^+(1^{-+})$ and $0^-(2^{--})$ $\Xi_{cc}\bar{\Xi}_{cc}^\ast$ as well as the $0^-(3^{--})$ $\Xi_{cc}^\ast\bar{\Xi}_{cc}^\ast$ can be regarded as the quadruply heavy counterparts of $X(3872)$, since they share the same effective potentials under the heavy diquark-antiquark symmetry [see Table IV of Ref.~\cite{Wang:2023hpp} for the effective potentials of $X(3872)$].

\begin{table*}[htbp]
\centering
\caption{The $I^{(G)}(J^{P(C)})$ quantum numbers, effective potentials, and bound/virtual state solutions of the $\Xi_{cc}^{(\ast)}\Xi_{cc}^{(\ast)}$ and $\Xi_{cc}^{(\ast)}\bar{\Xi}_{cc}^{(\ast)}$ systems. The notations are the same as those in Tables~\ref{tab:SigmacSigmac} and~\ref{tab:SigmacXic}.\label{tab:XiccXicc}}
\setlength{\tabcolsep}{1.6mm}
{
\begin{tabular}{cccccccc}
\hline 
\hline
Systems $[m_{\text{th}}]$ & $I^G(J^{PC})$ & $V_{\mathcal{H}_1\mathcal{H}_2}^{I,J}$ & $E_{B}/E_{V}$ & Systems & $I(J^{P})$ & $V_{\mathcal{H}_1\mathcal{H}_2}^{I,J}$ & $E_{B}/E_{V}$\tabularnewline
\hline
\multirow{4}{*}{$\Xi_{cc}\bar{\Xi}_{cc}$ $\left[7243.1\right]$} & $0^{+}(0^{-+})$ & $\frac{5}{6}\tilde{c}_{a}-\frac{5}{2}\tilde{c}_{s}$ & $\cdots$ & \multirow{4}{*}{$\Xi_{cc}\Xi_{cc}$} & \multirow{2}{*}{$0(1^{+})$} & \multirow{2}{*}{$-\frac{5}{18}c_{a}-\frac{5}{2}c_{s}$} & \multirow{2}{*}{$\left[7233.5_{-2.2}^{+2.1}\right]_{B}$}\tabularnewline
 & $0^{-}(1^{--})$ & $-\frac{5}{18}\tilde{c}_{a}-\frac{5}{2}\tilde{c}_{s}$ & $\left[7209.0,7243.1\right]_{V}$ &  &  &  & \tabularnewline
 & $1^{-}(0^{-+})$ & $-\frac{1}{2}\tilde{c}_{a}+\frac{3}{2}\tilde{c}_{s}$ & $\left[\sim7232.8\right]_{V}^{\sharp}$ &  & \multirow{2}{*}{$1(0^{+})$} & \multirow{2}{*}{$-\frac{1}{2}c_{a}+\frac{3}{2}c_{s}$} & \multirow{2}{*}{$\cdots$}\tabularnewline
 & $1^{+}(1^{--})$ & $\frac{1}{6}\tilde{c}_{a}+\frac{3}{2}\tilde{c}_{s}$ & $\cdots$ &  &  &  & \tabularnewline
\hline 
\multirow{8}{*}{$\Xi_{cc}\bar{\Xi}_{cc}^{\ast}$ $\left[7328.1\pm15\right]$} & $0^{-}(1^{--})$ & $-\frac{5}{18}\tilde{c}_{a}-\frac{5}{2}\tilde{c}_{s}$ & $\left[7311.7\pm16.4\pm15\right]_{V}$ & \multirow{8}{*}{$\Xi_{cc}\Xi_{cc}^{\ast}$} & \multirow{2}{*}{$0(1^{+})$} & \multirow{2}{*}{$-\frac{5}{18}c_{a}-\frac{5}{2}c_{s}$} & \multirow{2}{*}{$\left[7318.3_{-2.2}^{+2.1}\pm15\right]_{B}$}\tabularnewline
 & $0^{+}(1^{-+})$ & $-\frac{5}{2}\tilde{c}_{a}-\frac{5}{2}\tilde{c}_{s}$ & $\left[7310.7\pm0.3\pm15\right]_{B}$ &  &  &  & \tabularnewline
 & $0^{+}(2^{-+})$ & $\frac{25}{6}\tilde{c}_{a}-\frac{5}{2}\tilde{c}_{s}$ & $\cdots$ &  & \multirow{2}{*}{$0(2^{+})$} & \multirow{2}{*}{$-\frac{5}{2}c_{a}-\frac{5}{2}c_{s}$} & \multirow{2}{*}{$\left[7319.4_{-3.8}^{+3.5}\pm15\right]_{B}$}\tabularnewline
 & $0^{-}(2^{--})$ & $-\frac{5}{2}\tilde{c}_{a}-\frac{5}{2}\tilde{c}_{s}$ & $\left[7310.7\pm0.3\pm15\right]_{B}$ &  &  &  & \tabularnewline
 & $1^{+}(1^{--})$ & $\frac{1}{6}\tilde{c}_{a}+\frac{3}{2}\tilde{c}_{s}$ & $\cdots$ &  & \multirow{2}{*}{$1(1^{+})$} & \multirow{2}{*}{$\frac{3}{2}c_{a}+\frac{3}{2}c_{s}$} & \multirow{2}{*}{$\cdots$}\tabularnewline
 & $1^{-}(1^{-+})$ & $\frac{3}{2}\tilde{c}_{a}+\frac{3}{2}\tilde{c}_{s}$ & $\cdots$ &  &  &  & \tabularnewline
 & $1^{-}(2^{-+})$ & $-\frac{5}{2}\tilde{c}_{a}+\frac{3}{2}\tilde{c}_{s}$ & $\left[7318.6\pm7.0\pm15\right]_{B}$ &  & \multirow{2}{*}{$1(2^{+})$} & \multirow{2}{*}{$-\frac{5}{2}c_{a}+\frac{3}{2}c_{s}$} & \multirow{2}{*}{$\cdots$}\tabularnewline
 & $1^{+}(2^{--})$ & $\frac{3}{2}\tilde{c}_{a}+\frac{3}{2}\tilde{c}_{s}$ & $\cdots$ &  &  &  & \tabularnewline
\hline 
\multirow{8}{*}{$\Xi_{cc}^{\ast}\bar{\Xi}_{cc}^{\ast}$ $\left[7413.1\pm30\right]$} & $0^{+}(0^{-+})$ & $\frac{25}{6}\tilde{c}_{a}-\frac{5}{2}\tilde{c}_{s}$ & $\cdots$ & \multirow{8}{*}{$\Xi_{cc}^{\ast}\Xi_{cc}^{\ast}$} & \multirow{2}{*}{$0(1^{+})$} & \multirow{2}{*}{$\frac{55}{18}c_{a}-\frac{5}{2}c_{s}$} & \multirow{2}{*}{$\left[7401.3_{-4.5}^{+4.2}\pm30\right]_{B}$}\tabularnewline
 & $0^{-}(1^{--})$ & $\frac{55}{18}\tilde{c}_{a}-\frac{5}{2}\tilde{c}_{s}$ & $\cdots$ &  &  &  & \tabularnewline
 & $0^{+}(2^{-+})$ & $\frac{5}{6}\tilde{c}_{a}-\frac{5}{2}\tilde{c}_{s}$ & $\cdots$ &  & \multirow{2}{*}{$0(3^{+})$} & \multirow{2}{*}{$-\frac{5}{2}c_{a}-\frac{5}{2}c_{s}$} & \multirow{2}{*}{$\left[7404.2_{-3.9}^{+3.6}\pm30\right]_{B}$}\tabularnewline
 & $0^{-}(3^{--})$ & $-\frac{5}{2}\tilde{c}_{a}-\frac{5}{2}\tilde{c}_{s}$ & $\left[7395.5\pm0.3\pm30\right]_{B}$ &  &  &  & \tabularnewline
 & $1^{-}(0^{-+})$ & $-\frac{5}{2}\tilde{c}_{a}+\frac{3}{2}\tilde{c}_{s}$ & $\left[7403.4\pm7.0\pm30\right]_{B}$ &  & \multirow{2}{*}{$1(0^{+})$} & \multirow{2}{*}{$-\frac{5}{2}c_{a}+\frac{3}{2}c_{s}$} & \multirow{2}{*}{$\cdots$}\tabularnewline
 & $1^{+}(1^{--})$ & $-\frac{11}{6}\tilde{c}_{a}+\frac{3}{2}\tilde{c}_{s}$ & $\left[7409.2\pm3.9\pm30\right]_{B}^{\dagger}$ &  &  &  & \tabularnewline
 & $1^{-}(2^{-+})$ & $-\frac{1}{2}\tilde{c}_{a}+\frac{3}{2}\tilde{c}_{s}$ & $\left[\sim7403.8\pm30\right]_{V}^{\sharp}$ &  & \multirow{2}{*}{$1(2^{+})$} & \multirow{2}{*}{$-\frac{1}{2}c_{a}+\frac{3}{2}c_{s}$} & \multirow{2}{*}{$\cdots$}\tabularnewline
 & $1^{+}(3^{--})$ & $\frac{3}{2}\tilde{c}_{a}+\frac{3}{2}\tilde{c}_{s}$ & $\cdots$ &  &  &  & \tabularnewline
\hline 
\hline 
\end{tabular}
}
\end{table*}

\subsection{Hyperon-(anti)hyperon systems}\label{sec:HH}

The results of the hyperon-(anti)hyperon systems are shown in Tables~\ref{tab:SigmaSigma} and~\ref{tab:SigmaXi}. Under the \heavy quark symmetry, this type of system is very similar to the charmed baryon-(anti)charmed baryon ones. 

The study of dihyperon systems has a long history~\cite{Jaffe:1976yi,Balachandran:1983dj,Mackenzie:1985vv,Rosner:1985yh,Polinder:2007mp,NPLQCD:2010ocs,Inoue:2010es,Li:2018tbt,Chen:2019vdh,Clement:2020mab}. In the late 1970s, Jaffe predicted the existence of a state with $\Lambda\Lambda$ quantum numbers [$0(0^+)$], known as the H-dibaryon~\cite{Jaffe:1976yi}. It was described as a flavor singlet with quark composition $uuddss$ and a mass below the $\Lambda\Lambda$ threshold, around $80$ MeV. However, Mackenzie \etal~used lattice QCD calculations to suggest that the mass of the H-dibaryon is above the $\Lambda\Lambda$ threshold~\cite{Mackenzie:1985vv}.
In 2011, calculations by the HAL QCD Collaboration indicated that the binding energy of the H-dibaryon falls within the range of $30$$-$$40$ MeV when the pion mass is in the unphysical range of $673$$-$$1015$ MeV~\cite{Inoue:2011ai}. Additionally, the NPLQCD Collaboration showed that when the unphysical pion mass is around $389$ MeV, the binding energy of the H-dibaryon is approximately $20$ MeV~\cite{NPLQCD:2010ocs}. Recently, the lattice simulations with SU(3)-flavor symmetry at $m_\pi=m_K\approx$ 420 MeV indicated a bound H-dibaryons with binding energy of several MeVs~\cite{Green:2021qol}. However, it is crucial to note that these lattice QCD calculations were performed with significantly large unphysical pion masses. Therefore, it remains challenging to definitively conclude the existence of the H-dibaryon.
In our model, the $\Lambda\Lambda$ system lacks significant isospin-isospin coupling interactions, resulting in weak interactions that make it difficult to form bound states. Hence, we have focused our attention on the two-body systems containing $\Sigma^{(\ast)}$ and $\Xi^{(\ast)}$ hyperons.

In Ref.~\cite{Polinder:2007mp}, Polinder \etal~investigated the leading-order interaction of the $\Sigma\Sigma$ system using the chiral effective field theory. They obtained large scattering length in the $\Sigma\Sigma$ system when the effective potential is iterated into the LS equation. Additionally, they observed that the scattering cross-section in the $0(0^+)$ channel of the $\Sigma\Sigma$ system can become significantly large near the threshold. This implies that the $\Sigma\Sigma$ system exhibits strong attractive interactions in the $0(0^+)$ partial wave, which could potentially lead to the formation of a bound state close to the threshold. This implication is consistent with our calculations, for example, we obtained a bound state with binding energy about $20$ MeV in the $0(0^+)$ channel (see Table~\ref{tab:SigmaSigma}).

The mass spectra of other systems can be directly obtained from Tables~\ref{tab:SigmaSigma} and~\ref{tab:SigmaXi}.

\begin{table*}[htbp]
\centering
\caption{The $I^{(G)}(J^{P(C)})$ quantum numbers, effective potentials, and bound/virtual state solutions of the $\Sigma^{(\ast)}\Sigma^{(\ast)}$ and $\Sigma^{(\ast)}\bar{\Sigma}^{(\ast)}$ systems. The notations are the same as those in Tables~\ref{tab:SigmacSigmac} and~\ref{tab:SigmacXic}.\label{tab:SigmaSigma}}
\setlength{\tabcolsep}{3.0mm}
{
\begin{tabular}{cccccccc}
\hline 
\hline
Systems $[m_{\text{th}}]$ & $I^G(J^{PC})$ & $V_{\mathcal{H}_1\mathcal{H}_2}^{I,J}$ & $E_{B}/E_{V}$ & Systems & $I(J^{P})$ & $V_{\mathcal{H}_1\mathcal{H}_2}^{I,J}$ & $E_{B}/E_{V}$\tabularnewline
\hline
\multirow{6}{*}{$\Sigma\bar{\Sigma}$ $\left[2386.3\right]$} & $0^{+}(0^{-+})$ & $8\tilde{c}_{a}-6\tilde{c}_{s}$ & $\cdots$ & \multirow{6}{*}{$\Sigma\Sigma$} & \multirow{2}{*}{$0(0^{+})$} & \multirow{2}{*}{$8c_{a}-6c_{s}$} & \multirow{2}{*}{$\left[2368.5_{-10.1}^{+8.9}\right]_{B}$}\tabularnewline
 & $0^{-}(1^{--})$ & $-\frac{8}{3}\tilde{c}_{a}-6\tilde{c}_{s}$ & $\left[2382.5,2386.3\right]_{B}^{\dagger}$ &  &  &  & \tabularnewline
 & $1^{-}(0^{-+})$ & $\frac{8}{3}\tilde{c}_{a}-2\tilde{c}_{s}$ & $\cdots$ &  & \multirow{2}{*}{$1(1^{+})$} & \multirow{2}{*}{$-\frac{8}{9}c_{a}-2c_{s}$} & \multirow{2}{*}{$\left[2363.2_{-11.4}^{+7.9}\right]_{V}$}\tabularnewline
 & $1^{+}(1^{--})$ & $-\frac{8}{9}\tilde{c}_{a}-2\tilde{c}_{s}$ & $\left[2274.2,2336.6\right]_{V}$ &  &  &  & \tabularnewline
 & $2^{+}(0^{-+})$ & $-8\tilde{c}_{a}+6\tilde{c}_{s}$ & $\left[2331.3,2380.2\right]_{B}$ &  & \multirow{2}{*}{$2(0^{+})$} & \multirow{2}{*}{$-8c_{a}+6c_{s}$} & \multirow{2}{*}{$\cdots$}\tabularnewline
 & $2^{-}(1^{--})$ & $\frac{8}{3}\tilde{c}_{a}+6\tilde{c}_{s}$ & $\cdots$ &  &  &  & \tabularnewline
\hline 
\multirow{12}{*}{$\Sigma\bar{\Sigma}^{\ast}$ $\left[2577.7\right]$} & $0^{-}(1^{--})$ & $\frac{22}{3}\tilde{c}_{a}-6\tilde{c}_{s}$ & $\cdots$ & \multirow{12}{*}{$\Sigma\Sigma^{\ast}$} & \multirow{2}{*}{$0(1^{+})$} & \multirow{2}{*}{$6c_{a}-6c_{s}$} & \multirow{2}{*}{$\left[2557.8_{-8.9}^{+8.1}\right]_{B}$}\tabularnewline
 & $0^{+}(1^{-+})$ & $6\tilde{c}_{a}-6\tilde{c}_{s}$ & $\cdots$ &  &  &  & \tabularnewline
 & $0^{+}(2^{-+})$ & $-2\tilde{c}_{a}-6\tilde{c}_{s}$ & $\left[2574.6,2577.7\right]_{B}^{\dagger}$ &  & \multirow{2}{*}{$0(2^{+})$} & \multirow{2}{*}{$-2c_{a}-6c_{s}$} & \multirow{2}{*}{$\left[2561.7_{-5.7}^{+5.2}\right]_{B}$}\tabularnewline
 & $0^{-}(2^{--})$ & $-6\tilde{c}_{a}-6\tilde{c}_{s}$ & $\left[2543.6,2544.7\right]_{B}$ &  &  &  & \tabularnewline
 & $1^{+}(1^{--})$ & $\frac{22}{9}\tilde{c}_{a}-2\tilde{c}_{s}$ & $\cdots$ &  & \multirow{2}{*}{$1(1^{+})$} & \multirow{2}{*}{$\frac{22}{9}c_{a}-2c_{s}$} & \multirow{2}{*}{$\left[2566.5_{-11.3}^{+6.2}\right]_{V}$}\tabularnewline
 & $1^{-}(1^{-+})$ & $2\tilde{c}_{a}-2\tilde{c}_{s}$ & $\cdots$ &  &  &  & \tabularnewline
 & $1^{-}(2^{-+})$ & $-\frac{2}{3}\tilde{c}_{a}-2\tilde{c}_{s}$ & $\left[2409.1,2521.6\right]_{V}$ &  & \multirow{2}{*}{$1(2^{+})$} & \multirow{2}{*}{$-2c_{a}-2c_{s}$} & \multirow{2}{*}{$\left[2560.0_{-14.9}^{+8.7}\right]_{V}$}\tabularnewline
 & $1^{+}(2^{--})$ & $-2\tilde{c}_{a}-2\tilde{c}_{s}$ & $\left[2573.8,2574.2\right]_{V}$ &  &  &  & \tabularnewline
 & $2^{-}(1^{--})$ & $-\frac{22}{3}\tilde{c}_{a}+6\tilde{c}_{s}$ & $\left[2528.3,2573.7\right]_{B}$ &  & \multirow{2}{*}{$2(1^{+})$} & \multirow{2}{*}{$-6c_{a}+6c_{s}$} & \multirow{2}{*}{$\cdots$}\tabularnewline
 & $2^{+}(1^{-+})$ & $-6\tilde{c}_{a}+6\tilde{c}_{s}$ & $\left[2546.7,2577.7\right]_{B}^{\dagger}$ &  &  &  & \tabularnewline
 & $2^{+}(2^{-+})$ & $2\tilde{c}_{a}+6\tilde{c}_{s}$ & $\cdots$ &  & \multirow{2}{*}{$2(2^{+})$} & \multirow{2}{*}{$2c_{a}+6c_{s}$} & \multirow{2}{*}{$\cdots$}\tabularnewline
 & $2^{-}(2^{--})$ & $6\tilde{c}_{a}+6\tilde{c}_{s}$ & $\cdots$ &  &  &  & \tabularnewline
\hline 
\multirow{12}{*}{$\Sigma^{\ast}\bar{\Sigma}^{\ast}$ $\left[2769.2\right]$} & $0^{+}(0^{-+})$ & $10\tilde{c}_{a}-6\tilde{c}_{s}$ & $\cdots$ & \multirow{12}{*}{$\Sigma^{\ast}\Sigma^{\ast}$} & \multirow{2}{*}{$0(0^{+})$} & \multirow{2}{*}{$10c_{a}-6c_{s}$} & \multirow{2}{*}{$\left[2744.0_{-12.5}^{+11.4}\right]_{B}$}\tabularnewline
 & $0^{-}(1^{--})$ & $\frac{22}{3}\tilde{c}_{a}-6\tilde{c}_{s}$ & $\cdots$ &  &  &  & \tabularnewline
 & $0^{+}(2^{-+})$ & $2\tilde{c}_{a}-6\tilde{c}_{s}$ & $\cdots$ &  & \multirow{2}{*}{$0(2^{+})$} & \multirow{2}{*}{$2c_{a}-6c_{s}$} & \multirow{2}{*}{$\left[2748.0_{-6.0}^{+5.7}\right]_{B}$}\tabularnewline
 & $0^{-}(3^{--})$ & $-6\tilde{c}_{a}-6\tilde{c}_{s}$ & $\left[2731.4,2732.5\right]_{B}$ &  &  &  & \tabularnewline
 & $1^{-}(0^{-+})$ & $\frac{10}{3}\tilde{c}_{a}-2\tilde{c}_{s}$ & $\cdots$ &  & \multirow{2}{*}{$1(1^{+})$} & \multirow{2}{*}{$\frac{22}{9}c_{a}-2c_{s}$} & \multirow{2}{*}{$\left[2762.3_{-8.5}^{+4.4}\right]_{V}$}\tabularnewline
 & $1^{+}(1^{--})$ & $\frac{22}{9}\tilde{c}_{a}-2\tilde{c}_{s}$ & $\cdots$ &  &  &  & \tabularnewline
 & $1^{-}(2^{-+})$ & $\frac{2}{3}\tilde{c}_{a}-2\tilde{c}_{s}$ & $\cdots$ &  & \multirow{2}{*}{$1(3^{+})$} & \multirow{2}{*}{$-2c_{a}-2c_{s}$} & \multirow{2}{*}{$\left[2757.0_{-11.5}^{+6.4}\right]_{V}$}\tabularnewline
 & $1^{+}(3^{--})$ & $-2\tilde{c}_{a}-2\tilde{c}_{s}$ & $\left[2767.3,2767.6\right]_{V}$ &  &  &  & \tabularnewline
 & $2^{+}(0^{-+})$ & $-10\tilde{c}_{a}+6\tilde{c}_{s}$ & $\left[2676.1,2740.8\right]_{B}$ &  & \multirow{2}{*}{$2(0^{+})$} & \multirow{2}{*}{$-10c_{a}+6c_{s}$} & \multirow{2}{*}{$\cdots$}\tabularnewline
 & $2^{-}(1^{--})$ & $-\frac{22}{3}\tilde{c}_{a}+6\tilde{c}_{s}$ & $\left[2715.7,2763.2\right]_{B}$ &  &  &  & \tabularnewline
 & $2^{+}(2^{-+})$ & $-2\tilde{c}_{a}+6\tilde{c}_{s}$ & $\left[\sim2765.2\right]_{V}^{\sharp}$ &  & \multirow{2}{*}{$2(2^{+})$} & \multirow{2}{*}{$-2c_{a}+6c_{s}$} & \multirow{2}{*}{$\cdots$}\tabularnewline
 & $2^{-}(3^{--})$ & $6\tilde{c}_{a}+6\tilde{c}_{s}$ & $\cdots$ &  &  &  & \tabularnewline
\hline 
\hline 
\end{tabular}
}
\end{table*}

\begin{table*}[htbp]
\centering
\caption{The $I^{(G)}(J^{P(C)})$ quantum numbers, effective potentials, and bound/virtual state solutions of the $\Sigma^{(\ast)}\Xi^{(\ast)}$, $\Sigma^{(\ast)}\bar{\Xi}^{(\ast)}$, $\Xi^{(\ast)}\Xi^{(\ast)}$ and $\Xi^{(\ast)}\bar{\Xi}^{(\ast)}$ systems. The notations are the same as those in Tables~\ref{tab:SigmacSigmac} and~\ref{tab:SigmacXic}.\label{tab:SigmaXi}}
\setlength{\tabcolsep}{2.7mm}
{
\begin{tabular}{cccccccc}
\hline 
\hline
Systems $[m_{\text{th}}]$ & $I^{(G)}(J^{P(C)})$ & $V_{\mathcal{H}_1\mathcal{H}_2}^{I,J}$ & $E_{B}/E_{V}$ & Systems & $I(J^{P})$ & $V_{\mathcal{H}_1\mathcal{H}_2}^{I,J}$ & $E_{B}/E_{V}$\tabularnewline
\hline
\multirow{4}{*}{$\Sigma\bar{\Xi}$ $\left[2511.4\right]$} & $\frac{1}{2}(0^{-})$ & $-2\tilde{c}_{a}-3\tilde{c}_{s}$ & $\left[2506.9,2510.3\right]_{V}$ & \multirow{4}{*}{$\Sigma\Xi$} & $\frac{1}{2}(0^{+})$ & $-2c_{a}-3c_{s}$ & $\left[2510.2_{-3.3}^{+1.2}\right]_{V}$\tabularnewline
 & $\frac{1}{2}(1^{-})$ & $\frac{2}{3}\tilde{c}_{a}-3\tilde{c}_{s}$ & $\cdots$ &  & $\frac{1}{2}(1^{+})$ & $\frac{2}{3}c_{a}-3c_{s}$ & $\left[2510.8_{-1.6}^{+0.6}\right]_{V}$\tabularnewline
 & $\frac{3}{2}(0^{-})$ & $2\tilde{c}_{a}+3\tilde{c}_{s}$ & $\cdots$ &  & $\frac{3}{2}(0^{+})$ & $2c_{a}+3c_{s}$ & $\cdots$\tabularnewline
 & $\frac{3}{2}(1^{-})$ & $-\frac{2}{3}\tilde{c}_{a}+3\tilde{c}_{s}$ & $\cdots$ &  & $\frac{3}{2}(1^{+})$ & $-\frac{2}{3}c_{a}+3c_{s}$ & $\cdots$\tabularnewline
\hline 
\multirow{4}{*}{$\Sigma\bar{\Xi}^{\ast}$ $\left[2726.6\right]$} & $\frac{1}{2}(1^{-})$ & $\frac{10}{3}\tilde{c}_{a}-3\tilde{c}_{s}$ & $\cdots$ & \multirow{4}{*}{$\Sigma\Xi^{\ast}$} & $\frac{1}{2}(1^{+})$ & $\frac{10}{3}c_{a}-3c_{s}$ & $\left[2726.5_{-1.2}^{+0.1}\right]_{B}^{\dagger}$\tabularnewline
 & $\frac{1}{2}(2^{-})$ & $-2\tilde{c}_{a}-3\tilde{c}_{s}$ & $\left[2724.2,2726.3\right]_{V}$ &  & $\frac{1}{2}(2^{+})$ & $-2c_{a}-3c_{s}$ & $\left[2726.2_{-2.1}^{+0.3}\right]_{V}^{\dagger}$\tabularnewline
 & $\frac{3}{2}(1^{-})$ & $-\frac{10}{3}\tilde{c}_{a}+3\tilde{c}_{s}$ & $\left[2712.3,2726.6\right]_{V}$ &  & $\frac{3}{2}(1^{+})$ & $-\frac{10}{3}c_{a}+3c_{s}$ & $\cdots$\tabularnewline
 & $\frac{3}{2}(2^{-})$ & $2\tilde{c}_{a}+3\tilde{c}_{s}$ & $\cdots$ &  & $\frac{3}{2}(2^{+})$ & $2c_{a}+3c_{s}$ & $\cdots$\tabularnewline
\hline 
\multirow{4}{*}{$\Sigma^{\ast}\bar{\Xi}$ $\left[2702.9\right]$} & $\frac{1}{2}(1^{-})$ & $-\frac{5}{3}\tilde{c}_{a}-3\tilde{c}_{s}$ & $\left[2694.8,2701.3\right]_{V}$ & \multirow{4}{*}{$\Sigma^{\ast}\Xi$} & $\frac{1}{2}(1^{+})$ & $-\frac{5}{3}c_{a}-3c_{s}$ & $\left[2702.6_{-1.7}^{+0.3}\right]_{V}^{\dagger}$\tabularnewline
 & $\frac{1}{2}(2^{-})$ & $\tilde{c}_{a}-3\tilde{c}_{s}$ & $\cdots$ &  & $\frac{1}{2}(2^{+})$ & $c_{a}-3c_{s}$ & $\left[2702.8_{-0.8}^{+0.1}\right]_{V}^{\dagger}$\tabularnewline
 & $\frac{3}{2}(1^{-})$ & $\frac{5}{3}\tilde{c}_{a}+3\tilde{c}_{s}$ & $\cdots$ &  & $\frac{3}{2}(1^{+})$ & $\frac{5}{3}c_{a}+3c_{s}$ & $\cdots$\tabularnewline
 & $\frac{3}{2}(2^{-})$ & $-\tilde{c}_{a}+3\tilde{c}_{s}$ & $\left[\sim2636.9\right]_{V}^{\sharp}$ &  & $\frac{3}{2}(2^{+})$ & $-c_{a}+3c_{s}$ & $\cdots$\tabularnewline
\hline 
\multirow{8}{*}{$\Sigma^{\ast}\bar{\Xi}^{\ast}$ $\left[2918.0\right]$} & $\frac{1}{2}(0^{-})$ & $5\tilde{c}_{a}-3\tilde{c}_{s}$ & $\cdots$ & \multirow{8}{*}{$\Sigma^{\ast}\Xi^{\ast}$} & $\frac{1}{2}(0^{+})$ & $5c_{a}-3c_{s}$ & $\left[2917.4_{-2.7}^{+0.5}\right]_{B}^{\dagger}$\tabularnewline
 & $\frac{1}{2}(1^{-})$ & $\frac{11}{3}\tilde{c}_{a}-3\tilde{c}_{s}$ & $\cdots$ &  & $\frac{1}{2}(1^{+})$ & $\frac{11}{3}c_{a}-3c_{s}$ & $\left[2917.6_{-1.9}^{+0.4}\right]_{B}^{\dagger}$\tabularnewline
 & $\frac{1}{2}(2^{-})$ & $\tilde{c}_{a}-3\tilde{c}_{s}$ & $\cdots$ &  & $\frac{1}{2}(2^{+})$ & $c_{a}-3c_{s}$ & $\left[2917.9_{-0.7}^{+0.1}\right]_{B}^{\dagger}$\tabularnewline
 & $\frac{1}{2}(3^{-})$ & $-3\tilde{c}_{a}-3\tilde{c}_{s}$ & $\left[2914.7,2915.0\right]_{B}$ &  & $\frac{1}{2}(3^{+})$ & $-3c_{a}-3c_{s}$ & $\left[2917.9_{-1.6}^{+0.1}\right]_{V}^{\dagger}$\tabularnewline
 & $\frac{3}{2}(0^{-})$ & $-5\tilde{c}_{a}+3\tilde{c}_{s}$ & $\left[2893.8,2916.9\right]_{B}$ &  & $\frac{3}{2}(0^{+})$ & $-5c_{a}+3c_{s}$ & $\cdots$\tabularnewline
 & $\frac{3}{2}(1^{-})$ & $-\frac{11}{3}\tilde{c}_{a}+3\tilde{c}_{s}$ & $\left[2909.7,2918.0\right]_{B}^{\dagger}$ &  & $\frac{3}{2}(1^{+})$ & $-\frac{11}{3}c_{a}+3c_{s}$ & $\cdots$\tabularnewline
 & $\frac{3}{2}(2^{-})$ & $-\tilde{c}_{a}+3\tilde{c}_{s}$ & $\cdots$ &  & $\frac{3}{2}(2^{+})$ & $-c_{a}+3c_{s}$ & $\cdots$\tabularnewline
 & $\frac{3}{2}(3^{-})$ & $3\tilde{c}_{a}+3\tilde{c}_{s}$ & $\cdots$ &  & $\frac{3}{2}(3^{+})$ & $3c_{a}+3c_{s}$ & $\cdots$\tabularnewline
\hline 
\multirow{4}{*}{$\Xi\bar{\Xi}$ $\left[2636.6\right]$} & $0^{+}(0^{-+})$ & $\frac{5}{6}\tilde{c}_{a}-\frac{5}{2}\tilde{c}_{s}$ & $\cdots$ & \multirow{4}{*}{$\Xi\Xi$} & \multirow{2}{*}{$0(1^{+})$} & \multirow{2}{*}{$-\frac{5}{18}c_{a}-\frac{5}{2}c_{s}$} & \multirow{2}{*}{$\left[2633.1_{-2.8}^{+1.8}\right]_{V}$}\tabularnewline
 & $0^{-}(1^{--})$ & $-\frac{5}{18}\tilde{c}_{a}-\frac{5}{2}\tilde{c}_{s}$ & $\left[\sim2559.9\right]_{V}^{\sharp}$ &  &  &  & \tabularnewline
 & $1^{-}(0^{-+})$ & $-\frac{1}{2}\tilde{c}_{a}+\frac{3}{2}\tilde{c}_{s}$ & $\cdots$ &  & \multirow{2}{*}{$1(0^{+})$} & \multirow{2}{*}{$-\frac{1}{2}c_{a}+\frac{3}{2}c_{s}$} & \multirow{2}{*}{$\cdots$}\tabularnewline
 & $1^{+}(1^{--})$ & $\frac{1}{6}\tilde{c}_{a}+\frac{3}{2}\tilde{c}_{s}$ & $\cdots$ &  &  &  & \tabularnewline
\hline 
\multirow{8}{*}{$\Xi\bar{\Xi}^{\ast}$ $\left[2851.7\right]$} & $0^{-}(1^{--})$ & $-\frac{5}{18}\tilde{c}_{a}-\frac{5}{2}\tilde{c}_{s}$ & $\left[\sim2796.4\right]_{V}^{\sharp}$ & \multirow{8}{*}{$\Xi\Xi^{\ast}$} & \multirow{2}{*}{$0(1^{+})$} & \multirow{2}{*}{$-\frac{5}{18}c_{a}-\frac{5}{2}c_{s}$} & \multirow{2}{*}{$\left[2850.1_{-1.9}^{+1.1}\right]_{V}$}\tabularnewline
 & $0^{+}(1^{-+})$ & $-\frac{5}{2}\tilde{c}_{a}-\frac{5}{2}\tilde{c}_{s}$ & $\left[2851.4,2851.6\right]_{B}$ &  &  &  & \tabularnewline
 & $0^{+}(2^{-+})$ & $\frac{25}{6}\tilde{c}_{a}-\frac{5}{2}\tilde{c}_{s}$ & $\cdots$ &  & \multirow{2}{*}{$0(2^{+})$} & \multirow{2}{*}{$-\frac{5}{2}c_{a}-\frac{5}{2}c_{s}$} & \multirow{2}{*}{$\left[2849.1_{-5.4}^{+2.2}\right]_{V}$}\tabularnewline
 & $0^{-}(2^{--})$ & $-\frac{5}{2}\tilde{c}_{a}-\frac{5}{2}\tilde{c}_{s}$ & $\left[2851.4,2851.6\right]_{B}$ &  &  &  & \tabularnewline
 & $1^{+}(1^{--})$ & $\frac{1}{6}\tilde{c}_{a}+\frac{3}{2}\tilde{c}_{s}$ & $\cdots$ &  & \multirow{2}{*}{$1(1^{+})$} & \multirow{2}{*}{$\frac{3}{2}c_{a}+\frac{3}{2}c_{s}$} & \multirow{2}{*}{$\cdots$}\tabularnewline
 & $1^{-}(1^{-+})$ & $\frac{3}{2}\tilde{c}_{a}+\frac{3}{2}\tilde{c}_{s}$ & $\cdots$ &  &  &  & \tabularnewline
 & $1^{-}(2^{-+})$ & $-\frac{5}{2}\tilde{c}_{a}+\frac{3}{2}\tilde{c}_{s}$ & $\left[2834.6,2851.7\right]_{V}^{\dagger}$ &  & \multirow{2}{*}{$1(2^{+})$} & \multirow{2}{*}{$-\frac{5}{2}c_{a}+\frac{3}{2}c_{s}$} & \multirow{2}{*}{$\cdots$}\tabularnewline
 & $1^{+}(2^{--})$ & $\frac{3}{2}\tilde{c}_{a}+\frac{3}{2}\tilde{c}_{s}$ & $\cdots$ &  &  &  & \tabularnewline
\hline 
\multirow{8}{*}{$\Xi^{\ast}\bar{\Xi}^{\ast}$ $\left[3066.8\right]$} & $0^{+}(0^{-+})$ & $\frac{25}{6}\tilde{c}_{a}-\frac{5}{2}\tilde{c}_{s}$ & $\cdots$ & \multirow{8}{*}{$\Xi^{\ast}\Xi^{\ast}$} & \multirow{2}{*}{$0(1^{+})$} & \multirow{2}{*}{$\frac{55}{18}c_{a}-\frac{5}{2}c_{s}$} & \multirow{2}{*}{$\left[3066.7_{-1.8}^{+0.1}\right]_{V}^{\dagger}$}\tabularnewline
 & $0^{-}(1^{--})$ & $\frac{55}{18}\tilde{c}_{a}-\frac{5}{2}\tilde{c}_{s}$ & $\cdots$ &  &  &  & \tabularnewline
 & $0^{+}(2^{-+})$ & $\frac{5}{6}\tilde{c}_{a}-\frac{5}{2}\tilde{c}_{s}$ & $\cdots$ &  & \multirow{2}{*}{$0(3^{+})$} & \multirow{2}{*}{$-\frac{5}{2}c_{a}-\frac{5}{2}c_{s}$} & \multirow{2}{*}{$\left[3065.8_{-3.6}^{+1.0}\right]_{V}$}\tabularnewline
 & $0^{-}(3^{--})$ & $-\frac{5}{2}\tilde{c}_{a}-\frac{5}{2}\tilde{c}_{s}$ & $\left[3065.9,3066.1\right]_{B}$ &  &  &  & \tabularnewline
 & $1^{-}(0^{-+})$ & $-\frac{5}{2}\tilde{c}_{a}+\frac{3}{2}\tilde{c}_{s}$ & $\left[3055.4,3066.8\right]_{V}^{\dagger}$ &  & \multirow{2}{*}{$1(0^{+})$} & \multirow{2}{*}{$-\frac{5}{2}c_{a}+\frac{3}{2}c_{s}$} & \multirow{2}{*}{$\cdots$}\tabularnewline
 & $1^{+}(1^{--})$ & $-\frac{11}{6}\tilde{c}_{a}+\frac{3}{2}\tilde{c}_{s}$ & $\left[3016.7,3065.1\right]_{V}$ &  &  &  & \tabularnewline
 & $1^{-}(2^{-+})$ & $-\frac{1}{2}\tilde{c}_{a}+\frac{3}{2}\tilde{c}_{s}$ & $\cdots$ &  & \multirow{2}{*}{$1(2^{+})$} & \multirow{2}{*}{$-\frac{1}{2}c_{a}+\frac{3}{2}c_{s}$} & \multirow{2}{*}{$\cdots$}\tabularnewline
 & $1^{+}(3^{--})$ & $\frac{3}{2}\tilde{c}_{a}+\frac{3}{2}\tilde{c}_{s}$ & $\cdots$ &  &  &  & \tabularnewline
\hline 
\hline 
\end{tabular}
}
\end{table*}

\subsection{Charmed baryon-(anti)doubly charmed baryon systems}\label{sec:BQBQQ}

The spectra of charmed baryon-(anti)doubly charmed baryon systems are given in Tables~\ref{tab:SigmacXicc} and~\ref{tab:XicXicc}.

The doubly charmed baryon $\Xi_{cc}^{(\ast)}$ can be associated with the anticharm meson $\bar{D}^{(\ast)}$ with the heavy diquark-antiquark symmetry. Consequently, the $\Xi_{cc}^{(\ast)} \Sigma_c^{(\ast)}$ and $\Xi_{cc}^{(\ast)} \Xi_c^{(\ast)}$ systems correspond respectively to the $\bar{D}^{(\ast)} \Sigma_c^{(\ast)}$ and $\bar{D}^{(\ast)} \Xi_c^{(\ast)}$ systems. If the hidden-charm pentaquarks $P_{\psi}^N$ and $P_{\psi s}^\Lambda$ are indeed molecular states of $\bar{D}^{(\ast)} \Sigma_c$ and $\bar{D}^{(\ast)} \Xi_c$, it implies the presence of corresponding molecular states in the $\Xi_{cc}^{(\ast)} \Sigma_c$ and $\Xi_{cc}^{(\ast)} \Xi_c$ systems as well. A comprehensive analysis of the spin-flavor symmetry between these systems can be found in Ref.~\cite{Chen:2024tuu}.

In Ref.~\cite{Chen:2018pzd}, Chen \etal~investigated the $\Xi_{cc} \Sigma_c^{(\ast)}$ and $\Xi_{cc} \Xi_c^{(\ast)}$ systems using the OBE model, taking into account the coupled-channel effects. They obtained molecular states in both the highest isospin channels and the lowest isospin channels. The results in the lowest isospin channels align with our calculations. However, our calculations indicate the absence of bound states in the highest isospin channel in such systems.
Junnarkar \etal~obtained a molecular state with a binding energy of $-8\pm17$ MeV in the $\Xi_{cc}\Sigma_c$ system with $J^P=1^+$ through the lattice QCD calculations~\cite{Junnarkar:2019equ}. This result is in good agreement with our findings in the $\Xi_{cc}\Sigma_c$ system with $0(1^+)$, where we obtained a binding energy of $-6.7^{+2.2}_{-2.3}$ MeV. The results presented in Ref.~\cite{Junnarkar:2019equ} have also been used by Pan \etal~to determine the spin of the $P_\psi^N(4440)$ and $P_\psi^N(4457)$ states~\cite{Pan:2019skd}. The mass hierarchy of the $0^+$ and $1^+$ states in the $\Xi_{cc}\Sigma_c$ system exactly opposes that of the $\bar{D}^\ast\Sigma_c$ system, which consists of $\frac{1}{2}^-$ and $\frac{3}{2}^-$ states. Therefore, if the mass splitting of the $0^+$ and $1^+$ states in the $\Xi_{cc}\Sigma_c$ system can be accurately calculated on the lattice, it can be used to infer the spin of the $P_\psi^N(4440)$ and $P_\psi^N(4457)$ based on their masses. 

Furthermore, we also calculated the mass spectra of the $\Xi_{cc}^{(\ast)}\bar{\Sigma}_c^{(\ast)}$ and $\Xi_{cc}^{(\ast)}\bar{\Xi}_c^{(\prime,\ast)}$ systems. These systems will correspond to the $\bar{D}^{(\ast)}\bar{\Sigma}_c^{(\ast)}$ and $\bar{D}^{(\ast)}\bar{\Xi}_c^{(\prime,\ast)}$ systems, respectively, within the heavy diquark-antiquark symmetry. Therefore, if the $\bar{D}^{(\ast)}\bar{\Sigma}_c^{(\ast)}$ and $\bar{D}^{(\ast)}\bar{\Xi}_c^{(\prime,\ast)}$ systems both contain double-charm pentaquarks~\cite{Chen:2021htr,Wang:2023eng,Wang:2023ael}, it naturally follows that the $\Xi_{cc}^{(\ast)}\bar{\Sigma}_c^{(\ast)}$ and $\Xi_{cc}^{(\ast)}\bar{\Xi}_c^{(\prime,\ast)}$ systems also possess triple-charm hexaquarks.

\begin{table*}[htbp]
\centering
\caption{The $I(J^{P})$ quantum numbers, effective potentials, and bound/virtual state solutions of the $\Xi_{cc}^{(\ast)}\Sigma_c^{(\ast)}$, $\Xi_{cc}^{(\ast)}\bar{\Sigma}_c^{(\ast)}$, $\Xi_{cc}\Xi_c^{(\prime,\ast)}$ and $\Xi_{cc}\bar{\Xi}_c^{(\prime,\ast)}$ systems. The notations are the same as those in Tables~\ref{tab:SigmacSigmac} and~\ref{tab:SigmacXic}.\label{tab:SigmacXicc}}
\setlength{\tabcolsep}{1.9mm}
{
\begin{tabular}{cccccccc}
\hline 
\hline
Systems $[m_{\text{th}}]$ & $I(J^{P})$ & $V_{\mathcal{H}_1\mathcal{H}_2}^{I,J}$ & $E_{B}/E_{V}$ & Systems & $I(J^{P})$ & $V_{\mathcal{H}_1\mathcal{H}_2}^{I,J}$ & $E_{B}/E_{V}$\tabularnewline
\hline
\multirow{4}{*}{$\Xi_{cc}\bar{\Sigma}_{c}$ $\left[6075.0\right]$} & $\frac{1}{2}(0^{-})$ & $-2\tilde{c}_{a}-3\tilde{c}_{s}$ & $\left[6064.6,6068.4\right]_{B}$ & \multirow{4}{*}{$\Xi_{cc}\Sigma_{c}$} & $\frac{1}{2}(0^{+})$ & $-2c_{a}-3c_{s}$ & $\left[6064.9_{-3.8}^{+3.5}\right]_{B}$\tabularnewline
 & $\frac{1}{2}(1^{-})$ & $\frac{2}{3}\tilde{c}_{a}-3\tilde{c}_{s}$ & $\cdots$ &  & $\frac{1}{2}(1^{+})$ & $\frac{2}{3}c_{a}-3c_{s}$ & $\left[6063.5_{-2.8}^{+2.7}\right]_{B}$\tabularnewline
 & $\frac{3}{2}(0^{-})$ & $2\tilde{c}_{a}+3\tilde{c}_{s}$ & $\cdots$ &  & $\frac{3}{2}(0^{+})$ & $2c_{a}+3c_{s}$ & $\cdots$\tabularnewline
 & $\frac{3}{2}(1^{-})$ & $-\frac{2}{3}\tilde{c}_{a}+3\tilde{c}_{s}$ & $\left[\sim6064.8\right]_{V}^{\sharp}$ &  & $\frac{3}{2}(1^{+})$ & $-\frac{2}{3}c_{a}+3c_{s}$ & $\cdots$\tabularnewline
\hline 
\multirow{4}{*}{$\Xi_{cc}\bar{\Sigma}_{c}^{\ast}$ $\left[6139.7\right]$} & $\frac{1}{2}(1^{-})$ & $-\frac{5}{3}\tilde{c}_{a}-3\tilde{c}_{s}$ & $\left[6131.9,6136.3\right]_{B}$ & \multirow{4}{*}{$\Xi_{cc}\Sigma_{c}^{\ast}$} & $\frac{1}{2}(1^{+})$ & $-\frac{5}{3}c_{a}-3c_{s}$ & $\left[6129.1_{-3.6}^{+3.4}\right]_{B}$\tabularnewline
 & $\frac{1}{2}(2^{-})$ & $\tilde{c}_{a}-3\tilde{c}_{s}$ & $\cdots$ &  & $\frac{1}{2}(2^{+})$ & $c_{a}-3c_{s}$ & $\left[6127.7_{-3.1}^{+3.0}\right]_{B}$\tabularnewline
 & $\frac{3}{2}(1^{-})$ & $\frac{5}{3}\tilde{c}_{a}+3\tilde{c}_{s}$ & $\cdots$ &  & $\frac{3}{2}(1^{+})$ & $\frac{5}{3}c_{a}+3c_{s}$ & $\cdots$\tabularnewline
 & $\frac{3}{2}(2^{-})$ & $-\tilde{c}_{a}+3\tilde{c}_{s}$ & $\left[\sim6138.9\right]_{V}^{\sharp}$ &  & $\frac{3}{2}(2^{+})$ & $-c_{a}+3c_{s}$ & $\cdots$\tabularnewline
\hline 
\multirow{4}{*}{$\Xi_{cc}\bar{\Xi}_{c}$ $\left[6090.6\right]$} & $0(0^{-})$ & $-\frac{5}{2}\tilde{c}_{a}-\frac{5}{2}\tilde{c}_{s}$ & $\left[6077.4,6077.8\right]_{B}$ & \multirow{4}{*}{$\Xi_{cc}\Xi_{c}$} & $0(0^{+})$ & $-\frac{5}{2}c_{a}-\frac{5}{2}c_{s}$ & $\left[6085.4_{-3.4}^{+2.9}\right]_{B}$\tabularnewline
 & $0(1^{-})$ & $\frac{5}{6}\tilde{c}_{a}-\frac{5}{2}\tilde{c}_{s}$ & $\cdots$ &  & $0(1^{+})$ & $\frac{5}{6}c_{a}-\frac{5}{2}c_{s}$ & $\left[6083.9_{-2.3}^{+2.2}\right]_{B}$\tabularnewline
 & $1(0^{-})$ & $\frac{3}{2}\tilde{c}_{a}+\frac{3}{2}\tilde{c}_{s}$ & $\cdots$ &  & $1(0^{+})$ & $\frac{3}{2}c_{a}+\frac{3}{2}c_{s}$ & $\cdots$\tabularnewline
 & $1(1^{-})$ & $-\frac{1}{2}\tilde{c}_{a}+\frac{3}{2}\tilde{c}_{s}$ & $\left[\sim6067.2\right]_{V}^{\sharp}$ &  & $1(1^{+})$ & $-\frac{1}{2}c_{a}+\frac{3}{2}c_{s}$ & $\cdots$\tabularnewline
\hline 
\multirow{4}{*}{$\Xi_{cc}\bar{\Xi}_{c}^{\prime}$ $\left[6199.8\right]$} & $0(0^{-})$ & $\frac{5}{6}\tilde{c}_{a}-\frac{5}{2}\tilde{c}_{s}$ & $\cdots$ & \multirow{4}{*}{$\Xi_{cc}\Xi_{c}^{\prime}$} & $0(0^{+})$ & $\frac{5}{6}c_{a}-\frac{5}{2}c_{s}$ & $\left[6192.6_{-2.4}^{+2.2}\right]_{B}$\tabularnewline
 & $0(1^{-})$ & $-\frac{5}{18}\tilde{c}_{a}-\frac{5}{2}\tilde{c}_{s}$ & $\left[6136.4,6199.1\right]_{V}$ &  & $0(1^{+})$ & $-\frac{5}{18}c_{a}-\frac{5}{2}c_{s}$ & $\left[6193.1_{-2.0}^{+1.8}\right]_{B}$\tabularnewline
 & $1(0^{-})$ & $-\frac{1}{2}\tilde{c}_{a}+\frac{3}{2}\tilde{c}_{s}$ & $\left[\sim6178.5\right]_{V}^{\sharp}$ &  & $1(0^{+})$ & $-\frac{1}{2}c_{a}+\frac{3}{2}c_{s}$ & $\cdots$\tabularnewline
 & $1(1^{-})$ & $\frac{1}{6}\tilde{c}_{a}+\frac{3}{2}\tilde{c}_{s}$ & $\cdots$ &  & $1(1^{+})$ & $\frac{1}{6}c_{a}+\frac{3}{2}c_{s}$ & $\cdots$\tabularnewline
\hline 
\multirow{4}{*}{$\Xi_{cc}\bar{\Xi}_{c}^{\ast}$ $\left[6266.7\right]$} & $0(1^{-})$ & $-\frac{25}{18}\tilde{c}_{a}-\frac{5}{2}\tilde{c}_{s}$ & $\left[6262.4,6265.4\right]_{B}$ & \multirow{4}{*}{$\Xi_{cc}\Xi_{c}^{\ast}$} & $0(1^{+})$ & $-\frac{25}{18}c_{a}-\frac{5}{2}c_{s}$ & $\left[6260.3_{-2.7}^{+2.5}\right]_{B}$\tabularnewline
 & $0(2^{-})$ & $\frac{5}{6}\tilde{c}_{a}-\frac{5}{2}\tilde{c}_{s}$ & $\cdots$ &  & $0(2^{+})$ & $\frac{5}{6}c_{a}-\frac{5}{2}c_{s}$ & $\left[6259.3_{-2.4}^{+2.2}\right]_{B}$\tabularnewline
 & $1(1^{-})$ & $\frac{5}{6}\tilde{c}_{a}+\frac{3}{2}\tilde{c}_{s}$ & $\cdots$ &  & $1(1^{+})$ & $\frac{5}{6}c_{a}+\frac{3}{2}c_{s}$ & $\cdots$\tabularnewline
 & $1(2^{-})$ & $-\frac{1}{2}\tilde{c}_{a}+\frac{3}{2}\tilde{c}_{s}$ & $\left[\sim6245.9\right]_{V}^{\sharp}$ &  & $1(2^{+})$ & $-\frac{1}{2}c_{a}+\frac{3}{2}c_{s}$ & $\cdots$\tabularnewline
\hline 
\multirow{4}{*}{$\Xi_{cc}^{\ast}\bar{\Sigma}_{c}$ $\left[6160.0\pm15\right]$} & $\frac{1}{2}(1^{-})$ & $\frac{10}{3}\tilde{c}_{a}-3\tilde{c}_{s}$ & $\cdots$ & \multirow{4}{*}{$\Xi_{cc}^{\ast}\Sigma_{c}$} & $\frac{1}{2}(1^{+})$ & $\frac{10}{3}c_{a}-3c_{s}$ & $\left[6147.0_{-5.0}^{+4.7}\pm15\right]_{B}$\tabularnewline
 & $\frac{1}{2}(2^{-})$ & $-2\tilde{c}_{a}-3\tilde{c}_{s}$ & $\left[6151.3\pm1.9\pm15\right]_{B}$ &  & $\frac{1}{2}(2^{+})$ & $-2c_{a}-3c_{s}$ & $\left[6149.7_{-3.8}^{+3.6}\pm15\right]_{B}$\tabularnewline
 & $\frac{3}{2}(1^{-})$ & $-\frac{10}{3}\tilde{c}_{a}+3\tilde{c}_{s}$ & $\left[6147.4\pm10.7\pm15\right]_{B}$ &  & $\frac{3}{2}(1^{+})$ & $-\frac{10}{3}c_{a}+3c_{s}$ & $\cdots$\tabularnewline
 & $\frac{3}{2}(2^{-})$ & $2\tilde{c}_{a}+3\tilde{c}_{s}$ & $\cdots$ &  & $\frac{3}{2}(2^{+})$ & $2c_{a}+3c_{s}$ & $\cdots$\tabularnewline
\hline 
\multirow{8}{*}{$\Xi_{cc}^{\ast}\bar{\Sigma}_{c}^{\ast}$ $\left[6224.7\pm15\right]$} & $\frac{1}{2}(0^{-})$ & $5\tilde{c}_{a}-3\tilde{c}_{s}$ & $\cdots$ & \multirow{8}{*}{$\Xi_{cc}^{\ast}\Sigma_{c}^{\ast}$} & $\frac{1}{2}(0^{+})$ & $5c_{a}-3c_{s}$ & $\left[6210.4_{-6.4}^{+6.0}\pm15\right]_{B}$\tabularnewline
 & $\frac{1}{2}(1^{-})$ & $\frac{11}{3}\tilde{c}_{a}-3\tilde{c}_{s}$ & $\cdots$ &  & $\frac{1}{2}(1^{+})$ & $\frac{11}{3}c_{a}-3c_{s}$ & $\left[6211.1_{-5.3}^{+5.0}\pm15\right]_{B}$\tabularnewline
 & $\frac{1}{2}(2^{-})$ & $\tilde{c}_{a}-3\tilde{c}_{s}$ & $\cdots$ &  & $\frac{1}{2}(2^{+})$ & $c_{a}-3c_{s}$ & $\left[6212.5_{-3.1}^{+3.0}\pm15\right]_{B}$\tabularnewline
 & $\frac{1}{2}(3^{-})$ & $-3\tilde{c}_{a}-3\tilde{c}_{s}$ & $\left[6204.2\pm0.3\pm15\right]_{B}$ &  & $\frac{1}{2}(3^{+})$ & $-3c_{a}-3c_{s}$ & $\left[6214.6_{-4.6}^{+4.2}\pm15\right]_{B}$\tabularnewline
 & $\frac{3}{2}(0^{-})$ & $-5\tilde{c}_{a}+3\tilde{c}_{s}$ & $\left[6192.3\pm16.4\pm15\right]_{B}$ &  & $\frac{3}{2}(0^{+})$ & $-5c_{a}+3c_{s}$ & $\cdots$\tabularnewline
 & $\frac{3}{2}(1^{-})$ & $-\frac{11}{3}\tilde{c}_{a}+3\tilde{c}_{s}$ & $\left[6208.3\pm12.3\pm15\right]_{B}$ &  & $\frac{3}{2}(1^{+})$ & $-\frac{11}{3}c_{a}+3c_{s}$ & $\cdots$\tabularnewline
 & $\frac{3}{2}(2^{-})$ & $-\tilde{c}_{a}+3\tilde{c}_{s}$ & $\left[\sim6224.0\pm15\right]_{V}^{\dagger}$ &  & $\frac{3}{2}(2^{+})$ & $-c_{a}+3c_{s}$ & $\cdots$\tabularnewline
 & $\frac{3}{2}(3^{-})$ & $3\tilde{c}_{a}+3\tilde{c}_{s}$ & $\cdots$ &  & $\frac{3}{2}(3^{+})$ & $3c_{a}+3c_{s}$ & $\cdots$\tabularnewline
\hline 
\hline 
\end{tabular}
}
\end{table*}

\begin{table*}[htbp]
\centering
\caption{The $I(J^{P})$ quantum numbers, effective potentials, and bound/virtual state solutions of the $\Xi_{cc}^\ast\Xi_c^{(\prime,\ast)}$ and $\Xi_{cc}^\ast\bar{\Xi}_c^{(\prime,\ast)}$ systems. The notations are the same as those in Tables~\ref{tab:SigmacSigmac} and~\ref{tab:SigmacXic}.\label{tab:XicXicc}}
\setlength{\tabcolsep}{2.0mm}
{
\begin{tabular}{cccccccc}
\hline 
\hline
Systems $[m_{\text{th}}]$ & $I(J^{P})$ & $V_{\mathcal{H}_1\mathcal{H}_2}^{I,J}$ & $E_{B}/E_{V}$ & Systems & $I(J^{P})$ & $V_{\mathcal{H}_1\mathcal{H}_2}^{I,J}$ & $E_{B}/E_{V}$\tabularnewline
\hline
\multirow{4}{*}{$\Xi_{cc}^{\ast}\bar{\Xi}_{c}$ $\left[6175.6\pm15\right]$} & $0(1^{-})$ & $\frac{25}{6}\tilde{c}_{a}-\frac{5}{2}\tilde{c}_{s}$ & $\cdots$ & \multirow{4}{*}{$\Xi_{cc}^{\ast}\Xi_{c}$} & $0(1^{+})$ & $\frac{25}{6}c_{a}-\frac{5}{2}c_{s}$ & $\left[6167.2_{-4.9}^{+4.4}\pm15\right]_{B}$\tabularnewline
 & $0(2^{-})$ & $-\frac{5}{2}\tilde{c}_{a}-\frac{5}{2}\tilde{c}_{s}$ & $\left[6162.4\pm0.2\pm15\right]_{B}$ &  & $0(2^{+})$ & $-\frac{5}{2}c_{a}-\frac{5}{2}c_{s}$ & $\left[6170.3_{-3.4}^{+2.9}\pm15\right]_{B}$\tabularnewline
 & $1(1^{-})$ & $-\frac{5}{2}\tilde{c}_{a}+\frac{3}{2}\tilde{c}_{s}$ & $\left[6169.1\pm5.8\pm15\right]_{B}$ &  & $1(1^{+})$ & $-\frac{5}{2}c_{a}+\frac{3}{2}c_{s}$ & $\cdots$\tabularnewline
 & $1(2^{-})$ & $\frac{3}{2}\tilde{c}_{a}+\frac{3}{2}\tilde{c}_{s}$ & $\cdots$ &  & $1(2^{+})$ & $\frac{3}{2}c_{a}+\frac{3}{2}c_{s}$ & $\cdots$\tabularnewline
\hline 
\multirow{4}{*}{$\Xi_{cc}^{\ast}\bar{\Xi}_{c}^{\prime}$ $\left[6284.8\pm15\right]$} & $0(1^{-})$ & $-\frac{25}{18}\tilde{c}_{a}-\frac{5}{2}\tilde{c}_{s}$ & $\left[6282.1\pm1.5\pm15\right]_{B}$ & \multirow{4}{*}{$\Xi_{cc}^{\ast}\Xi_{c}^{\prime}$} & $0(1^{+})$ & $-\frac{25}{18}c_{a}-\frac{5}{2}c_{s}$ & $\left[6278.5_{-2.7}^{+2.5}\pm15\right]_{B}$\tabularnewline
 & $0(2^{-})$ & $\frac{5}{6}\tilde{c}_{a}-\frac{5}{2}\tilde{c}_{s}$ & $\cdots$ &  & $0(2^{+})$ & $\frac{5}{6}c_{a}-\frac{5}{2}c_{s}$ & $\left[6277.5_{-2.4}^{+2.2}\pm15\right]_{B}$\tabularnewline
 & $1(1^{-})$ & $\frac{5}{6}\tilde{c}_{a}+\frac{3}{2}\tilde{c}_{s}$ & $\cdots$ &  & $1(1^{+})$ & $\frac{5}{6}c_{a}+\frac{3}{2}c_{s}$ & $\cdots$\tabularnewline
 & $1(2^{-})$ & $-\frac{1}{2}\tilde{c}_{a}+\frac{3}{2}\tilde{c}_{s}$ & $\left[\sim6264.4\pm15\right]_{V}^{\sharp}$ &  & $1(2^{+})$ & $-\frac{1}{2}c_{a}+\frac{3}{2}c_{s}$ & $\cdots$\tabularnewline
\hline 
\multirow{8}{*}{$\Xi_{cc}^{\ast}\bar{\Xi}_{c}^{\ast}$ $\left[6351.7\pm15\right]$} & $0(0^{-})$ & $\frac{25}{6}\tilde{c}_{a}-\frac{5}{2}\tilde{c}_{s}$ & $\cdots$ & \multirow{8}{*}{$\Xi_{cc}^{\ast}\Xi_{c}^{\ast}$} & $0(0^{+})$ & $\frac{25}{6}c_{a}-\frac{5}{2}c_{s}$ & $\left[6342.5_{-5.0}^{+4.5}\pm15\right]_{B}$\tabularnewline
 & $0(1^{-})$ & $\frac{55}{18}\tilde{c}_{a}-\frac{5}{2}\tilde{c}_{s}$ & $\cdots$ &  & $0(1^{+})$ & $\frac{55}{18}c_{a}-\frac{5}{2}c_{s}$ & $\left[6343.0_{-4.1}^{+3.8}\pm15\right]_{B}$\tabularnewline
 & $0(2^{-})$ & $\frac{5}{6}\tilde{c}_{a}-\frac{5}{2}\tilde{c}_{s}$ & $\cdots$ &  & $0(2^{+})$ & $\frac{5}{6}c_{a}-\frac{5}{2}c_{s}$ & $\left[6344.1_{-2.4}^{+2.3}\pm15\right]_{B}$\tabularnewline
 & $0(3^{-})$ & $-\frac{5}{2}\tilde{c}_{a}-\frac{5}{2}\tilde{c}_{s}$ & $\left[6337.6\pm0.2\pm15\right]_{B}$ &  & $0(3^{+})$ & $-\frac{5}{2}c_{a}-\frac{5}{2}c_{s}$ & $\left[6345.7_{-3.5}^{+3.1}\pm15\right]_{B}$\tabularnewline
 & $1(0^{-})$ & $-\frac{5}{2}\tilde{c}_{a}+\frac{3}{2}\tilde{c}_{s}$ & $\left[6344.6\pm6.2\pm15\right]_{B}$ &  & $1(0^{+})$ & $-\frac{5}{2}c_{a}+\frac{3}{2}c_{s}$ & $\cdots$\tabularnewline
 & $1(1^{-})$ & $-\frac{11}{6}\tilde{c}_{a}+\frac{3}{2}\tilde{c}_{s}$ & $\left[6349.2\pm2.5\pm15\right]_{B}^{\dagger}$ &  & $1(1^{+})$ & $-\frac{11}{6}c_{a}+\frac{3}{2}c_{s}$ & $\cdots$\tabularnewline
 & $1(2^{-})$ & $-\frac{1}{2}\tilde{c}_{a}+\frac{3}{2}\tilde{c}_{s}$ & $\left[\sim6332.3\pm15\right]_{V}^{\dagger}$ &  & $1(2^{+})$ & $-\frac{1}{2}c_{a}+\frac{3}{2}c_{s}$ & $\cdots$\tabularnewline
 & $1(3^{-})$ & $\frac{3}{2}\tilde{c}_{a}+\frac{3}{2}\tilde{c}_{s}$ & $\cdots$ &  & $1(3^{+})$ & $\frac{3}{2}c_{a}+\frac{3}{2}c_{s}$ & $\cdots$\tabularnewline
\hline 
\hline 
\end{tabular}
}
\end{table*}

\subsection{Charmed baryon-(anti)hyperon systems}\label{sec:BQH}

The results for charmed baryon-(anti)hyperon systems are listed in Tables~\ref{tab:SigmacSigma}, \ref{tab:SigmacXi}, and \ref{tab:XicXi}. 

This type of systems such as $\Sigma_c^{(\ast)}\Sigma^{(\ast)}$, $\Sigma_c^{(\ast)}\bar{\Sigma}^{(\ast)}$, $\Xi_c^{(\prime,\ast)}\Xi^{(\ast)}$, and $\Xi_c^{(\prime,\ast)}\bar{\Xi}^{(\ast)}$ were investigated in literature~\cite{Kong:2022rvd,Kong:2023dwz,Wu:2024trh}. In Ref.~\cite{Kong:2022rvd}, Kong \etal~studied the possible molecular states in $\Sigma_c^{(\ast)}\Sigma^{(\ast)}$ and $\Sigma_c^{(\ast)}\bar{\Sigma}^{(\ast)}$ systems within the OBE model, and they obtained a series of bound states in the isoscalar, isovector, and isotensor channels. Their results in the isoscalar $\Sigma_c^{(\ast)}\Sigma^{(\ast)}$ systems are consistent with ours, while in our work, the isovector states become virtual states due to the weaker attractions, and no bound/virtual states can be obtained in the isotensor channels. In Ref.~\cite{Wu:2024trh}, Wu \etal~noticed that the $\Sigma_c^\ast\bar{\Sigma}$ system in $0(2^-)$ and $1(2^-)$ channels are likely to form bound states. We obtained the bound and virtual state solutions in these two channels, respectively, and we also found that the more attractive interaction in $2(1^-)$ channel gives rise to deeper bound state. 

\begin{table*}[htbp]
\centering
\caption{The $I(J^{P})$ quantum numbers, effective potentials, and bound/virtual state solutions of the $\Sigma_c^{(\ast)}\Sigma^{(\ast)}$, $\Sigma_c^{(\ast)}\bar{\Sigma}^{(\ast)}$, $\Sigma_c\Xi^{(\ast)}$ and $\Sigma_c\bar{\Xi}^{(\ast)}$ systems. The notations are the same as those in Tables~\ref{tab:SigmacSigmac} and~\ref{tab:SigmacXic}.\label{tab:SigmacSigma}}
\setlength{\tabcolsep}{3.3mm}
{
\begin{tabular}{cccccccc}
\hline 
\hline
Systems $[m_{\text{th}}]$ & $I(J^{P})$ & $V_{\mathcal{H}_1\mathcal{H}_2}^{I,J}$ & $E_{B}/E_{V}$ & Systems & $I(J^{P})$ & $V_{\mathcal{H}_1\mathcal{H}_2}^{I,J}$ & $E_{B}/E_{V}$\tabularnewline
\hline
\multirow{6}{*}{$\Sigma_{c}\bar{\Sigma}$ $\left[3646.6\right]$} & $0(0^{-})$ & $8\tilde{c}_{a}-6\tilde{c}_{s}$ & $\cdots$ & \multirow{6}{*}{$\Sigma_{c}\Sigma$} & $0(0^{+})$ & $8c_{a}-6c_{s}$ & $\left[3616.4_{-11.4}^{+10.8}\right]_{B}$\tabularnewline
 & $0(1^{-})$ & $-\frac{8}{3}\tilde{c}_{a}-6\tilde{c}_{s}$ & $\left[3634.4,3644.0\right]_{B}$ &  & $0(1^{+})$ & $-\frac{8}{3}c_{a}-6c_{s}$ & $\left[3622.2_{-7.0}^{+6.6}\right]_{B}$\tabularnewline
 & $1(0^{-})$ & $\frac{8}{3}\tilde{c}_{a}-2\tilde{c}_{s}$ & $\cdots$ &  & $1(0^{+})$ & $\frac{8}{3}c_{a}-2c_{s}$ & $\left[3644.7_{-4.7}^{+1.7}\right]_{V}$\tabularnewline
 & $1(1^{-})$ & $-\frac{8}{9}\tilde{c}_{a}-2\tilde{c}_{s}$ & $\left[3608.5,3633.3\right]_{V}$ &  & $1(1^{+})$ & $-\frac{8}{9}c_{a}-2c_{s}$ & $\left[3642.7_{-3.8}^{+2.3}\right]_{V}$\tabularnewline
 & $2(0^{-})$ & $-8\tilde{c}_{a}+6\tilde{c}_{s}$ & $\left[3576.0,3631.0\right]_{B}$ &  & $2(0^{+})$ & $-8c_{a}+6c_{s}$ & $\cdots$\tabularnewline
 & $2(1^{-})$ & $\frac{8}{3}\tilde{c}_{a}+6\tilde{c}_{s}$ & $\cdots$ &  & $2(1^{+})$ & $\frac{8}{3}c_{a}+6c_{s}$ & $\cdots$\tabularnewline
\hline 
\multirow{6}{*}{$\Sigma_{c}\bar{\Sigma}^{\ast}$ $\left[3838.0\right]$} & $0(1^{-})$ & $\frac{20}{3}\tilde{c}_{a}-6\tilde{c}_{s}$ & $\cdots$ & \multirow{6}{*}{$\Sigma_{c}\Sigma^{\ast}$} & $0(1^{+})$ & $\frac{20}{3}c_{a}-6c_{s}$ & $\left[3804.8_{-10.5}^{+10.1}\right]_{B}$\tabularnewline
 & $0(2^{-})$ & $-4\tilde{c}_{a}-6\tilde{c}_{s}$ & $\left[3810.0,3818.4\right]_{B}$ &  & $0(2^{+})$ & $-4c_{a}-6c_{s}$ & $\left[3810.7_{-8.1}^{+7.8}\right]_{B}$\tabularnewline
 & $1(1^{-})$ & $\frac{20}{9}\tilde{c}_{a}-2\tilde{c}_{s}$ & $\cdots$ &  & $1(1^{+})$ & $\frac{20}{9}c_{a}-2c_{s}$ & $\left[3837.4_{-2.5}^{+0.6}\right]_{V}$\tabularnewline
 & $1(2^{-})$ & $-\frac{4}{3}\tilde{c}_{a}-2\tilde{c}_{s}$ & $\left[3833.3,3836.5\right]_{V}$ &  & $1(2^{+})$ & $-\frac{4}{3}c_{a}-2c_{s}$ & $\left[3836.3_{-3.1}^{+1.4}\right]_{V}$\tabularnewline
 & $2(1^{-})$ & $-\frac{20}{3}\tilde{c}_{a}+6\tilde{c}_{s}$ & $\left[3782.9,3830.6\right]_{B}$ &  & $2(1^{+})$ & $-\frac{20}{3}c_{a}+6c_{s}$ & $\cdots$\tabularnewline
 & $2(2^{-})$ & $4\tilde{c}_{a}+6\tilde{c}_{s}$ & $\cdots$ &  & $2(2^{+})$ & $4c_{a}+6c_{s}$ & $\cdots$\tabularnewline
\hline 
\multirow{4}{*}{$\Sigma_{c}\bar{\Xi}$ $\left[3771.7\right]$} & $\frac{1}{2}(0^{-})$ & $-2\tilde{c}_{a}-3\tilde{c}_{s}$ & $\left[3770.7,3771.7\right]_{B}$ & \multirow{4}{*}{$\Sigma_{c}\Xi$} & $\frac{1}{2}(0^{+})$ & $-2c_{a}-3c_{s}$ & $\left[3770.8_{-1.9}^{+0.9}\right]_{B}$\tabularnewline
 & $\frac{1}{2}(1^{-})$ & $\frac{2}{3}\tilde{c}_{a}-3\tilde{c}_{s}$ & $\cdots$ &  & $\frac{1}{2}(1^{+})$ & $\frac{2}{3}c_{a}-3c_{s}$ & $\left[3770.2_{-1.5}^{+1.1}\right]_{B}$\tabularnewline
 & $\frac{3}{2}(0^{-})$ & $2\tilde{c}_{a}+3\tilde{c}_{s}$ & $\cdots$ &  & $\frac{3}{2}(0^{+})$ & $2c_{a}+3c_{s}$ & $\cdots$\tabularnewline
 & $\frac{3}{2}(1^{-})$ & $-\frac{2}{3}\tilde{c}_{a}+3\tilde{c}_{s}$ & $\left[\sim3688.7\right]_{V}^{\sharp}$ &  & $\frac{3}{2}(1^{+})$ & $-\frac{2}{3}c_{a}+3c_{s}$ & $\cdots$\tabularnewline
\hline 
\multirow{4}{*}{$\Sigma_{c}\bar{\Xi}^{\ast}$ $\left[3986.9\right]$} & $\frac{1}{2}(1^{-})$ & $\frac{10}{3}\tilde{c}_{a}-3\tilde{c}_{s}$ & $\cdots$ & \multirow{4}{*}{$\Sigma_{c}\Xi^{\ast}$} & $\frac{1}{2}(1^{+})$ & $\frac{10}{3}c_{a}-3c_{s}$ & $\left[3983.0_{-3.6}^{+2.7}\right]_{B}$\tabularnewline
 & $\frac{1}{2}(2^{-})$ & $-2\tilde{c}_{a}-3\tilde{c}_{s}$ & $\left[3984.5,3986.3\right]_{B}$ &  & $\frac{1}{2}(2^{+})$ & $-2c_{a}-3c_{s}$ & $\left[3984.7_{-2.4}^{+1.7}\right]_{B}$\tabularnewline
 & $\frac{3}{2}(1^{-})$ & $-\frac{10}{3}\tilde{c}_{a}+3\tilde{c}_{s}$ & $\left[3975.0,3986.9\right]_{B}^{\dagger}$ &  & $\frac{3}{2}(1^{+})$ & $-\frac{10}{3}c_{a}+3c_{s}$ & $\cdots$\tabularnewline
 & $\frac{3}{2}(2^{-})$ & $2\tilde{c}_{a}+3\tilde{c}_{s}$ & $\cdots$ &  & $\frac{3}{2}(2^{+})$ & $2c_{a}+3c_{s}$ & $\cdots$\tabularnewline
\hline 
\multirow{6}{*}{$\Sigma_{c}^{\ast}\bar{\Sigma}$ $\left[3711.3\right]$} & $0(1^{-})$ & $\frac{20}{3}\tilde{c}_{a}-6\tilde{c}_{s}$ & $\cdots$ & \multirow{6}{*}{$\Sigma_{c}^{\ast}\Sigma$} & $0(1^{+})$ & $\frac{20}{3}c_{a}-6c_{s}$ & $\left[3681.4_{-10.3}^{+9.8}\right]_{B}$\tabularnewline
 & $0(2^{-})$ & $-4\tilde{c}_{a}-6\tilde{c}_{s}$ & $\left[3686.6,3694.6\right]_{B}$ &  & $0(2^{+})$ & $-4c_{a}-6c_{s}$ & $\left[3687.2_{-8.0}^{+7.5}\right]_{B}$\tabularnewline
 & $1(1^{-})$ & $\frac{20}{9}\tilde{c}_{a}-2\tilde{c}_{s}$ & $\cdots$ &  & $1(1^{+})$ & $\frac{20}{9}c_{a}-2c_{s}$ & $\left[3709.3_{-4.2}^{+1.7}\right]_{V}$\tabularnewline
 & $1(2^{-})$ & $-\frac{4}{3}\tilde{c}_{a}-2\tilde{c}_{s}$ & $\left[3702.7,3707.6\right]_{V}$ &  & $1(2^{+})$ & $-\frac{4}{3}c_{a}-2c_{s}$ & $\left[3707.3_{-4.7}^{+2.5}\right]_{V}$\tabularnewline
 & $2(1^{-})$ & $-\frac{20}{3}\tilde{c}_{a}+6\tilde{c}_{s}$ & $\left[3660.0,3705.9\right]_{B}$ &  & $2(1^{+})$ & $-\frac{20}{3}c_{a}+6c_{s}$ & $\cdots$\tabularnewline
 & $2(2^{-})$ & $4\tilde{c}_{a}+6\tilde{c}_{s}$ & $\cdots$ &  & $2(2^{+})$ & $4c_{a}+6c_{s}$ & $\cdots$\tabularnewline
\hline 
\multirow{12}{*}{$\Sigma_{c}^{\ast}\bar{\Sigma}^{\ast}$ $\left[3902.7\right]$} & $0(0^{-})$ & $10\tilde{c}_{a}-6\tilde{c}_{s}$ & $\cdots$ & \multirow{12}{*}{$\Sigma_{c}^{\ast}\Sigma^{\ast}$} & $0(0^{+})$ & $10c_{a}-6c_{s}$ & $\left[3867.2_{-13.4}^{+12.8}\right]_{B}$\tabularnewline
 & $0(1^{-})$ & $\frac{22}{3}\tilde{c}_{a}-6\tilde{c}_{s}$ & $\cdots$ &  & $0(1^{+})$ & $\frac{22}{3}c_{a}-6c_{s}$ & $\left[3868.7_{-11.1}^{+10.7}\right]_{B}$\tabularnewline
 & $0(2^{-})$ & $2\tilde{c}_{a}-6\tilde{c}_{s}$ & $\cdots$ &  & $0(2^{+})$ & $2c_{a}-6c_{s}$ & $\left[3871.7_{-6.6}^{+6.4}\right]_{B}$\tabularnewline
 & $0(3^{-})$ & $-6\tilde{c}_{a}-6\tilde{c}_{s}$ & $\left[3853.6,3854.8\right]_{B}$ &  & $0(3^{+})$ & $-6c_{a}-6c_{s}$ & $\left[3876.1_{-9.7}^{+9.2}\right]_{B}$\tabularnewline
 & $1(0^{-})$ & $\frac{10}{3}\tilde{c}_{a}-2\tilde{c}_{s}$ & $\cdots$ &  & $1(0^{+})$ & $\frac{10}{3}c_{a}-2c_{s}$ & $\left[3902.4_{-2.9}^{+0.3}\right]_{V}^{\dagger}$\tabularnewline
 & $1(1^{-})$ & $\frac{22}{9}\tilde{c}_{a}-2\tilde{c}_{s}$ & $\cdots$ &  & $1(1^{+})$ & $\frac{22}{9}c_{a}-2c_{s}$ & $\left[3902.2_{-2.5}^{+0.4}\right]_{V}^{\dagger}$\tabularnewline
 & $1(2^{-})$ & $\frac{2}{3}\tilde{c}_{a}-2\tilde{c}_{s}$ & $\cdots$ &  & $1(2^{+})$ & $\frac{2}{3}c_{a}-2c_{s}$ & $\left[3901.8_{-1.6}^{+0.7}\right]_{V}$\tabularnewline
 & $1(3^{-})$ & $-2\tilde{c}_{a}-2\tilde{c}_{s}$ & $\left[3902.4,3902.5\right]_{B}$ &  & $1(3^{+})$ & $-2c_{a}-2c_{s}$ & $\left[3900.8_{-4.2}^{+1.6}\right]_{V}$\tabularnewline
 & $2(0^{-})$ & $-10\tilde{c}_{a}+6\tilde{c}_{s}$ & $\left[3796.5,3863.7\right]_{B}$ &  & $2(0^{+})$ & $-10c_{a}+6c_{s}$ & $\cdots$\tabularnewline
 & $2(1^{-})$ & $-\frac{22}{3}\tilde{c}_{a}+6\tilde{c}_{s}$ & $\left[3837.2,3889.5\right]_{B}$ &  & $2(1^{+})$ & $-\frac{22}{3}c_{a}+6c_{s}$ & $\cdots$\tabularnewline
 & $2(2^{-})$ & $-2\tilde{c}_{a}+6\tilde{c}_{s}$ & $\left[\sim3902.6\right]_{V}^{\sharp}$ &  & $2(2^{+})$ & $-2c_{a}+6c_{s}$ & $\cdots$\tabularnewline
 & $2(3^{-})$ & $6\tilde{c}_{a}+6\tilde{c}_{s}$ & $\cdots$ &  & $2(3^{+})$ & $6c_{a}+6c_{s}$ & $\cdots$\tabularnewline
\hline 
\hline 
\end{tabular}
}
\end{table*}

\begin{table*}[htbp]
\centering
\caption{The $I(J^{P})$ quantum numbers, effective potentials, and bound/virtual state solutions of the $\Sigma_c^{\ast}\Xi^{(\ast)}$, $\Sigma_c^{\ast}\bar{\Xi}^{(\ast)}$, $\Xi_c\Sigma^{(\ast)}$, $\Xi_c\bar{\Sigma}^{(\ast)}$, $\Xi_c\Xi^{(\ast)}$, $\Xi_c\bar{\Xi}^{(\ast)}$, $\Xi_c^\prime\Sigma$ and $\Xi_c^\prime\bar{\Sigma}$ systems. The notations are the same as those in Tables~\ref{tab:SigmacSigmac} and~\ref{tab:SigmacXic}.\label{tab:SigmacXi}}
\setlength{\tabcolsep}{3.4mm}
{
\begin{tabular}{cccccccc}
\hline 
\hline
Systems $[m_{\text{th}}]$ & $I(J^{P})$ & $V_{\mathcal{H}_1\mathcal{H}_2}^{I,J}$ & $E_{B}/E_{V}$ & Systems & $I(J^{P})$ & $V_{\mathcal{H}_1\mathcal{H}_2}^{I,J}$ & $E_{B}/E_{V}$\tabularnewline
\hline
\multirow{4}{*}{$\Sigma_{c}^{\ast}\bar{\Xi}$ $\left[3836.4\right]$} & $\frac{1}{2}(1^{-})$ & $-\frac{5}{3}\tilde{c}_{a}-3\tilde{c}_{s}$ & $\left[3835.6,3836.4\right]_{V}^{\dagger}$ & \multirow{4}{*}{$\Sigma_{c}^{\ast}\Xi$} & $\frac{1}{2}(1^{+})$ & $-\frac{5}{3}c_{a}-3c_{s}$ & $\left[3835.3_{-1.8}^{+1.0}\right]_{B}$\tabularnewline
 & $\frac{1}{2}(2^{-})$ & $\tilde{c}_{a}-3\tilde{c}_{s}$ & $\cdots$ &  & $\frac{1}{2}(2^{+})$ & $c_{a}-3c_{s}$ & $\left[3834.6_{-1.7}^{+1.2}\right]_{B}$\tabularnewline
 & $\frac{3}{2}(1^{-})$ & $\frac{5}{3}\tilde{c}_{a}+3\tilde{c}_{s}$ & $\cdots$ &  & $\frac{3}{2}(1^{+})$ & $\frac{5}{3}c_{a}+3c_{s}$ & $\cdots$\tabularnewline
 & $\frac{3}{2}(2^{-})$ & $-\tilde{c}_{a}+3\tilde{c}_{s}$ & $\left[\sim3812.2\right]_{V}^{\sharp}$ &  & $\frac{3}{2}(2^{+})$ & $-c_{a}+3c_{s}$ & $\cdots$\tabularnewline
\hline 
\multirow{8}{*}{$\Sigma_{c}^{\ast}\bar{\Xi}^{\ast}$ $\left[4051.5\right]$} & $\frac{1}{2}(0^{-})$ & $5\tilde{c}_{a}-3\tilde{c}_{s}$ & $\cdots$ & \multirow{8}{*}{$\Sigma_{c}^{\ast}\Xi^{\ast}$} & $\frac{1}{2}(0^{+})$ & $5c_{a}-3c_{s}$ & $\left[4046.8_{-4.8}^{+3.6}\right]_{B}$\tabularnewline
 & $\frac{1}{2}(1^{-})$ & $\frac{11}{3}\tilde{c}_{a}-3\tilde{c}_{s}$ & $\cdots$ &  & $\frac{1}{2}(1^{+})$ & $\frac{11}{3}c_{a}-3c_{s}$ & $\left[4047.3_{-3.9}^{+2.9}\right]_{B}$\tabularnewline
 & $\frac{1}{2}(2^{-})$ & $\tilde{c}_{a}-3\tilde{c}_{s}$ & $\cdots$ &  & $\frac{1}{2}(2^{+})$ & $c_{a}-3c_{s}$ & $\left[4048.2_{-2.1}^{+1.7}\right]_{B}$\tabularnewline
 & $\frac{1}{2}(3^{-})$ & $-3\tilde{c}_{a}-3\tilde{c}_{s}$ & $\left[4041.9,4042.4\right]_{B}$ &  & $\frac{1}{2}(3^{+})$ & $-3c_{a}-3c_{s}$ & $\left[4049.5_{-2.9}^{+1.8}\right]_{B}$\tabularnewline
 & $\frac{3}{2}(0^{-})$ & $-5\tilde{c}_{a}+3\tilde{c}_{s}$ & $\left[4017.0,4045.6\right]_{B}$ &  & $\frac{3}{2}(0^{+})$ & $-5c_{a}+3c_{s}$ & $\cdots$\tabularnewline
 & $\frac{3}{2}(1^{-})$ & $-\frac{11}{3}\tilde{c}_{a}+3\tilde{c}_{s}$ & $\left[4035.2,4051.5\right]_{B}^{\dagger}$ &  & $\frac{3}{2}(1^{+})$ & $-\frac{11}{3}c_{a}+3c_{s}$ & $\cdots$\tabularnewline
 & $\frac{3}{2}(2^{-})$ & $-\tilde{c}_{a}+3\tilde{c}_{s}$ & $\left[\sim4035.1\right]_{V}^{\dagger}$ &  & $\frac{3}{2}(2^{+})$ & $-c_{a}+3c_{s}$ & $\cdots$\tabularnewline
 & $\frac{3}{2}(3^{-})$ & $3\tilde{c}_{a}+3\tilde{c}_{s}$ & $\cdots$ &  & $\frac{3}{2}(3^{+})$ & $3c_{a}+3c_{s}$ & $\cdots$\tabularnewline
\hline
\multirow{4}{*}{$\Xi_{c}\bar{\Sigma}$ $\left[3662.2\right]$} & $\frac{1}{2}(0^{-})$ & $6\tilde{c}_{a}-3\tilde{c}_{s}$ & $\cdots$ & \multirow{4}{*}{$\Xi_{c}\Sigma$} & $\frac{1}{2}(0^{+})$ & $6c_{a}-3c_{s}$ & $\left[3660.2_{-4.3}^{+2.0}\right]_{B}$\tabularnewline
 & $\frac{1}{2}(1^{-})$ & $-2\tilde{c}_{a}-3\tilde{c}_{s}$ & $\left[3661.8,3662.2\right]_{B}^{\dagger}$ &  & $\frac{1}{2}(1^{+})$ & $-2c_{a}-3c_{s}$ & $\left[3661.9_{-1.4}^{+0.3}\right]_{B}^{\dagger}$\tabularnewline
 & $\frac{3}{2}(0^{-})$ & $-6\tilde{c}_{a}+3\tilde{c}_{s}$ & $\left[3620.1,3653.0\right]_{B}$ &  & $\frac{3}{2}(0^{+})$ & $-6c_{a}+3c_{s}$ & $\cdots$\tabularnewline
 & $\frac{3}{2}(1^{-})$ & $2\tilde{c}_{a}+3\tilde{c}_{s}$ & $\cdots$ &  & $\frac{3}{2}(1^{+})$ & $2c_{a}+3c_{s}$ & $\cdots$\tabularnewline
\hline 
\multirow{4}{*}{$\Xi_{c}\bar{\Sigma}^{\ast}$ $\left[3853.7\right]$} & $\frac{1}{2}(1^{-})$ & $5\tilde{c}_{a}-3\tilde{c}_{s}$ & $\cdots$ & \multirow{4}{*}{$\Xi_{c}\Sigma^{\ast}$} & $\frac{1}{2}(1^{+})$ & $5c_{a}-3c_{s}$ & $\left[3850.3_{-4.4}^{+2.9}\right]_{B}$\tabularnewline
 & $\frac{1}{2}(2^{-})$ & $-3\tilde{c}_{a}-3\tilde{c}_{s}$ & $\left[3845.9,3846.3\right]_{B}$ &  & $\frac{1}{2}(2^{+})$ & $-3c_{a}-3c_{s}$ & $\left[3852.5_{-2.5}^{+1.1}\right]_{B}$\tabularnewline
 & $\frac{3}{2}(1^{-})$ & $-5\tilde{c}_{a}+3\tilde{c}_{s}$ & $\left[3821.8,3849.3\right]_{B}$ &  & $\frac{3}{2}(1^{+})$ & $-5c_{a}+3c_{s}$ & $\cdots$\tabularnewline
 & $\frac{3}{2}(2^{-})$ & $3\tilde{c}_{a}+3\tilde{c}_{s}$ & $\cdots$ &  & $\frac{3}{2}(2^{+})$ & $3c_{a}+3c_{s}$ & $\cdots$\tabularnewline
\hline 
\multirow{4}{*}{$\Xi_{c}\bar{\Xi}$ $\left[3787.4\right]$} & $0(0^{-})$ & $-\frac{5}{2}\tilde{c}_{a}-\frac{5}{2}\tilde{c}_{s}$ & $\left[3784.9,3785.2\right]_{B}$ & \multirow{4}{*}{$\Xi_{c}\Xi$} & $0(0^{+})$ & $-\frac{5}{2}c_{a}-\frac{5}{2}c_{s}$ & $\left[3787.3_{-1.6}^{+0.1}\right]_{V}^{\dagger}$\tabularnewline
 & $0(1^{-})$ & $\frac{5}{6}\tilde{c}_{a}-\frac{5}{2}\tilde{c}_{s}$ & $\cdots$ &  & $0(1^{+})$ & $\frac{5}{6}c_{a}-\frac{5}{2}c_{s}$ & $\left[3787.3_{-0.5}^{+0.1}\right]_{B}^{\dagger}$\tabularnewline
 & $1(0^{-})$ & $\frac{3}{2}\tilde{c}_{a}+\frac{3}{2}\tilde{c}_{s}$ & $\cdots$ &  & $1(0^{+})$ & $\frac{3}{2}c_{a}+\frac{3}{2}c_{s}$ & $\cdots$\tabularnewline
 & $1(1^{-})$ & $-\frac{1}{2}\tilde{c}_{a}+\frac{3}{2}\tilde{c}_{s}$ & $\left[\sim3635.3\right]_{V}^{\sharp}$ &  & $1(1^{+})$ & $-\frac{1}{2}c_{a}+\frac{3}{2}c_{s}$ & $\cdots$\tabularnewline
\hline 
\multirow{4}{*}{$\Xi_{c}\bar{\Xi}^{\ast}$ $\left[4002.5\right]$} & $0(1^{-})$ & $\frac{25}{6}\tilde{c}_{a}-\frac{5}{2}\tilde{c}_{s}$ & $\cdots$ & \multirow{4}{*}{$\Xi_{c}\Xi^{\ast}$} & $0(1^{+})$ & $\frac{25}{6}c_{a}-\frac{5}{2}c_{s}$ & $\left[4001.3_{-2.9}^{+1.2}\right]_{B}$\tabularnewline
 & $0(2^{-})$ & $-\frac{5}{2}\tilde{c}_{a}-\frac{5}{2}\tilde{c}_{s}$ & $\left[3998.3,3998.6\right]_{B}$ &  & $0(2^{+})$ & $-\frac{5}{2}c_{a}-\frac{5}{2}c_{s}$ & $\left[4002.4_{-1.2}^{+0.1}\right]_{B}^{\dagger}$\tabularnewline
 & $1(1^{-})$ & $-\frac{5}{2}\tilde{c}_{a}+\frac{3}{2}\tilde{c}_{s}$ & $\left[3999.0,4002.5\right]_{B}^{\dagger}$ &  & $1(1^{+})$ & $-\frac{5}{2}c_{a}+\frac{3}{2}c_{s}$ & $\cdots$\tabularnewline
 & $1(2^{-})$ & $\frac{3}{2}\tilde{c}_{a}+\frac{3}{2}\tilde{c}_{s}$ & $\cdots$ &  & $1(2^{+})$ & $\frac{3}{2}c_{a}+\frac{3}{2}c_{s}$ & $\cdots$\tabularnewline
\hline 
\multirow{4}{*}{$\Xi_{c}^{\prime}\bar{\Sigma}$ $\left[3771.4\right]$} & $\frac{1}{2}(0^{-})$ & $-2\tilde{c}_{a}-3\tilde{c}_{s}$ & $\left[3770.8,3771.4\right]_{B}^{\dagger}$ & \multirow{4}{*}{$\Xi_{c}^{\prime}\Sigma$} & $\frac{1}{2}(0^{+})$ & $-2c_{a}-3c_{s}$ & $\left[3770.9_{-1.5}^{+0.5}\right]_{B}$\tabularnewline
 & $\frac{1}{2}(1^{-})$ & $\frac{2}{3}\tilde{c}_{a}-3\tilde{c}_{s}$ & $\cdots$ &  & $\frac{1}{2}(1^{+})$ & $\frac{2}{3}c_{a}-3c_{s}$ & $\left[3770.4_{-1.3}^{+0.8}\right]_{B}$\tabularnewline
 & $\frac{3}{2}(0^{-})$ & $2\tilde{c}_{a}+3\tilde{c}_{s}$ & $\cdots$ &  & $\frac{3}{2}(0^{+})$ & $2c_{a}+3c_{s}$ & $\cdots$\tabularnewline
 & $\frac{3}{2}(1^{-})$ & $-\frac{2}{3}\tilde{c}_{a}+3\tilde{c}_{s}$ & $\left[\sim3674.7\right]_{V}^{\dagger}$ &  & $\frac{3}{2}(1^{+})$ & $-\frac{2}{3}c_{a}+3c_{s}$ & $\cdots$\tabularnewline
\hline 
\hline 
\end{tabular}
}
\end{table*}

\begin{table*}[htbp]
\centering
\caption{The $I(J^{P})$ quantum numbers, effective potentials, and bound/virtual state solutions of the $\Xi_c^{\prime}\Sigma^{\ast}$, $\Xi_c^{\prime}\bar{\Sigma}^{\ast}$, $\Xi_c^\prime\Xi^{(\ast)}$, $\Xi_c^\prime\bar{\Xi}^{(\ast)}$, $\Xi_c^\ast\Sigma^{(\ast)}$, $\Xi_c^\ast\bar{\Sigma}^{(\ast)}$, $\Xi_c^\ast\Xi^{(\ast)}$ and $\Xi_c^\ast\bar{\Xi}^{(\ast)}$ systems. The notations are the same as those in Tables~\ref{tab:SigmacSigmac} and~\ref{tab:SigmacXic}.\label{tab:XicXi}}
\setlength{\tabcolsep}{3.3mm}
{
\begin{tabular}{cccccccc}
\hline 
\hline
Systems $[m_{\text{th}}]$ & $I(J^{P})$ & $V_{\mathcal{H}_1\mathcal{H}_2}^{I,J}$ & $E_{B}/E_{V}$ & Systems & $I(J^{P})$ & $V_{\mathcal{H}_1\mathcal{H}_2}^{I,J}$ & $E_{B}/E_{V}$\tabularnewline
\hline
\multirow{4}{*}{$\Xi_{c}^{\prime}\bar{\Sigma}^{\ast}$ $\left[3962.8\right]$} & $\frac{1}{2}(1^{-})$ & $-\frac{5}{3}\tilde{c}_{a}-3\tilde{c}_{s}$ & $\left[3962.3,3962.8\right]_{V}^{\dagger}$ & \multirow{4}{*}{$\Xi_{c}^{\prime}\Sigma^{\ast}$} & $\frac{1}{2}(1^{+})$ & $-\frac{5}{3}c_{a}-3c_{s}$ & $\left[3961.2_{-2.0}^{+1.3}\right]_{B}$\tabularnewline
 & $\frac{1}{2}(2^{-})$ & $\tilde{c}_{a}-3\tilde{c}_{s}$ & $\cdots$ &  & $\frac{1}{2}(2^{+})$ & $c_{a}-3c_{s}$ & $\left[3960.4_{-1.9}^{+1.5}\right]_{B}$\tabularnewline
 & $\frac{3}{2}(1^{-})$ & $\frac{5}{3}\tilde{c}_{a}+3\tilde{c}_{s}$ & $\cdots$ &  & $\frac{3}{2}(1^{+})$ & $\frac{5}{3}c_{a}+3c_{s}$ & $\cdots$\tabularnewline
 & $\frac{3}{2}(2^{-})$ & $-\tilde{c}_{a}+3\tilde{c}_{s}$ & $\left[\sim3941.7\right]_{V}^{\sharp}$ &  & $\frac{3}{2}(2^{+})$ & $-c_{a}+3c_{s}$ & $\cdots$\tabularnewline
\hline 
\multirow{4}{*}{$\Xi_{c}^{\prime}\bar{\Xi}$ $\left[3896.5\right]$} & $0(0^{-})$ & $\frac{5}{6}\tilde{c}_{a}-\frac{5}{2}\tilde{c}_{s}$ & $\cdots$ & \multirow{4}{*}{$\Xi_{c}^{\prime}\Xi$} & $0(0^{+})$ & $\frac{5}{6}c_{a}-\frac{5}{2}c_{s}$ & $\left[3896.4_{-0.6}^{+0.1}\right]_{B}^{\dagger}$\tabularnewline
 & $0(1^{-})$ & $-\frac{5}{18}\tilde{c}_{a}-\frac{5}{2}\tilde{c}_{s}$ & $\left[\sim3872.2\right]_{V}^{\sharp}$ &  & $0(1^{+})$ & $-\frac{5}{18}c_{a}-\frac{5}{2}c_{s}$ & $\left[3896.4_{-0.4}^{+0.1}\right]_{B}^{\dagger}$\tabularnewline
 & $1(0^{-})$ & $-\frac{1}{2}\tilde{c}_{a}+\frac{3}{2}\tilde{c}_{s}$ & $\left[\sim3754.3\right]_{V}^{\dagger}$ &  & $1(0^{+})$ & $-\frac{1}{2}c_{a}+\frac{3}{2}c_{s}$ & $\cdots$\tabularnewline
 & $1(1^{-})$ & $\frac{1}{6}\tilde{c}_{a}+\frac{3}{2}\tilde{c}_{s}$ & $\cdots$ &  & $1(1^{+})$ & $\frac{1}{6}c_{a}+\frac{3}{2}c_{s}$ & $\cdots$\tabularnewline
\hline 
\multirow{4}{*}{$\Xi_{c}^{\prime}\bar{\Xi}^{\ast}$ $\left[4111.6\right]$} & $0(1^{-})$ & $-\frac{25}{18}\tilde{c}_{a}-\frac{5}{2}\tilde{c}_{s}$ & $\left[4109.9,4111.6\right]_{V}$ & \multirow{4}{*}{$\Xi_{c}^{\prime}\Xi^{\ast}$} & $0(1^{+})$ & $-\frac{25}{18}c_{a}-\frac{5}{2}c_{s}$ & $\left[4111.3_{-1.1}^{+0.3}\right]_{B}^{\dagger}$\tabularnewline
 & $0(2^{-})$ & $\frac{5}{6}\tilde{c}_{a}-\frac{5}{2}\tilde{c}_{s}$ & $\cdots$ &  & $0(2^{+})$ & $\frac{5}{6}c_{a}-\frac{5}{2}c_{s}$ & $\left[4110.9_{-1.1}^{+0.6}\right]_{B}$\tabularnewline
 & $1(1^{-})$ & $\frac{5}{6}\tilde{c}_{a}+\frac{3}{2}\tilde{c}_{s}$ & $\cdots$ &  & $1(1^{+})$ & $\frac{5}{6}c_{a}+\frac{3}{2}c_{s}$ & $\cdots$\tabularnewline
 & $1(2^{-})$ & $-\frac{1}{2}\tilde{c}_{a}+\frac{3}{2}\tilde{c}_{s}$ & $\left[\sim4008.6\right]_{V}^{\sharp}$ &  & $1(2^{+})$ & $-\frac{1}{2}c_{a}+\frac{3}{2}c_{s}$ & $\cdots$\tabularnewline
\hline 
\multirow{4}{*}{$\Xi_{c}^{\ast}\bar{\Sigma}$ $\left[3838.3\right]$} & $\frac{1}{2}(1^{-})$ & $\frac{10}{3}\tilde{c}_{a}-3\tilde{c}_{s}$ & $\cdots$ & \multirow{4}{*}{$\Xi_{c}^{\ast}\Sigma$} & $\frac{1}{2}(1^{+})$ & $\frac{10}{3}c_{a}-3c_{s}$ & $\left[3836.6_{-2.8}^{+1.5}\right]_{B}$\tabularnewline
 & $\frac{1}{2}(2^{-})$ & $-2\tilde{c}_{a}-3\tilde{c}_{s}$ & $\left[3837.6,3838.3\right]_{B}^{\dagger}$ &  & $\frac{1}{2}(2^{+})$ & $-2c_{a}-3c_{s}$ & $\left[3837.7_{-1.6}^{+0.5}\right]_{B}^{\dagger}$\tabularnewline
 & $\frac{3}{2}(1^{-})$ & $-\frac{10}{3}\tilde{c}_{a}+3\tilde{c}_{s}$ & $\left[3830.1,3838.3\right]_{B}^{\dagger}$ &  & $\frac{3}{2}(1^{+})$ & $-\frac{10}{3}c_{a}+3c_{s}$ & $\cdots$\tabularnewline
 & $\frac{3}{2}(2^{-})$ & $2\tilde{c}_{a}+3\tilde{c}_{s}$ & $\cdots$ &  & $\frac{3}{2}(2^{+})$ & $2c_{a}+3c_{s}$ & $\cdots$\tabularnewline
\hline 
\multirow{8}{*}{$\Xi_{c}^{\ast}\bar{\Sigma}^{\ast}$ $\left[4029.7\right]$} & $\frac{1}{2}(0^{-})$ & $5\tilde{c}_{a}-3\tilde{c}_{s}$ & $\cdots$ & \multirow{8}{*}{$\Xi_{c}^{\ast}\Sigma^{\ast}$} & $\frac{1}{2}(0^{+})$ & $5c_{a}-3c_{s}$ & $\left[4025.9_{-4.6}^{+3.1}\right]_{B}$\tabularnewline
 & $\frac{1}{2}(1^{-})$ & $\frac{11}{3}\tilde{c}_{a}-3\tilde{c}_{s}$ & $\cdots$ &  & $\frac{1}{2}(1^{+})$ & $\frac{11}{3}c_{a}-3c_{s}$ & $\left[4026.4_{-3.6}^{+2.6}\right]_{B}$\tabularnewline
 & $\frac{1}{2}(2^{-})$ & $\tilde{c}_{a}-3\tilde{c}_{s}$ & $\cdots$ &  & $\frac{1}{2}(2^{+})$ & $c_{a}-3c_{s}$ & $\left[4027.2_{-1.9}^{+1.5}\right]_{B}$\tabularnewline
 & $\frac{1}{2}(3^{-})$ & $-3\tilde{c}_{a}-3\tilde{c}_{s}$ & $\left[4021.3,4021.7\right]_{B}$ &  & $\frac{1}{2}(3^{+})$ & $-3c_{a}-3c_{s}$ & $\left[4028.3_{-2.6}^{+1.4}\right]_{B}$\tabularnewline
 & $\frac{3}{2}(0^{-})$ & $-5\tilde{c}_{a}+3\tilde{c}_{s}$ & $\left[3996.9,4024.8\right]_{B}$ &  & $\frac{3}{2}(0^{+})$ & $-5c_{a}+3c_{s}$ & $\cdots$\tabularnewline
 & $\frac{3}{2}(1^{-})$ & $-\frac{11}{3}\tilde{c}_{a}+3\tilde{c}_{s}$ & $\left[4014.8,4029.7\right]_{B}^{\dagger}$ &  & $\frac{3}{2}(1^{+})$ & $-\frac{11}{3}c_{a}+3c_{s}$ & $\cdots$\tabularnewline
 & $\frac{3}{2}(2^{-})$ & $-\tilde{c}_{a}+3\tilde{c}_{s}$ & $\left[\sim4010.0\right]_{V}^{\sharp}$ &  & $\frac{3}{2}(2^{+})$ & $-c_{a}+3c_{s}$ & $\cdots$\tabularnewline
 & $\frac{3}{2}(3^{-})$ & $3\tilde{c}_{a}+3\tilde{c}_{s}$ & $\cdots$ &  & $\frac{3}{2}(3^{+})$ & $3c_{a}+3c_{s}$ & $\cdots$\tabularnewline
\hline 
\multirow{4}{*}{$\Xi_{c}^{\ast}\bar{\Xi}$ $\left[3963.4\right]$} & $0(1^{-})$ & $-\frac{25}{18}\tilde{c}_{a}-\frac{5}{2}\tilde{c}_{s}$ & $\left[3959.6,3963.1\right]_{V}$ & \multirow{4}{*}{$\Xi_{c}^{\ast}\Xi$} & $0(1^{+})$ & $-\frac{25}{18}c_{a}-\frac{5}{2}c_{s}$ & $\left[3963.3_{-0.5}^{+0.1}\right]_{B}^{\dagger}$\tabularnewline
 & $0(2^{-})$ & $\frac{5}{6}\tilde{c}_{a}-\frac{5}{2}\tilde{c}_{s}$ & $\cdots$ &  & $0(2^{+})$ & $\frac{5}{6}c_{a}-\frac{5}{2}c_{s}$ & $\left[3963.3_{-0.7}^{+0.1}\right]_{B}^{\dagger}$\tabularnewline
 & $1(1^{-})$ & $\frac{5}{6}\tilde{c}_{a}+\frac{3}{2}\tilde{c}_{s}$ & $\cdots$ &  & $1(1^{+})$ & $\frac{5}{6}c_{a}+\frac{3}{2}c_{s}$ & $\cdots$\tabularnewline
 & $1(2^{-})$ & $-\frac{1}{2}\tilde{c}_{a}+\frac{3}{2}\tilde{c}_{s}$ & $\left[\sim3825.9\right]_{V}^{\sharp}$ &  & $1(2^{+})$ & $-\frac{1}{2}c_{a}+\frac{3}{2}c_{s}$ & $\cdots$\tabularnewline
\hline 
\multirow{8}{*}{$\Xi_{c}^{\ast}\bar{\Xi}^{\ast}$ $\left[4178.5\right]$} & $0(0^{-})$ & $\frac{25}{6}\tilde{c}_{a}-\frac{5}{2}\tilde{c}_{s}$ & $\cdots$ & \multirow{8}{*}{$\Xi_{c}^{\ast}\Xi^{\ast}$} & $0(0^{+})$ & $\frac{25}{6}c_{a}-\frac{5}{2}c_{s}$ & $\left[4177.0_{-3.1}^{+1.5}\right]_{B}$\tabularnewline
 & $0(1^{-})$ & $\frac{55}{18}\tilde{c}_{a}-\frac{5}{2}\tilde{c}_{s}$ & $\cdots$ &  & $0(1^{+})$ & $\frac{55}{18}c_{a}-\frac{5}{2}c_{s}$ & $\left[4177.3_{-2.4}^{+1.2}\right]_{B}$\tabularnewline
 & $0(2^{-})$ & $\frac{5}{6}\tilde{c}_{a}-\frac{5}{2}\tilde{c}_{s}$ & $\cdots$ &  & $0(2^{+})$ & $\frac{5}{6}c_{a}-\frac{5}{2}c_{s}$ & $\left[4177.7_{-1.2}^{+0.7}\right]_{B}$\tabularnewline
 & $0(3^{-})$ & $-\frac{5}{2}\tilde{c}_{a}-\frac{5}{2}\tilde{c}_{s}$ & $\left[4173.9,4174.2\right]_{B}$ &  & $0(3^{+})$ & $-\frac{5}{2}c_{a}-\frac{5}{2}c_{s}$ & $\left[4178.3_{-1.4}^{+0.2}\right]_{B}^{\dagger}$\tabularnewline
 & $1(0^{-})$ & $-\frac{5}{2}\tilde{c}_{a}+\frac{3}{2}\tilde{c}_{s}$ & $\left[4174.6,4178.5\right]_{B}^{\dagger}$ &  & $1(0^{+})$ & $-\frac{5}{2}c_{a}+\frac{3}{2}c_{s}$ & $\cdots$\tabularnewline
 & $1(1^{-})$ & $-\frac{11}{6}\tilde{c}_{a}+\frac{3}{2}\tilde{c}_{s}$ & $\left[4159.1,4178.5\right]_{V}^{\dagger}$ &  & $1(1^{+})$ & $-\frac{11}{6}c_{a}+\frac{3}{2}c_{s}$ & $\cdots$\tabularnewline
 & $1(2^{-})$ & $-\frac{1}{2}\tilde{c}_{a}+\frac{3}{2}\tilde{c}_{s}$ & $\left[\sim4078.6\right]_{V}^{\dagger}$ &  & $1(2^{+})$ & $-\frac{1}{2}c_{a}+\frac{3}{2}c_{s}$ & $\cdots$\tabularnewline
 & $1(3^{-})$ & $\frac{3}{2}\tilde{c}_{a}+\frac{3}{2}\tilde{c}_{s}$ & $\cdots$ &  & $1(3^{+})$ & $\frac{3}{2}c_{a}+\frac{3}{2}c_{s}$ & $\cdots$\tabularnewline
\hline 
\hline 
\end{tabular}
}
\end{table*}

\subsection{Doubly charmed baryon-(anti)hyperon systems}\label{sec:BQQH}

The results for doubly charmed baryon-(anti)hyperon systems are given in Table~\ref{tab:XiccH}. We obtained several spin multiplets in the lowest isospin $\Xi_{cc}^{(\ast)}\Sigma^{(\ast)}$ and $\Xi_{cc}^{(\ast)}\Xi^{(\ast)}$ systems, and they can be related to the anticharmed strange pentaquarks predicted in Ref.~\cite{Wang:2023eng} within the heavy diquark-antiquark symmetry. A series of bound/virtual states were also obtained in $\Xi_{cc}^{(\ast)}\bar{\Sigma}^{(\ast)}$ and $\Xi_{cc}^{(\ast)}\bar{\Xi}^{(\ast)}$ systems.

\begin{table*}[htbp]
\centering
\caption{The $I(J^{P})$ quantum numbers, effective potentials, and bound/virtual state solutions of the $\Xi_{cc}^{(\ast)}\Sigma^{(\ast)}$, $\Xi_{cc}^{(\ast)}\bar{\Sigma}^{(\ast)}$, $\Xi_{cc}^{(\ast)}\Xi^{(\ast)}$ and $\Xi_{cc}^{(\ast)}\bar{\Xi}^{(\ast)}$ systems. The notations are the same as those in Tables~\ref{tab:SigmacSigmac} and~\ref{tab:SigmacXic}.\label{tab:XiccH}}
\setlength{\tabcolsep}{1.9mm}
{
\begin{tabular}{cccccccc}
\hline 
\hline
Systems $[m_{\text{th}}]$ & $I(J^{P})$ & $V_{\mathcal{H}_1\mathcal{H}_2}^{I,J}$ & $E_{B}/E_{V}$ & Systems & $I(J^{P})$ & $V_{\mathcal{H}_1\mathcal{H}_2}^{I,J}$ & $E_{B}/E_{V}$\tabularnewline
\hline
\multirow{4}{*}{$\Xi_{cc}\bar{\Sigma}$ $\left[4814.7\right]$} & $\frac{1}{2}(0^{-})$ & $-2\tilde{c}_{a}-3\tilde{c}_{s}$ & $\left[4813.1,4814.5\right]_{B}$ & \multirow{4}{*}{$\Xi_{cc}\Sigma$} & $\frac{1}{2}(0^{+})$ & $-2c_{a}-3c_{s}$ & $\left[4813.2_{-2.2}^{+1.3}\right]_{B}$\tabularnewline
 & $\frac{1}{2}(1^{-})$ & $\frac{2}{3}\tilde{c}_{a}-3\tilde{c}_{s}$ & $\cdots$ &  & $\frac{1}{2}(1^{+})$ & $\frac{2}{3}c_{a}-3c_{s}$ & $\left[4812.5_{-1.7}^{+1.3}\right]_{B}$\tabularnewline
 & $\frac{3}{2}(0^{-})$ & $2\tilde{c}_{a}+3\tilde{c}_{s}$ & $\cdots$ &  & $\frac{3}{2}(0^{+})$ & $2c_{a}+3c_{s}$ & $\cdots$\tabularnewline
 & $\frac{3}{2}(1^{-})$ & $-\frac{2}{3}\tilde{c}_{a}+3\tilde{c}_{s}$ & $\left[\sim4745.4\right]_{V}^{\sharp}$ &  & $\frac{3}{2}(1^{+})$ & $-\frac{2}{3}c_{a}+3c_{s}$ & $\cdots$\tabularnewline
\hline 
\multirow{4}{*}{$\Xi_{cc}\bar{\Sigma}^{\ast}$ $\left[5006.1\right]$} & $\frac{1}{2}(1^{-})$ & $-\frac{5}{3}\tilde{c}_{a}-3\tilde{c}_{s}$ & $\left[5004.7,5006.0\right]_{B}$ & \multirow{4}{*}{$\Xi_{cc}\Sigma^{\ast}$} & $\frac{1}{2}(1^{+})$ & $-\frac{5}{3}c_{a}-3c_{s}$ & $\left[5002.9_{-2.5}^{+2.0}\right]_{B}$\tabularnewline
 & $\frac{1}{2}(2^{-})$ & $\tilde{c}_{a}-3\tilde{c}_{s}$ & $\cdots$ &  & $\frac{1}{2}(2^{+})$ & $c_{a}-3c_{s}$ & $\left[5002.0_{-2.3}^{+1.9}\right]_{B}$\tabularnewline
 & $\frac{3}{2}(1^{-})$ & $\frac{5}{3}\tilde{c}_{a}+3\tilde{c}_{s}$ & $\cdots$ &  & $\frac{3}{2}(1^{+})$ & $\frac{5}{3}c_{a}+3c_{s}$ & $\cdots$\tabularnewline
 & $\frac{3}{2}(2^{-})$ & $-\tilde{c}_{a}+3\tilde{c}_{s}$ & $\left[\sim4993.7\right]_{V}^{\dagger}$ &  & $\frac{3}{2}(2^{+})$ & $-c_{a}+3c_{s}$ & $\cdots$\tabularnewline
\hline 
\multirow{4}{*}{$\Xi_{cc}\bar{\Xi}$ $\left[4939.8\right]$} & $0(0^{-})$ & $\frac{5}{6}\tilde{c}_{a}-\frac{5}{2}\tilde{c}_{s}$ & $\cdots$ & \multirow{4}{*}{$\Xi_{cc}\Xi$} & $0(0^{+})$ & $\frac{5}{6}c_{a}-\frac{5}{2}c_{s}$ & $\left[4939.1_{-1.1}^{+0.6}\right]_{B}$\tabularnewline
 & $0(1^{-})$ & $-\frac{5}{18}\tilde{c}_{a}-\frac{5}{2}\tilde{c}_{s}$ & $\left[\sim4924.4\right]_{V}^{\sharp}$ &  & $0(1^{+})$ & $-\frac{5}{18}c_{a}-\frac{5}{2}c_{s}$ & $\left[4939.3_{-0.8}^{+0.5}\right]_{B}$\tabularnewline
 & $1(0^{-})$ & $-\frac{1}{2}\tilde{c}_{a}+\frac{3}{2}\tilde{c}_{s}$ & $\left[\sim4838.5\right]_{V}^{\sharp}$ &  & $1(0^{+})$ & $-\frac{1}{2}c_{a}+\frac{3}{2}c_{s}$ & $\cdots$\tabularnewline
 & $1(1^{-})$ & $\frac{1}{6}\tilde{c}_{a}+\frac{3}{2}\tilde{c}_{s}$ & $\cdots$ &  & $1(1^{+})$ & $\frac{1}{6}c_{a}+\frac{3}{2}c_{s}$ & $\cdots$\tabularnewline
\hline 
\multirow{4}{*}{$\Xi_{cc}\bar{\Xi}^{\ast}$ $\left[5155.0\right]$} & $0(1^{-})$ & $-\frac{25}{18}\tilde{c}_{a}-\frac{5}{2}\tilde{c}_{s}$ & $\left[5154.6,5155.0\right]_{B}^{\dagger}$ & \multirow{4}{*}{$\Xi_{cc}\Xi^{\ast}$} & $0(1^{+})$ & $-\frac{25}{18}c_{a}-\frac{5}{2}c_{s}$ & $\left[5153.6_{-1.7}^{+1.1}\right]_{B}$\tabularnewline
 & $0(2^{-})$ & $\frac{5}{6}\tilde{c}_{a}-\frac{5}{2}\tilde{c}_{s}$ & $\cdots$ &  & $0(2^{+})$ & $\frac{5}{6}c_{a}-\frac{5}{2}c_{s}$ & $\left[5153.0_{-1.6}^{+1.2}\right]_{B}$\tabularnewline
 & $1(1^{-})$ & $\frac{5}{6}\tilde{c}_{a}+\frac{3}{2}\tilde{c}_{s}$ & $\cdots$ &  & $1(1^{+})$ & $\frac{5}{6}c_{a}+\frac{3}{2}c_{s}$ & $\cdots$\tabularnewline
 & $1(2^{-})$ & $-\frac{1}{2}\tilde{c}_{a}+\frac{3}{2}\tilde{c}_{s}$ & $\left[\sim5083.8\right]_{V}^{\sharp}$ &  & $1(2^{+})$ & $-\frac{1}{2}c_{a}+\frac{3}{2}c_{s}$ & $\cdots$\tabularnewline
\hline 
\multirow{4}{*}{$\Xi_{cc}^{\ast}\bar{\Sigma}$ $\left[4899.7\pm15\right]$} & $\frac{1}{2}(1^{-})$ & $\frac{10}{3}\tilde{c}_{a}-3\tilde{c}_{s}$ & $\cdots$ & \multirow{4}{*}{$\Xi_{cc}^{\ast}\Sigma$} & $\frac{1}{2}(1^{+})$ & $\frac{10}{3}c_{a}-3c_{s}$ & $\left[4896.6_{-3.4}^{+2.4}\pm15\right]_{B}$\tabularnewline
 & $\frac{1}{2}(2^{-})$ & $-2\tilde{c}_{a}-3\tilde{c}_{s}$ & $\left[4898.7\pm0.7\pm15\right]_{B}$ &  & $\frac{1}{2}(2^{+})$ & $-2c_{a}-3c_{s}$ & $\left[4898.1_{-2.2}^{+1.3}\pm15\right]_{B}$\tabularnewline
 & $\frac{3}{2}(1^{-})$ & $-\frac{10}{3}\tilde{c}_{a}+3\tilde{c}_{s}$ & $\left[4894.4\pm4.7\pm15\right]_{B}^{\dagger}$ &  & $\frac{3}{2}(1^{+})$ & $-\frac{10}{3}c_{a}+3c_{s}$ & $\cdots$\tabularnewline
 & $\frac{3}{2}(2^{-})$ & $2\tilde{c}_{a}+3\tilde{c}_{s}$ & $\cdots$ &  & $\frac{3}{2}(2^{+})$ & $2c_{a}+3c_{s}$ & $\cdots$\tabularnewline
\hline 
\multirow{8}{*}{$\Xi_{cc}^{\ast}\bar{\Sigma}^{\ast}$ $\left[5091.1\pm15\right]$} & $\frac{1}{2}(0^{-})$ & $5\tilde{c}_{a}-3\tilde{c}_{s}$ & $\cdots$ & \multirow{8}{*}{$\Xi_{cc}^{\ast}\Sigma^{\ast}$} & $\frac{1}{2}(0^{+})$ & $5c_{a}-3c_{s}$ & $\left[5085.3_{-5.1}^{+4.0}\pm15\right]_{B}$\tabularnewline
 & $\frac{1}{2}(1^{-})$ & $\frac{11}{3}\tilde{c}_{a}-3\tilde{c}_{s}$ & $\cdots$ &  & $\frac{1}{2}(1^{+})$ & $\frac{11}{3}c_{a}-3c_{s}$ & $\left[5085.9_{-4.2}^{+3.3}\pm15\right]_{B}$\tabularnewline
 & $\frac{1}{2}(2^{-})$ & $\tilde{c}_{a}-3\tilde{c}_{s}$ & $\cdots$ &  & $\frac{1}{2}(2^{+})$ & $c_{a}-3c_{s}$ & $\left[5086.9_{-2.3}^{+2.0}\pm15\right]_{B}$\tabularnewline
 & $\frac{1}{2}(3^{-})$ & $-3\tilde{c}_{a}-3\tilde{c}_{s}$ & $\left[5080.4\pm0.3\pm15\right]_{B}$ &  & $\frac{1}{2}(3^{+})$ & $-3c_{a}-3c_{s}$ & $\left[5088.3_{-3.2}^{+2.3}\pm15\right]_{B}$\tabularnewline
 & $\frac{3}{2}(0^{-})$ & $-5\tilde{c}_{a}+3\tilde{c}_{s}$ & $\left[5069.3\pm15.1\pm15\right]_{B}$ &  & $\frac{3}{2}(0^{+})$ & $-5c_{a}+3c_{s}$ & $\cdots$\tabularnewline
 & $\frac{3}{2}(1^{-})$ & $-\frac{11}{3}\tilde{c}_{a}+3\tilde{c}_{s}$ & $\left[5082.2\pm9.0\pm15\right]_{B}$ &  & $\frac{3}{2}(1^{+})$ & $-\frac{11}{3}c_{a}+3c_{s}$ & $\cdots$\tabularnewline
 & $\frac{3}{2}(2^{-})$ & $-\tilde{c}_{a}+3\tilde{c}_{s}$ & $\left[\sim5079.1\pm15\right]_{V}^{\sharp}$ &  & $\frac{3}{2}(2^{+})$ & $-c_{a}+3c_{s}$ & $\cdots$\tabularnewline
 & $\frac{3}{2}(3^{-})$ & $3\tilde{c}_{a}+3\tilde{c}_{s}$ & $\cdots$ &  & $\frac{3}{2}(3^{+})$ & $3c_{a}+3c_{s}$ & $\cdots$\tabularnewline
\hline 
\multirow{4}{*}{$\Xi_{cc}^{\ast}\bar{\Xi}$ $\left[5024.8\pm15\right]$} & $0(1^{-})$ & $-\frac{25}{18}\tilde{c}_{a}-\frac{5}{2}\tilde{c}_{s}$ & $\left[5024.1\pm0.7\pm15\right]_{V}$ & \multirow{4}{*}{$\Xi_{cc}^{\ast}\Xi$} & $0(1^{+})$ & $-\frac{25}{18}c_{a}-\frac{5}{2}c_{s}$ & $\left[5024.4_{-1.2}^{+0.4}\pm15\right]_{B}$\tabularnewline
 & $0(2^{-})$ & $\frac{5}{6}\tilde{c}_{a}-\frac{5}{2}\tilde{c}_{s}$ & $\cdots$ &  & $0(2^{+})$ & $\frac{5}{6}c_{a}-\frac{5}{2}c_{s}$ & $\left[5024.1_{-1.2}^{+0.7}\pm15\right]_{B}$\tabularnewline
 & $1(1^{-})$ & $\frac{5}{6}\tilde{c}_{a}+\frac{3}{2}\tilde{c}_{s}$ & $\cdots$ &  & $1(1^{+})$ & $\frac{5}{6}c_{a}+\frac{3}{2}c_{s}$ & $\cdots$\tabularnewline
 & $1(2^{-})$ & $-\frac{1}{2}\tilde{c}_{a}+\frac{3}{2}\tilde{c}_{s}$ & $\left[\sim4925.6\pm15\right]_{V}^{\sharp}$ &  & $1(2^{+})$ & $-\frac{1}{2}c_{a}+\frac{3}{2}c_{s}$ & $\cdots$\tabularnewline
\hline 
\multirow{8}{*}{$\Xi_{cc}^{\ast}\bar{\Xi}^{\ast}$ $\left[5240.0\pm15\right]$} & $0(0^{-})$ & $\frac{25}{6}\tilde{c}_{a}-\frac{5}{2}\tilde{c}_{s}$ & $\cdots$ & \multirow{8}{*}{$\Xi_{cc}^{\ast}\Xi^{\ast}$} & $0(0^{+})$ & $\frac{25}{6}c_{a}-\frac{5}{2}c_{s}$ & $\left[5236.9_{-3.8}^{+2.6}\pm15\right]_{B}$\tabularnewline
 & $0(1^{-})$ & $\frac{55}{18}\tilde{c}_{a}-\frac{5}{2}\tilde{c}_{s}$ & $\cdots$ &  & $0(1^{+})$ & $\frac{55}{18}c_{a}-\frac{5}{2}c_{s}$ & $\left[5237.3_{-3.0}^{+2.1}\pm15\right]_{B}$\tabularnewline
 & $0(2^{-})$ & $\frac{5}{6}\tilde{c}_{a}-\frac{5}{2}\tilde{c}_{s}$ & $\cdots$ &  & $0(2^{+})$ & $\frac{5}{6}c_{a}-\frac{5}{2}c_{s}$ & $\left[5238.0_{-1.6}^{+1.2}\pm15\right]_{B}$\tabularnewline
 & $0(3^{-})$ & $-\frac{5}{2}\tilde{c}_{a}-\frac{5}{2}\tilde{c}_{s}$ & $\left[5233.3\pm0.2\pm15\right]_{B}$ &  & $0(3^{+})$ & $-\frac{5}{2}c_{a}-\frac{5}{2}c_{s}$ & $\left[5238.8_{-2.2}^{+1.1}\pm15\right]_{B}$\tabularnewline
 & $1(0^{-})$ & $-\frac{5}{2}\tilde{c}_{a}+\frac{3}{2}\tilde{c}_{s}$ & $\left[5236.8\pm3.0\pm15\right]_{B}^{\dagger}$ &  & $1(0^{+})$ & $-\frac{5}{2}c_{a}+\frac{3}{2}c_{s}$ & $\cdots$\tabularnewline
 & $1(1^{-})$ & $-\frac{11}{6}\tilde{c}_{a}+\frac{3}{2}\tilde{c}_{s}$ & $\left[5234.3\pm5.7\pm15\right]_{V}^{\dagger}$ &  & $1(1^{+})$ & $-\frac{11}{6}c_{a}+\frac{3}{2}c_{s}$ & $\cdots$\tabularnewline
 & $1(2^{-})$ & $-\frac{1}{2}\tilde{c}_{a}+\frac{3}{2}\tilde{c}_{s}$ & $\left[\sim5171.0\pm15\right]_{V}^{\sharp}$ &  & $1(2^{+})$ & $-\frac{1}{2}c_{a}+\frac{3}{2}c_{s}$ & $\cdots$\tabularnewline
 & $1(3^{-})$ & $\frac{3}{2}\tilde{c}_{a}+\frac{3}{2}\tilde{c}_{s}$ & $\cdots$ &  & $1(3^{+})$ & $\frac{3}{2}c_{a}+\frac{3}{2}c_{s}$ & $\cdots$\tabularnewline
\hline 
\hline 
\end{tabular}
}
\end{table*}

\section{Summary and outlook}\label{sec:sum}

In the past two decades, numerous heavy flavor near-threshold hadrons have been discovered in experiments, which are considered as promising candidates for molecular tetraquarks and pentaquarks. In this paper, we systematically study the possible molecular hexaquarks in the dibaryon systems. These dibaryon systems are composed of charmed baryons $[\Sigma_c^{(\ast)}$, $\Xi_c^{(\prime,\ast)}]$, doubly charmed baryons $[\Xi_{cc}^{(\ast)}]$, and hyperons $[\Sigma^{(\ast)}$, $\Xi^{(\ast)}]$ combined pairwise. We consider all possible particle-particle and particle-antiparticle combinations. We calculate the mass spectrum of molecular hexaquarks in different systems, which can be regarded as the \heavy flavor counterparts of the deuteron.

Just as the mass spectrum of conventional hadrons can be described using a set of parameters within the quark potential models, in the molecular picture, there shall also exist an underlying relation between the hadronic molecules of different dihadron systems, rather than each hadronic molecule being an independent entity. Based on our previous research~\cite{Wang:2023hpp}, we use the effective potentials at the quark level to describe the residual strong interactions of the S-wave dihadron systems. The interactions in the baryon-baryon and baryon-antibaryon systems are respectively governed by the correlations of $q_1q_2$ and $\bar{q}_1q_2$ ($q=u,d$), for which we only need two parameters to describe the corresponding interaction strengths, and these two parameters are determined by the well-established states, such as $P_\psi^N$, $T_{cc}$, and $X(3872)$, $Z_c(3900)$, respectively.

It can be easily inferred within our model that if the molecular tetraquarks and pentaquarks exist, then the molecular hexaquarks must also exist. Experimental searches for these hexaquark states will help reveal whether the nature prefers to construct higher-level structural units, i.e., hadronic molecular states, using the color-singlet hadrons, or whether it merely favors putting the quarks into a bag, i.e., compact multiquark states.

\section*{Acknowledgement}
This work is supported by the National Natural Science Foundation of China under Grants No.~12105072, No.~12305090, No.~11975033 and No.~12070131001. B. Wang was also supported by the Start-up Funds for Young Talents of Hebei University (No.~521100221021). This project is also funded by the Deutsche Forschungsgemeinschaft (DFG, German Research Foundation, Project ID 196253076-TRR 110).

\bibliography{refs}
\end{document}